\documentclass{aa}

\usepackage{graphicx}
\usepackage{txfonts}
\usepackage{newtxtext,newtxmath}
\usepackage[T1]{fontenc}
\usepackage[export]{adjustbox}
\usepackage{graphicx}	
\usepackage{amsmath}	
\usepackage{amssymb}	

\usepackage{etoolbox}
\usepackage{xcolor, ulem}

\providetoggle{long}
\settoggle{long}{true}
\usepackage{subcaption}

\providetoggle{test}
\settoggle{test}{true}

\newcommand{\sfrd}{$\Sigma_{\rm SFR}$\xspace}
\newcommand{\mfsfr}{$B$--$\Sigma_{\rm SFR}$\xspace}
\newcommand{\mfhi}{$B$--$\Sigma_{\rm H\,I}$\xspace}
\newcommand{\mfhtwo}{$B$--$\Sigma_{\rm H_2}$\xspace}
\newcommand{\mfgas}{$B$--$\Sigma_{\rm H\,{\sc I}+H_2}$\xspace}
\newcommand{\mhid}{$\Sigma_{\rm H\,{\sc I}}$\xspace}
\newcommand{\mhtwod}{$\Sigma_{\rm H_2}$\xspace}
\newcommand{\mgasd}{$\Sigma_{\rm H\,{\sc I}+H_2}$\xspace}

\usepackage[colorlinks,linkcolor=blue,anchorcolor=blue,citecolor=blue]{hyperref}
\newcommand{\map}{Equipartition magnetic field strength as grey-scale map (a) and as contour map (b).}

\newcommand{\corr}{Overlay of the equipartition magnetic field strength as contours on the H\,{\sc i} mass surface density (a), $\rm H_2$ mass surface density (c) and combined H\,{\sc i} and $\rm H_2$ mass surface density (e). Spatially resolved \mfhi relation (b), \mfhtwo relation (d) and \mfgas relation (f).}

\begin{document}

\title{Nearby galaxies in the LOFAR Two-metre Sky Survey}
\subtitle{II. The magnetic field--gas relation}

    \author{V.~Heesen\inst{1}
    \and
    T.-L. Klocke\inst{1}
    \and
    M.~Br\"{u}ggen\inst{1}
    \and
    F.~S.~Tabatabaei\inst{2}
    \and
    A.~Basu\inst{3}
    \and
    R.~Beck\inst{4}
    \and
    A.~Drabent\inst{3}
    \and
    B.~Nikiel-Wroczy\'nski\inst{5}
    \and
    R.~Paladino\inst{6}
    \and
    S.~Schulz\inst{1}
    \and
    M.~Stein\inst{7}
    }

    \institute{University of Hamburg, Hamburger Sternwarte, Gojenbergsweg 112, 21029 Hamburg, Germany\\
    \email{volker.heesen@hs.uni-hamburg.de}
    \and
    School of Astronomy, Institute for Research in Fundamental Sciences, 19395-5531 Tehran, Iran
    \and
    Th\"uringer Landessternwarte, Sternwarte 5, 07778 Tautenburg, Germany 
    \and
    Max-Planck-Institute for Radioastronomy, Auf dem H\"ugel 69, 53121 Bonn, Germany
    \and
    Astronomical Observatory of the Jagiellonian University, ul. Orla 171, 30-244 Krak\'ow, Poland
    \and
    INAF–Istituto di Radioastronomia, Via Gobetti 101, 40129 Bologna, Italy
    \and
    Ruhr University Bochum, Faculty of Physics and Astronomy, Astronomical Institute, 44780 Bochum, Germany
    }

\abstract
 {Magnetic fields are key to understand galaxy evolution, regulating stellar feedback and star formation in galaxies.}
{We probe the origin of magnetic fields in late-type galaxies, measuring magnetic field strengths, exploring whether magnetic fields are only passive constituents of the interstellar medium, or whether they are active constituents being part of the local energy equilibrium.}
{We measure equipartition magnetic field strengths in 39 galaxies from LoTSS-DR2 using LOFAR observations at 144~MHz with 6~arcsec angular resolution which  ($0.1$--$0.7$~kpc). For a subset of 9 galaxies, we obtain atomic and molecular mass surface densities using H\,{\sc i} and CO(2-1) data, from the THINGS and HERACLES surveys, respectively. These data are at 13~arcsec angular resolution, which corresponds to $0.3$--$1.2$~kpc at the distances of our galaxies. We measure kinetic energy densities using H\,{\sc i} and CO velocity dispersions.}
{We found a mean magnetic field strength of $3.6$--$12.5$~$\upmu$G with a mean of $7.9\pm 2.0~\rm\upmu G$ across the full sample. The magnetic field strength has the tightest and steepest relation with the total gas surface density with $B\propto \Sigma_{\rm H\,I+H_2}^{0.309\pm 0.006}$. The relation with the star-formation rate surface density and molecular gas surface density has significantly flatter slopes. After accounting for the influence of cosmic-ray transport, we found an even steeper relation of $B\propto \Sigma_{\rm H\,I+H_2}^{0.393\pm 0.009}$.}
{These results suggest that the magnetic field is regulated by a $B$--$\rho$ relation, which has its origin in the saturation of the small-scale dynamo. This is borne out by an agreement of kinetic and magnetic energy densities although local deviations do exist in particular in areas of high kinetic energy densities where the magnetic field is sub-dominant.}


\maketitle

\section{Introduction}
Magnetic fields are a key ingredient for galaxy formation and evolution. They regulate the transport of highly energetic particles, the cosmic rays \citep{BeckerTjus2020}, and are essential in any theory of star formation \citep{Krumholz2019} as they can decelerate the formation of massive stars \citep{Tabatabaei2018}. Magnetic fields need to be taken into account for modelling stellar feedback, as cosmic ray-driven winds are now thought to be behind galactic winds that are very important for galaxy evolution \citep{Veilleux2020}, so that steady state models have been developed \citep[e.g.][]{Breitschwerdt1991,Everett2008,Recchia2016,Yu2020} and simulations include them now as well \citep{Salem2014,Pakmor2016,Girichidis2018,Jacob2018}. Radio continuum observations of nearby galaxies show that high-energy cosmic rays are scattered in the tangled magnetic field near star-forming regions causing winds and outflows \citep{Tabatabaei2017}. Hence, the question of origin and what regulates magnetic fields in galaxies is of wide-ranging importance for understanding galaxies as they appear in today's Universe \citep{Vogelsberger2020}. In particular, it is now acknowledged that magnetic fields in galaxies have a high field strength of the order of  10~$\upmu\rm G$ and so the magnetic field becomes dynamically important in comparison with the other phases of the interstellar medium \citep{Beck2007}.

Observations have been able to reveal some of the structure and role that magnetic fields play in galaxies \citep[see][for an overview]{Beck2015}. There is a variety of processes able to generate the necessary seed fields, which can be distinguished between primordial and astrophysical sources. The amplification of magnetic fields in galaxies is thought to be related to the turbulence in the interstellar medium, which is then referred to as the small-scale dynamo as the amplification happens on a scale smaller than the turbulent injection scale \citep{Beck2012,Rieder2016,Pakmor2017}. The magnetic field strength is expected to saturate following a phase of exponential amplification \citep[e.g.][]{Bhat2019}. This phase is then followed by the ordering of fields due to differential rotation. In order to generate so-called regular fields, the mean-field galactic dynamo is the leading theory which operates at scales larger than the turbulence injection scale \citep[e.g.][]{Brandenburg2005,Chamandy2016}. Observations indicate approximate energy equipartition between the magnetic energy density and the kinetic energy density of neutral gas in the interstellar medium \citep{Beck2007,Tabatabaei2008b,Basu2013,Beck2015a}.

The magnetic field is closely related to star formation, and \citet{Schleicher2013} propose a theoretical explanation for the observed relation between the magnetic field strength ($B$) and the star-formation rate surface density (\sfrd). This $B$--SFR relation is $B\propto \Sigma_{\rm SFR} ^{1/3}$. They assumed that the kinetic energy density of the gas is regulated by the star formation and so this relation is the result of a saturating small-scale dynamo. However, as \citet{Steinwandel2020} point out, a similar relation is also found when the magnetic field is amplified by adiabatic compression only with the Kennicutt--Schmidt slopes of $1.4$ for the atomic gas and $1.0$ for the molecular gas.  In galaxy samples, the magnetic field strength has so far been only analysed for global measurements \citep{Niklas1995,Chyzy2011,Tabatabaei2017}, although there are a few case studies for spatially resolved observations of individual galaxies \citep[e.g.][]{Tabatabaei2008b,Chyzy2008,Tabatabaei2013} and several more with reviews of such observations presented in \citet{Fletcher2010} and \citet{Beck2019}. A previous study using LOFAR data for M~51 was already presented by \citet{Mulcahy2014} who found magnetic fields with a strength up of to 30 $\upmu$G in the central region, 10--20 $\upmu$G in the spiral arms, and 10--15 $\upmu$G in the inter-arm regions. What is so far missing, however, is a survey of galaxies, where the magnetic field strength has been calculated on a spatially resolved basis with a consistent approach. With this work, we attempt to fill this gap. We use low-frequency radio continuum observations obtained with the LOw Frequency ARay \citep[LOFAR;][]{vanHaarlem2013}, which has the advantage of low thermal emission, and take advantage of an interferometric telescope giving us good angular and spatial resolution. Our work analyses the first statistically meaningful sample of galaxies to estimate the magnetic field strength and study both the \mfsfr and \mfgas relations at $0.3$--$1.2$~kpc spatial scales. 

Our galaxy sample consists of a mix of galaxies from `The SIRTF Nearby Galaxies Survey'
\citep[SINGS;][]{Kennicutt2003}, the `Continuum Halos in Nearby Galaxies: An EVLA Survey'  \citep[CHANG-ES][]{Irwin2012} and `Key Insights on Nearby Galaxies: A Far-Infrared Survey with {\it Herschel}' \citep[KINGFISH;][]{Kennicutt2011}. We calculate magnetic field strength maps from LOFAR 144-MHz intensity maps and spectral index maps from \citet[in\ prep.]{Heesen2022}. The galaxies were chosen to have rich ancillary data, including observations from the H\,{\sc i} Nearby Galaxy Survey (THINGS; \citealp{Walter2008}) and the HERA CO-Line Extragalactic Survey (HERACLES; \citealp{Leroy2009}). Both are surveys, which trace the atomic (H\,{\sc i}) and molecular (H\textsubscript{2}) hydrogen in the ISM, respectively, and we will use their results for the comparison of the magnetic field to the surface and energy densities of our sample. We also use maps of the star formation rate (SFR) by \citet{Leroy2008}, who calculated the SFR surface density in 23 nearby galaxies using far-ultraviolet data from the {\it Galaxy Evolution Explorer} \citep[{\it GALEX};][]{GildePaz2007} SINGS 24-$\upmu$m maps. 

This paper is organised as follows. In Section~\ref{s:data}, we present an overview of both the radio continuum and ancillary data. Section~\ref{s:methodology} summarises the methodology used including the calculation of equipartition magnetic field strengths, a consideration of thermal emission and accounting for cosmic-ray transport. We present out results in Section~\ref{s:results} and the discussion in Section~\ref{s:discussion}. We conclude in Section~\ref{s:conclusions}. The Appendix contains more information about the correct approach to calculate equipartition magnetic field strengths (Appendix~\ref{as:equipartition_method}), the atlas of magnetic fields in galaxies (Appendix~\ref{as:atlas_of_magnetic_fields_in_galaxies}), the magnetic field--gas relations in individual sample galaxies (Appendix~\ref{as:correlations}) and radial magnetic field profiles for moderately inclined galaxies (Appendix~\ref{as:radial_profile}).

\begin{table*}
	\centering
	\caption{Magnetic field strengths and ancillary data in the LoTSS-DR2 sample.}
	\label{tab:data-table}
	\begin{tabular}{l ccc cc cc ccc} 
		\hline
                Galaxy   & $i$         & $d$   & $l_{\mathrm{eff}}$ & $S_{144~\rm MHz}$ & $\alpha_{\mathrm{int}}$ & $\langle B_{\mathrm{eq}}\rangle$ & $B_{\mathrm{eq}}^{\rm max}$ & $M_{\rm HI}$ & $M_{\rm H_2}$ & $\langle \varv_{\rm t}\rangle^b$ \\
                         & ($\degr$)   & (Mpc) & (kpc)         & (Jy)     &                    & ($\upmu$G)            & ($\upmu$G)         & ($\rm 10^8~M_\sun$)  & ($\rm 10^8~M_\sun$) & ($\rm km\,s^{-1}$)    \\
                (1)      & (2)         & (3)   & (4)           & (5)      & (6)                & (7)                 & (8)              & (9)        &  (10)     &   (11)     \\ \hline
                NGC 855  & 74          & 9.73  & 5.1           & 0.007    & $-0.24 \pm 0.09$   & $8.1 \pm 4.1$       & $11.6 \pm 6.0$   &            &           &            \\
		NGC 891  & 90          & 9.1   & 12.6          & 2.985    & $-0.63 \pm 0.04$   & $7.9 \pm 1.9$       & $18.4 \pm 4.0$   &            &           &            \\
		NGC 925  & 50          & 9.12  & 2.2           & 0.197    & $-0.46 \pm 0.06$   & $6.2 \pm 2.2$       & $20.3 \pm 6.3$   &  45.8      &  1.8      &  11.0/6.0  \\
		NGC 2683 & 79          & 6.27  & 4.7           & 0.284    & $-0.66 \pm 0.06$   & $7.2 \pm 2.1$       & $15.7 \pm 3.4$   &            &           &            \\
		NGC 2798 & 75          & 25.8  & 5.4           & 0.220    & $-0.46 \pm 0.06$   & $12.5 \pm 4.5$      & $33.8 \pm 12.2$  &            &           &            \\
		NGC 2820 & 90          & 26.5  & 6.9           & 0.251    & $-0.64 \pm 0.04$   & $6.6 \pm 1.5$       & $15.9 \pm 3.4$   &            &           &            \\
		NGC 2841 & 69          & 14.1  & 3.9           & 0.463    & $-0.72 \pm 0.05$   & $6.4 \pm 1.5$       & $20.2 \pm 6.2$   &  85.8      &  11.2     &  10.0/8.0  \\
		NGC 2976 & 54          & 3.55  & 2.4           & 0.109    & $-0.49 \pm 0.05$   & $8.3 \pm 2.5$       & $14.9 \pm 5.6$   &   1.36     &  0.7      &  11.0/6.0  \\
		NGC 3003 & 90          & 25.4  & 13.9          & 0.132    & $-0.59 \pm 0.06$   & $5.3 \pm 1.9$       & $10.7 \pm 2.3$   &            &           &            \\
		NGC 3077 & 38          & 3.83  & 1.8           & 0.059    & $-0.31 \pm 0.06$   & $11.0 \pm 3.9$      & $25.4 \pm 9.2$   &   8.81$^a$ &  0.03$^a$ &            \\
		NGC 3079 & 88          & 20.6  & 13.0          & 3.707    & $-0.67 \pm 0.04$   & $8.7 \pm 1.9$       & $41.0 \pm 8.9$   &            &           &            \\
		NGC 3184 & 29          & 11.7  & 1.6           & 0.337    & $-0.62 \pm 0.05$   & $7.5 \pm 2.1$       & $14.9 \pm 4.7$   &   30.7     &  12.0     &  10.0/8.0  \\
		NGC 3198 & 72          & 14.1  & 4.5           & 0.099    & $-0.46 \pm 0.06$   & $6.0 \pm 2.1$       & $13.8 \pm 4.3$   &  101.7     &  4.1      &  17.0/11.0 \\
		NGC 3265 & 47          & 19.6  & 2.1           & 0.023    & $-0.34 \pm 0.06$   & $10.3 \pm 3.7$      & $22.7 \pm 8.2$   &            &           &            \\
		NGC 3432 & 85          & 9.42  & 5.0           & 0.274    & $-0.58 \pm 0.04$   & $7.4 \pm 1.9$       & $12.9 \pm 2.8$   &            &           &            \\
		NGC 3448 & 78          & 24.5  & 6.3           & 0.173    & $-0.62 \pm 0.04$   & $9.2 \pm 2.2$       & $19.6 \pm 4.2$   &            &           &            \\
		NGC 3556 & 81          & 14.09 & 12.5          & 1.065    & $-0.62 \pm 0.04$   & $7.1 \pm 1.7$       & $14.1 \pm 3.0$   &            &           &            \\
		NGC 3877 & 85          & 17.7  & 9.3           & 0.147    & $-0.55 \pm 0.04$   & $6.8 \pm 1.8$       & $12.4 \pm 2.7$   &            &           &            \\
		NGC 3938 & 36          & 17.9  & 1.8           & 0.334    & $-0.67 \pm 0.05$   & $8.9 \pm 2.2$       & $12.9 \pm 4.1$   &            &           &            \\
		NGC 4013 & 88          & 16.0  & 8.1           & 0.144    & $-0.60 \pm 0.04$   & $7.1 \pm 1.9$       & $18.4 \pm 4.0$   &            &           &            \\
		NGC 4096 & 82          & 10.32 & 5.9           & 0.169    & $-0.60 \pm 0.04$   & $7.1 \pm 1.8$       & $11.2 \pm 2.4$   &            &           &            \\
		NGC 4125 & 36          & 24.77 & 1.7           & 0.012    & $-0.84 \pm 0.07$   & $9.0 \pm 2.4$       & $15.5 \pm 3.5$   &            &           &            \\
		NGC 4157 & 83          & 15.6  & 8.3           & 0.921    & $-0.69 \pm 0.04$   & $7.6 \pm 1.6$       & $15.4 \pm 3.2$   &            &           &            \\
		NGC 4217 & 86          & 20.6  & 11.7          & 0.427    & $-0.63 \pm 0.04$   & $7.4 \pm 1.7$       & $17.4 \pm 3.7$   &            &           &            \\
		NGC 4244 & 90          & 4.4   & 7.2           & 0.049    & $-0.57 \pm 0.05$   & $3.6 \pm 1.1$       & $10.0 \pm 2.6$   &            &           &            \\
		NGC 4449 & 54          & 4.02  & 2.4           & 0.863    & $-0.49 \pm 0.06$   & $10.0 \pm 3.6$      & $30.9 \pm 11.2$  &   11.0$^a$ & 0.06$^a$  &            \\
		NGC 4625 & 23          & 9.3   & 1.5           & 0.017    & $-0.35 \pm 0.15$   & $8.1 \pm 6.7$       & $11.4 \pm 9.5$   &            &           &            \\
		NGC 4631 & 85          & 7.62  & 11.7          & 4.031    & $-0.65 \pm 0.04$   & $6.6 \pm 1.5$       & $24.8 \pm 5.4$   &            &           &            \\
		NGC 4725 & 43          & 11.9  & 1.9           & 0.154    & $-0.56 \pm 0.06$   & $6.5 \pm 2.3$       & $23.9 \pm 7.4$   &            &           &            \\
		NGC 4736 & 44          & 4.66  & 1.9           & 0.803    & $-0.45 \pm 0.05$   & $12.1 \pm 3.7$      & $33.5 \pm 7.7$   &   4.00     & 4.5       & 15.0/10.0  \\
		NGC 5033 & 64          & 17.13 & 3.2           & 1.133    & $-0.66 \pm 0.05$   & $7.4 \pm 1.9$       & $31.1 \pm 9.6$   &            &           &            \\
		NGC 5055 & 51          & 7.94  & 2.2           & 2.262    & $-0.75 \pm 0.05$   & $9.1 \pm 2.1$       & $24.4 \pm 5.4$   &   91.0     & 7.3       & 11.0/9.0   \\
		NGC 5194 & 20          & 8.0   & 1.5           & 6.707    & $-0.67 \pm 0.04$   & $12.0 \pm 2.6$      & $41.1 \pm 9.2$   &   25.4     & 25.1      &            \\
		NGC 5195 & 20          & 8.0   & 1.5           & 1.084    & $-0.71 \pm 0.06$   & $11.6 \pm 3.1$      & $29.9 \pm 8.7$   &            &           &            \\
		NGC 5297 & 89          & 40.4  & 13.0          & 0.107    & $-0.72 \pm 0.04$   & $4.8 \pm 1.0$       & $7.5 \pm 1.5$    &            &           &            \\
		NGC 5457 & 30          & 6.7   & 1.6           & 2.493    & $-0.55 \pm 0.06$   & $8.0 \pm 2.9$       & $28.8 \pm 10.4$  &  141.7$^a$ &           &            \\
		NGC 5474 & 37          & 6.8   & 1.8           & 0.008    & $-0.22 \pm 0.06$   & $7.5 \pm 2.7$       & $10.1 \pm 3.6$   &            &           &            \\
		NGC 5907 & 90          & 16.8  & 18.0          & 0.492    & $-0.69 \pm 0.04$   & $5.1 \pm 1.1$       & $21.2 \pm 3.3$   &            &           &            \\
		NGC 7331 & 77          & 14.5  & 6.2           & 2.077    & $-0.72 \pm 0.02$   & $8.0 \pm 1.3$       & $16.3 \pm 5.0$   &   91.3     & 45.7      &            \\
                \hline
        \end{tabular}
        \flushleft
            {\small {\bf Notes.} Column (1) Name of the galaxy; (2) inclination angle with respect to the line-of-sight; (3) distance; (4) path length; (5) integrated flux density at 144~MHz; (6) integrated radio spectral index; (7) mean magnetic field strength; (8) maximum magnetic field strength; (9) H\,{\sc i} mass from \citet{Walter2008}; (10) $\rm H_2$ mass from \citet{Leroy2009}; (11) mean turbulent velocity dispersion for H\,{\sc i}/$\rm H_2$ gas from \citet{Mogotsi2016}. Other data are from \citet{Heesen2022}. $^a$ the CO detections of NGC 3077 and 4449 were marginal, so that these galaxies were left out of the analysis of the ancillary data (Sect.~\ref{ss:ancillary_data}); for NGC 5474 we have only an H\,{\sc i} map, so omitted this galaxy as well for consistency. $^b$ velocity dispersion for the atomic/molecular gas, respectively.}
\end{table*}

\section{Data}
\label{s:data}

\subsection{LOFAR Two-metre Sky Survey}
The Low Frequency Array (LOFAR) is a radio interferometer which consists of multiple stations with combined thousands of radio antennas spread across Europe with its core in the Netherlands. The LOFAR Two-metre Sky Survey (LoTSS; \citealp{Shimwell2017}) is a survey observing the northern sky at a frequency of 120-168 MHz. It uses the LOFAR High Band Antennas (HBA), which are sensitive in a range of 120-240 MHz. Currently, only the stations in the Netherlands are included in the measurements, which results in a resolution limit of $\sim 6$~arcsec, but in the future, with the so-called LOFAR-VLBI, the resolution could be improved up to 0.3~arcsec for small sources. The first LoTSS data release, which is described by \citet{Shimwell2019}, was released in 2019 and included around of 20\,\% of the northern sky. In this paper, we will use data of the second LoTSS data release \citep[LoTSS-DR2;][]{Shimwell2022}. The data consists of maps of 39 nearby galaxies at a frequency of 144 MHz, which were presented by \citet{Heesen2022}, and we will use the data to calculate spatially resolved magnetic field strength maps.

We used the LoTSS-DR2 144-MHz maps at 6~arcsec angular resolution as presented by \citet{Heesen2022}. We also use their radio spectral index maps, or if not available, the integrated radio spectral index. Since the equipartition formula is only valid for a spectral index in the range of $-0.6 \lessapprox \alpha \lessapprox -1.1$, values outside this range were clipped to the boundaries (see Appendix~\ref{as:equipartition_method} for a discussion). In some galaxies the radio spectral index error is larger than the allowed spectral index range, so we clipped the error at a value of $1.0$. The radio continuum intensity maps were masked below $3\sigma_{6\arcsec}$, where $\sigma_{6\arcsec}$ is the rms noise of the 6-arcsec radio continuum map. We note that the radio spectral index maps have a lower angular resolution of 20~arcsec (for a few galaxies, slightly worse), which is not matched to the intensity maps.

\subsection{Ancillary data}
\label{ss:ancillary_data}

For a subset of 9 galaxies (NGC 925, 2841, 2976, 3184, 3198, 4736, 5055, 5194 and 7331), we use a range of ancillary data for further analysis. We use H\,{\sc i} maps from the THINGS survey \citep{Walter2008} including moment 0 (H\,{\sc i} surface mass density) and moment 2 (velocity dispersion). We also use SFR surface density maps created from {\it Spitzer} 24-$\upmu\rm m$ and {\it GALEX} 156-nm emission \citep{Leroy2008}. To calculate $\rm H_2$ masses, we use CO ($J=2\rightarrow 1$) maps from the HERACLES survey \citep{Leroy2009}. For details of the ancillary data see Table~\ref{tab:data-table}. Since the angular resolution of the THINGS maps is approximately 6~arcsec, the full resolution of the LoTSS data can be exploited. The \sfrd and CO maps have a resolution of $\approx$13~arcsec, so that we need to lower the resolution of the LoTSS maps accordingly.


 
 The 21-cm line emission is then converted into the surface mass density of the atomic gas \mhid where we used a correction factor of 1.36 to account for helium. The CO(2-1) emission is converted to the mass surface density of the molecular gas \mhtwod again correcting with a factor of 1.36 to account for helium \mbox{\citep{Leroy2009}}. For the CO data we assumed a 20\,\% calibration error. We calculate the kinetic energy density of the combined atomic and molecular gas as $u_{\rm H\,I + \mathrm{H}_2}=1/2\,\rho\,\varv_{\rm t}^2$. The velocity dispersion $\varv_\mathrm{t}$ is the average of the CO and H\,{\sc i} velocity dispersion in each individual galaxy by \citet[][see Table~\ref{tab:data-table} for values]{Mogotsi2016}. NGC~5194 and 7331 were not included in their sample so we use their mean velocity dispersions of $11.7$ and $7.3~\rm km\,s^{-1}$ for CO and H\,{\sc i}, respectively. We assume an error of 25\,\% for the velocity dispersion. In order to derive the gas density $\rho$, we assume a path length of $400\pm 40$~pc for the H\,{\sc i} gas and $100\pm 10$~pc for the $\rm H_2$ gas, which we then correct for with the inclination angle using a factor of $1/\cos(i)$. The scale height of the H\,{\sc i} disc in THINGS galaxies was measured by \citet{Bagetakos2011} and a scale height of $\approx$200~pc is a reasonable approximation across the star-forming disc.  This is further motivated by \citet{Imamura1997}, who found that in the Milky Way the transition from $\rm H_2$ near the mid-plane to H\,{\sc i} in the halo happens at a height of $\approx$50~pc, where the H\,{\sc i} gas extends a few hundred parsec away from the mid-plane. 
 
 All our galaxies with ancillary data have angular sizes smaller than 15 arcmin, so that missing spacings for the interferometric H\,{\sc i} observations from THINGS are not an issue \citep{Walter2008}. The CO data from the HERACLES survey were obtained with the 30-m IRAM telescope, so that missing spacings do not have to be considered.

\section{Methodology}
\label{s:methodology}

\subsection{Equipartition magnetic field strength}
\label{ss:equipartition_magnetic_field_strength}

In order to estimate the total magnetic field strength from the intensity of radio continuum emission, one has to assume a relation between the energy densities of the cosmic rays ($u_{\rm CR}$) and the magnetic field ($u_{B}$). In this work, we assume energy equipartition, which ensures that the total energy density $u_{B} + u_{\mathrm{CR}}$ is close to the minimum value possible for a given radio continuum luminosity. The equipartition assumption may fail, in particular on scales smaller than the cosmic ray diffusion length \citep{Seta2019}. Hence, we will attempt a correction for diffusion by using either integrated relations or by smoothing by the diffusion length. We will use the revised equipartition formula from \citet{Beck2005}:\footnote{We use the radio spectral index convention $I_\nu\propto \nu^\alpha$.}

\begin{equation}
	B_{\mathrm{eq,\perp}} = \left( \frac{ 4\pi (2\alpha - 1) (K_0 + 1) I_\nu  E_p^{1+2\alpha} \left(\frac{\nu}{2c_1}\right)^{-\alpha} }{ (2\alpha + 1) c_2(\alpha) l_{\mathrm{eff}}(i) \cdot c_4(i) } \right)^{\frac{1}{3 - \alpha}}, \label{eqn:B_eq}
\end{equation}
with $B_{\mathrm{eq,\perp}}$ the total magnetic field strength in the sky plane, the constant $c_1$, the parameter $c_2$ which depends on spectral index, the proton rest mass $E_p$, and the effective path length $l_{\mathrm{eff}}(i)$ through the source.
The parameter $c_4$ depends on inclination and spectral index. We assume isotropic magnetic fields, so that:
\begin{equation}
	c_4 = \left(\frac{2}{3}\right)^{(1 - \alpha)/2}.
\end{equation}

Equation~\eqref{eqn:B_eq} is valid under the assumption of a constant ratio of protons to electrons, $K_0$. We assume a constant $K_0$,
which should be valid at particle energies of a few GeV. If cosmic-ray electrons suffer from energy losses, $K_0$ is a function of particle energy
at lower as well as at higher energies (see Appendix~\ref{as:equipartition_method}). Also, magnetic field fluctuations lead to an overestimate of the total magnetic field strength \citep{Beck2003}.




For the path length we assume $l_{\mathrm{eff}}$ = 1.4 kpc/$\cos i$ for the mildly inclined galaxies ($i < 78\degr$), which is motivated by the results of \citet{Krause2018}, who found a radio scale height of $1.4\pm0.7$ kpc at $1.5$~GHz (see Sect.~\ref{ss:magnetic_field_strength} for details). $i$ is the inclination angle of the galaxy disc relative to the sky plane ($i=0\degr$ is face-on).
For the edge-on galaxies, i.e. the galaxies from the CHANG-ES survey, $l_{\mathrm{eff}}$ is assumed to be the star-forming radius, as defined by the extent of the 24-$\upmu$m emission \citep{Wiegert2015}.
Since
$l_{\mathrm{eff}}$ and $K_0$ are only estimates, we are using a 50\,\% uncertainty on both values to calculate the error of the magnetic field. For the calculation of a galaxy-wide average of the magnetic field strength $\langle B_{\rm eq}\rangle$, we use the integrated flux density $S_{6\arcsec}$ from the 6~arcsec data and calculate the mean intensity using the 6-arcsec major axis as radius.


\begin{table}
	\centering
	\caption{Influence of the correction for thermal emission on the magnetic field strength.}
        \label{tab:therm}
	\begin{tabular}{lccc}
	        \hline
		Galaxy   & $\sigma_{B}$ & $\langle \Delta B \rangle$ & $B^{\rm max}_{\rm eq}-B^{\rm max}_{\rm nt}$ \\
		         & ($\upmu$G) & ($\upmu$G) & ($\upmu$G)                     \\
		\hline
		NGC 2976 & $2.5$  & $-0.2\pm 1.5$  & $0.0 \pm 7.4$   \\
		NGC 3184 & $2.1$  & $-1.0\pm 1.5$  & $-0.1 \pm 6.6$   \\
		NGC 4736 & $3.7$  & $-2.7\pm 3.5$  & $0.1 \pm 13.0$   \\
		NGC 5055 & $2.1$  & $0.1\pm 4.2$   & $1.7 \pm 9.7$    \\
		NGC 5194 & $2.6$  & $-0.8\pm 5.4$  & $0.1 \pm 15.8$   \\
		\hline
	\end{tabular}
        \flushleft
	    {\small {\bf Notes.} Column (1) name of the galaxy; (2) standard deviation of the magnetic field strength; (3) mean and standard deviation of the difference $\Delta B=B_{\rm eq}-B_{\rm nt}$ between the magnetic field strength without and with thermal correction; (4) difference between the maximum value of the magnetic field strength without and with thermal correction.}
\end{table}

\subsection{Thermal emission}
\label{ss:thermal_emission}

The low-frequency radio continuum emission is dominated by non-thermal synchrotron emission, so we did not correct for thermal emission when calculating equipartition magnetic field strengths. In this Section, we justify this assumption, by exploring the influence of thermal emission for a subset of five galaxies (see Table~\ref{tab:therm}). The thermal and non-thermal components of the radio continuum emission were separated using the thermal radio tracer (TRT) approach in which  the H\,$\alpha$ line emission is used as a template for the free-- free emission after correcting for extinction \citep{Tabatabaei2007,Tabatabaei2013,Tabatabaei2018}. Hence, extinction maps are first constructed for galaxies using the {\it Herschel} PACS 70-$\upmu$m and 160-$\upmu$m data taken as part of the KINGFISH project \citep{Kennicutt2011}. Following \citet{Tabatabaei2013}, the brightness temperature of the free--free radio continuum emission, $T_{\rm b}$, is given by:

\iftoggle{test}{
\begin{figure}[htb]
	\includegraphics{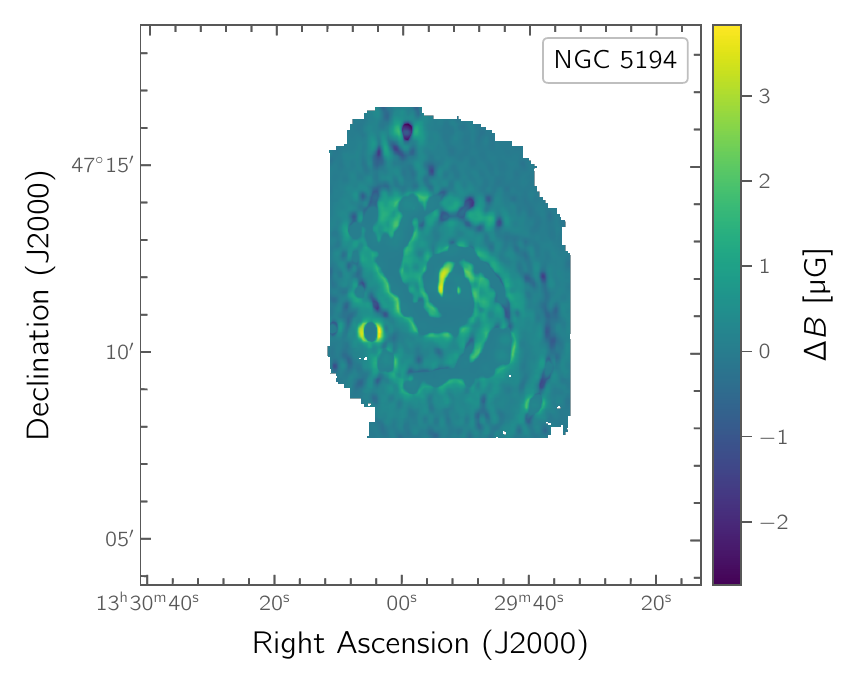}
	\caption{Difference in magnetic field strength $\Delta B = B_{\mathrm{eq}} - B_{\mathrm{nt}}$ due to the thermal correction in NGC~5194/5.}
\label{fig:thermal_diff}
\end{figure}
}

\begin{equation}
{T_{\rm b}=T_{\rm e}(1-e^{-A\,I_{{\rm H}\,\alpha}})}~\rm K,
\end{equation}
where the factor $A$ is:
\begin{equation}
A=3.763\,\left (\frac{\nu}{\rm GHz}\right )^{-2.1}\, \left (\frac{T_{\rm e}}{\rm 10^4~K}\right ) ^{-0.3}\, 10^{\frac{290~\rm K}{T_{\rm e}}}.
\end{equation}
Here, $I_{{\rm H}\alpha}$ is the de-reddened H\,$\alpha$ intensity in units of $\rm erg\,cm^{-2}\,s^{-1}\, sr^{-1}$ and $T_{\rm e}$ the electron temperature. The brightness temperature is derived for $T_{\rm e}=10^4$~K. A 30\,\% variation in $T_{\rm e}$ would change the thermal fraction by about 23\,\%. The thermal free-free emission, obtained after the Kelvin-to-Jy/beam conversion of $T_{\rm b}$, is subtracted from the observed radio continuum emission resulting in a map of the synchrotron emission for each galaxy. We then also re-calculated non-thermal radio spectral index maps between 144 and 1365~MHz using the maps from the WSRT--SINGS survey \citep{Braun2007}. With the non-thermal radio continuum map at 144~MHz and the non-thermal radio spectral index map, we re-calculated the non-thermal magnetic field strength $B_{\rm nt}$.

 In Fig.~\ref{fig:thermal_diff} the difference between the magnetic field strengths $\Delta B = B_{\mathrm{eq}} - B_{\mathrm{nt}}$ is shown for the specific case of NGC~5194 (M~51). The difference in the estimate of the magnetic field is below the 1$\sigma$ difference and $B_{\mathrm{eq}}$ is smaller than $B_{\mathrm{nt}}$ for most parts of the galaxy. Thus, the non-thermally corrected estimate on the magnetic field strength gives a lower bound of the real strength, but lies within the uncertainties. The mean of the difference (see Table~\ref{tab:therm}) is for all galaxies smaller than $\sigma_{B}$, hence it is within the magnetic field uncertainties. Notably also the maximum magnetic field strength in the map is hardly changed by the thermal correction. We hence can justify to neglect thermal correction for the remainder of the sample.




\begin{table}[htp]
	\centering
	\caption{Diffusion lengths used to account for cosmic-ray transport.}
        \label{tab:diffusion}
	\begin{tabular}{lc}
          \hline
	  Galaxy   & {$l_\mathrm{CR}$} \\
		   & [kpc] \\
	  \hline
		NGC 925  & 3.25                \\
		NGC 2841 & 3.23                \\
		NGC 3184 & 3.79                \\
		NGC 3198 & 3.17                \\
		NGC 4736 & 1.23                \\
		NGC 5055 & 3.17                \\
		NGC 5194 & 3.45                \\
                \hline
	\end{tabular}
        \flushleft
            {\small (1) Name of the galaxy; (2) diffusion length.}
\end{table}

\subsection{Correlations}
\label{ss:correlations}

We correlate the maps of the magnetic field strength with our ancillary data. Maps that need to be correlated are transformed to the same coordinate system with the \texttt{imrebin} function of the Common Astronomy Software Applications package \citep[{\sc casa};][]{McMullin2007}. The pixel size is increased to be equal or slightly larger than the FWHM of the lowest resolution map with {\sc casa}'s \texttt{imrebin} function. Maps are masked below $3 \sigma$, although we also tested the influence of using either $4\sigma$ or $5\sigma$ as a cut-off level and found no significant difference with regard to the best-fitting correlations. The retained pixels are displayed in a log-log diagram (Fig.~\ref{fig:relations}), where data points are coloured according to the radio spectral index: data points with $\alpha < -0.85$, a spectral index steeper than the injection index corresponding to aged CRs, are coloured dark blue; data points with $-0.85 \leq \alpha < -0.65$, a spectral index near the injection index corresponding to young CRs, are coloured in light blue; and data points with $-0.65 \leq \alpha$, a spectral index flatter than the injection spectrum, possibly due to e.g. thermal emission, synchrotron self-absorption, or thermal absorption, are coloured in green.  Since the number of data points for the \mfhi correlation is very large, we chose to color-code the point density instead.




We fit a power-law with the following parametrisation:
\begin{equation}
    y = a\left (\frac{x}{c}\right )^b,
\end{equation}
where we consider errors in both axes using orthogonal distance regression (ODR). On the fitted graphs, the 95\,\% (2$\sigma$) confidence interval is highlighted in grey. The theoretical prediction is indicated by a red dashed line and their origin is explained in the corresponding section. Because the spatially resolved relations are dominated by the gas rich `grand spiral' galaxies NGC 5055, 5194/5 and 7331 that each span two orders of magnitude in gas mass surface densities, one may speculate that these galaxies entirely dominate the relations. Hence, we also investigated relations using only the remaining relatively gas poor galaxies NGC 925, 2841, 2976, 3184, 3198 and 4736 shown in Fig.~\ref{fig:relations_poor}. As we will explain in more detail below, these galaxies show relations that are broadly in line with the full sample.






We also present global correlation, where we plot the mean quantities for individual galaxies. For the mean magnetic field strength, we use the mean intensity which is then converted as described in Section~\ref{ss:equipartition_magnetic_field_strength}. For the ancillary data, we then use the integrated values from Table~\ref{tab:data-table}, which are then divided by the surface area to obtain an average value for the galaxy. For the error, we use the standard deviation of the maps. The resulting global relations are presented in Fig.~\ref{fig:mean_relations}.




\subsection{Accounting for the influence of cosmic-ray transport}
\label{ss:correcting_for_cosmic_ray_transport}

At the low frequencies of LOFAR, the effect of cosmic-ray transport can be significant, resulting in the smoothing the radio continuum maps. This is because the cosmic-ray electron lifetime increases at low frequencies and so the cosmic rays travel further from their sites of acceleration. When one studies the spatially resolved radio--SFR relation, the relation becomes sub-linear unlike the integrated radio--SFR relation which is approximately linear when one accounts for the influence of cosmic-ray transport \citep{Smith2021,Heesen2022}. This influence of the cosmic-ray transport on spatially resolved observations \citep{Heesen2014,Heesen2019} is supported further by \citet{Mulcahy2016} who analysed the radial variation of the radio spectral index in NGC~5194. In order to measure the cosmic-ray transport length, we convolved the \sfrd-maps in order to linearize the radio--SFR relation \citep{Berkhuijsen2013}. The process is explained in more detail in \citet{Heesen2019} and we present in Table~\ref{tab:diffusion} the resulting cosmic-ray transport lengths for seven galaxies.


To account for the cosmic-ray transport effect, we convolve the \sfrd- and \mgasd-maps with a Gaussian kernel, converting the cosmic-ray transport length into the standard deviation, $\sigma_{xy}$, as follows:
\begin{equation}
	l_\mathrm{CR} = \frac{1}{2} \left( 2\sqrt{2\ln2} \right) \sigma_{xy}.
\end{equation}
We use {\sc AstroPy}'s \texttt{convolve\_fft} for the convolution and analyse the correlations as described in Section~\ref{ss:correlations}. The resulting relations where we accounted for the influence of cosmic-ray transport are presented in Fig.~\ref{fig:smooth_relations}.


\begin{table}[!htbp]
	\centering
	\caption{Radial scale length of the magnetic field.}
        \label{tab:radial}
	\begin{tabular}{l ccc}
	        \hline
		Galaxy   & $B_0$ & $r_{\rm B}$ & $R_{25} $ \\
		         & ($\upmu$G) & (kpc) & (kpc) \\
		\hline
                NGC 855  & $11.1 \pm 0.2$ & $3.9 \pm 0.6$      & 13.9 \\
                NGC 925  & $11.1 \pm 0.2$ & $>100$             & 15.0 \\
                NGC 2798 & $25.8 \pm 3.7$ & $3.1 \pm 0.6$      & 10.1 \\
                NGC 2841 & $9.0 \pm 0.3$  & $56.9 \pm 15.0$    & 13.5 \\
                NGC 2976 & $9.8 \pm 0.2$  & N/A                & 2.6 \\
                NGC 3077 & $12.5 \pm 3.2$ & N/A                & 3.0 \\
                NGC 3184 & $11.0 \pm 0.2$ & $73.7 \pm 18.3$    & 12.8 \\
                NGC 3198 & $10.2 \pm 0.5$ & $33.0 \pm 14.4$    & 18.0 \\
                NGC 3265 & $12.9 \pm 5.6$ & N/A                & 27.9 \\
                NGC 3448 & $18.0 \pm 0.4$ & $6.5 \pm 0.2$      & 17.5 \\
                NGC 3938 & $11.1 \pm 0.2$ & $>100$             & 12.8 \\
                NGC 4125 & $14.4 \pm 1.0$ & $3.2 \pm 0.6$      & 23.8 \\
                NGC 4449 & $14.9 \pm 0.5$ & $9.8 \pm 1.7$      & 3.2  \\
                NGC 4625 & $11.1 \pm 0.2$ & $>100$             & 2.2 \\
                NGC 4725 & $11.0 \pm 0.3$ & $88.6 \pm 21.3$    & 18.2 \\
                NGC 4736 & $19.4 \pm 0.8$ & $4.1 \pm 0.4$      & 8.3 \\
                NGC 5033 & $11.9 \pm 0.8$ & $31.8 \pm 6.6$     & 25.2 \\ 
                NGC 5055 & $19.8 \pm 0.5$ & $10.4 \pm 0.4$     & 15.0 \\
                NGC 5194 & $21.5 \pm 0.6$ & $14.0 \pm 0.6$     & 15.8 \\
                NGC 5195 & $15.4 \pm 0.9$ & $10.2 \pm 2.5$     & 7.3 \\
                NGC 5457 & $11.7 \pm 0.2$ & N/A                & 5.8 \\
                \hline
	\end{tabular}
        \flushleft
	    {\small {\bf Notes.} Column (1) Name of the galaxy; (2) magnetic field strength in centre; (3) radial scale length; (4) radius as projected $D_{25}/2$.}
\end{table}

\iftoggle{test}{
\begin{figure*}[htp]
\centering
    \begin{subfigure}[t]{0.02\textwidth}
        \textbf{(a)}    
    \end{subfigure}
    \begin{subfigure}[t]{0.47\linewidth}
        \includegraphics[width=\linewidth,valign=t]{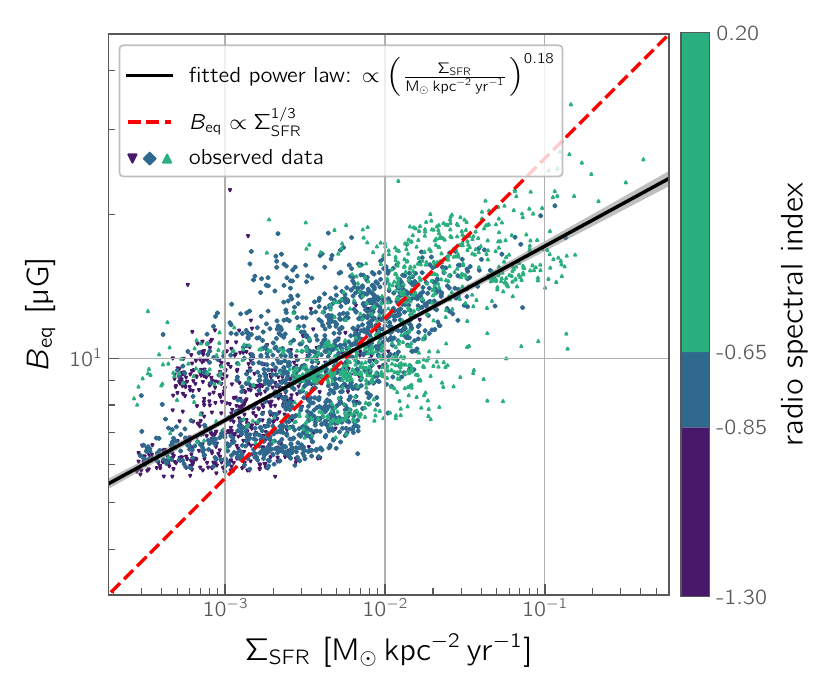}
    \end{subfigure}
    \begin{subfigure}[t]{0.02\textwidth}
        \textbf{(b)}    
    \end{subfigure}
    \begin{subfigure}[t]{0.47\linewidth}
        \includegraphics[width=0.94\linewidth,valign=t]{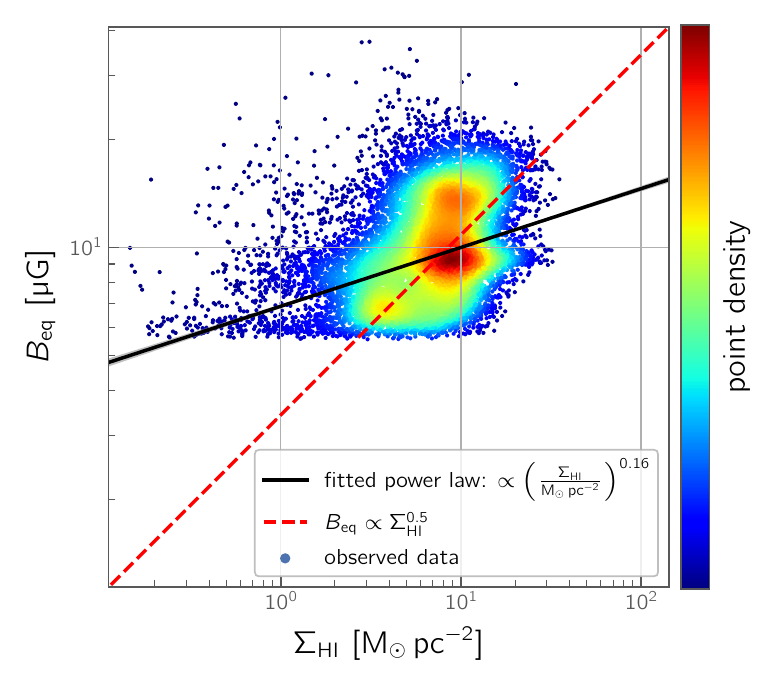}
    \end{subfigure}
    \\
    \begin{subfigure}[t]{0.02\textwidth}
        \textbf{(c)}    
    \end{subfigure}
    \begin{subfigure}[t]{0.47\linewidth}
        \includegraphics[width=\linewidth,valign=t]{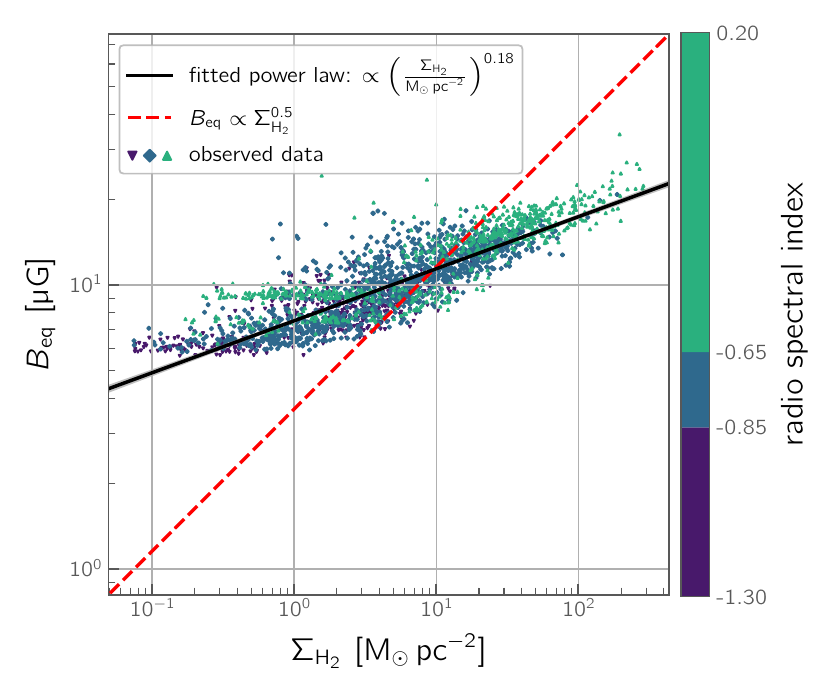}
    \end{subfigure}
    \begin{subfigure}[t]{0.02\textwidth}
        \textbf{(d)}    
    \end{subfigure}
    \begin{subfigure}[t]{0.47\linewidth}
        \includegraphics[width=\linewidth,valign=t]{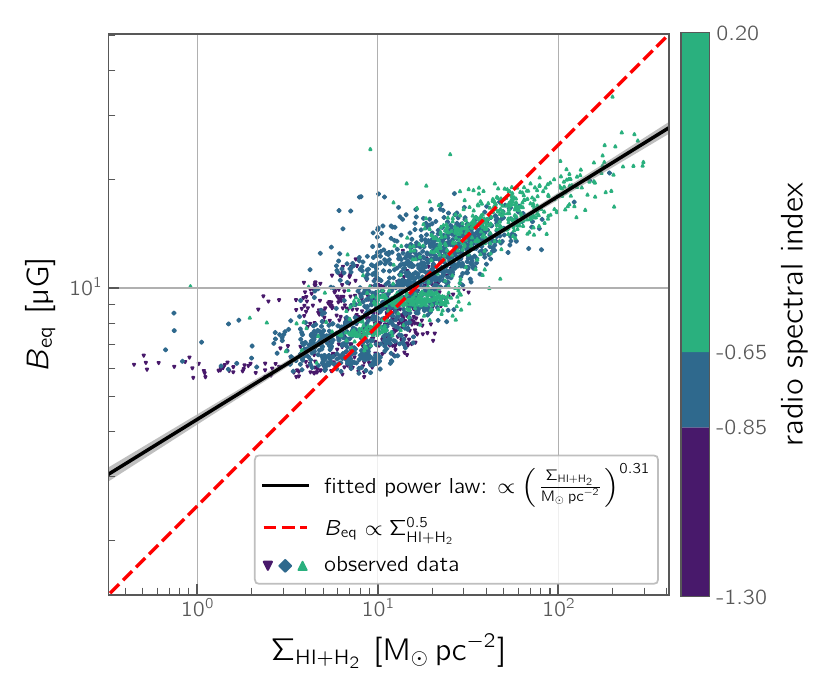}
    \end{subfigure}
    \caption{Spatially resolved relations between the magnetic field strength and the star-formation rate surface density (\mfsfr) (a), the atomic gas mass surface density (\mfhi) (b), molecular gas mass surface density (\mfhtwo) (c), and the combined atomic and molecular gas mass surface density (\mfgas) (d). Data points are coloured according to the radio spectral index and in panel (b) according to the point density. Best-fitting relations are shown as sold lines and the theoretical expectations as dashed lines.}
	\label{fig:relations}
\end{figure*}%

\begin{figure*}[htp]
\centering
    \begin{subfigure}[t]{0.02\textwidth}
        \textbf{(a)}    
    \end{subfigure}
    \begin{subfigure}[t]{0.47\linewidth}
        \includegraphics[width=\linewidth,valign=t]{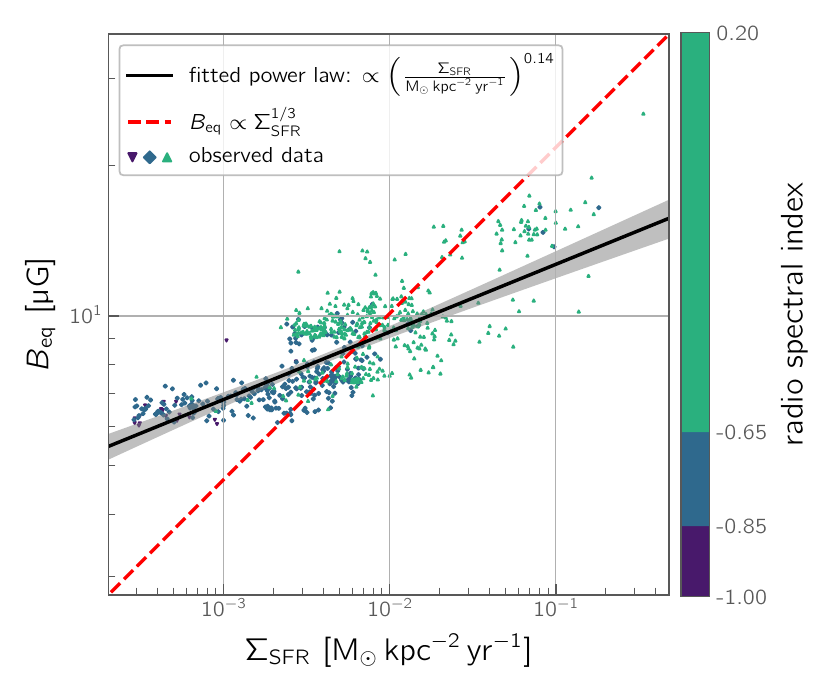}
    \end{subfigure}
    \begin{subfigure}[t]{0.02\textwidth}
        \textbf{(b)}    
    \end{subfigure}
    \begin{subfigure}[t]{0.47\linewidth}
        \includegraphics[width=0.94\linewidth,valign=t]{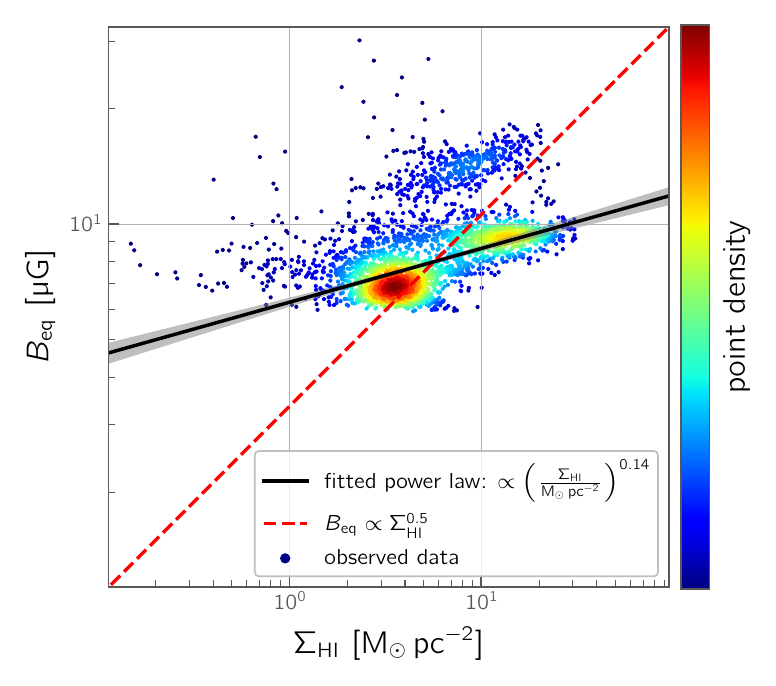}
    \end{subfigure}
    \\
    \begin{subfigure}[t]{0.02\textwidth}
        \textbf{(c)}    
    \end{subfigure}
    \begin{subfigure}[t]{0.47\linewidth}
        \includegraphics[width=\linewidth,valign=t]{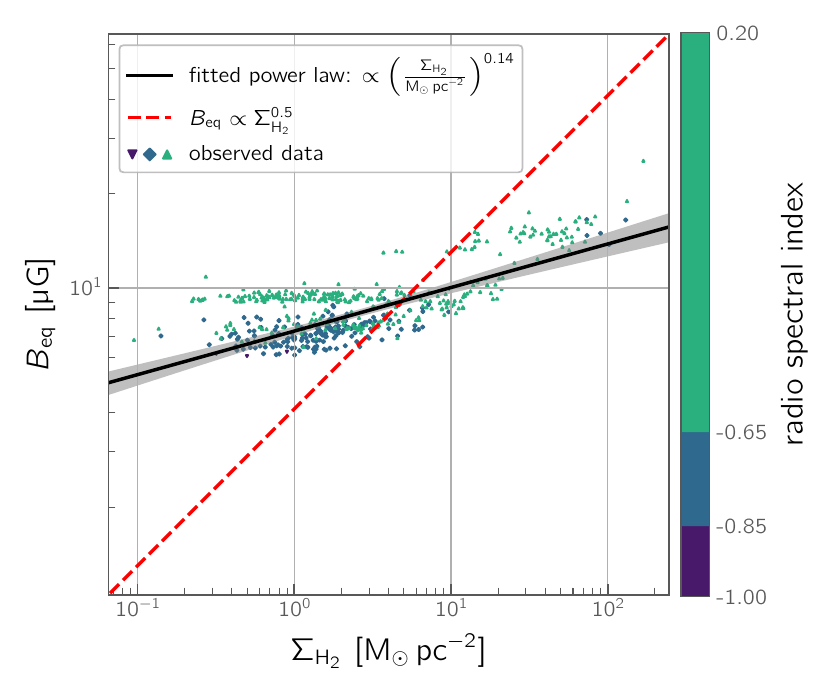}
    \end{subfigure}
    \begin{subfigure}[t]{0.02\textwidth}
        \textbf{(d)}    
    \end{subfigure}
    \begin{subfigure}[t]{0.47\linewidth}
        \includegraphics[width=\linewidth,valign=t]{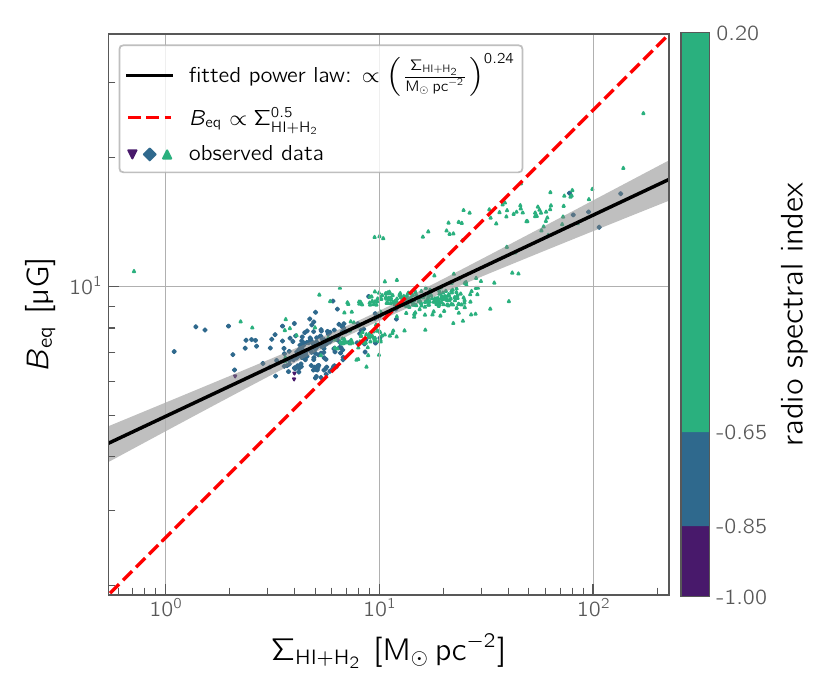}
    \end{subfigure}
    \caption{As Fig.~\ref{fig:relations}, but only for galaxies that are relatively gas poor. These are all galaxies except the grand spiral galaxies NGC~5055, 5194/5 and 7331. As can be seen these galaxies approximately conform with the relations that are found for the complete sample.}
	\label{fig:relations_poor}
\end{figure*}%

\begin{figure*}[htp]
\centering
    \begin{subfigure}[t]{0.02\textwidth}
        \textbf{(a)}    
    \end{subfigure}
    \begin{subfigure}[t]{0.47\linewidth}
        \includegraphics[width=1.0\linewidth,valign=t]{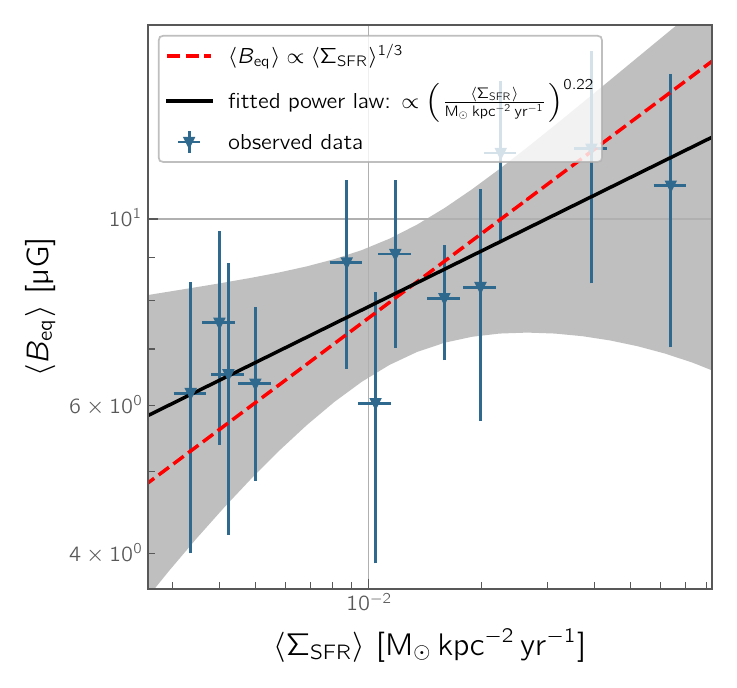}
    \end{subfigure}
    \begin{subfigure}[t]{0.02\textwidth}
        \textbf{(b)}    
    \end{subfigure}
    \begin{subfigure}[t]{0.47\linewidth}
        \includegraphics[width=1.0\linewidth,valign=t]{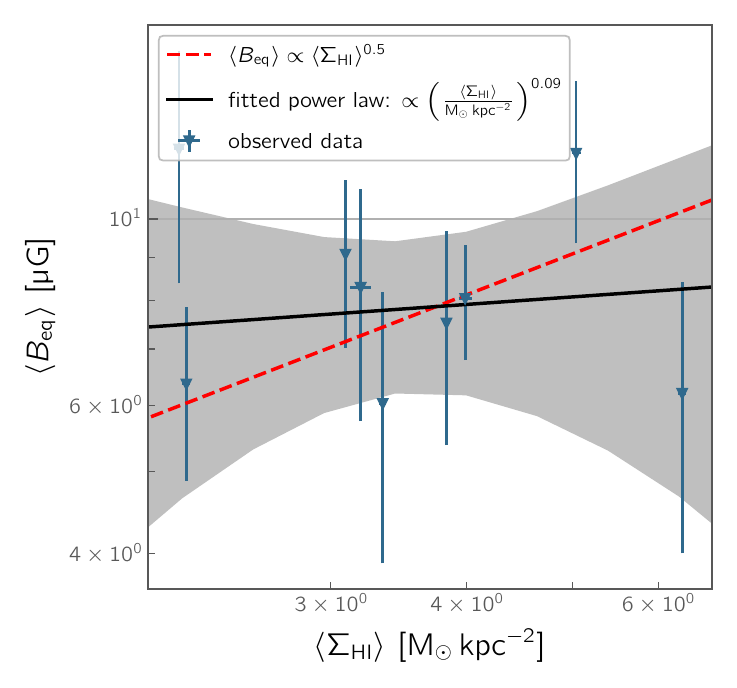}
    \end{subfigure}
    \\
    \begin{subfigure}[t]{0.02\textwidth}
        \textbf{(c)}    
    \end{subfigure}
    \begin{subfigure}[t]{0.47\linewidth}
        \includegraphics[width=1.0\linewidth,valign=t]{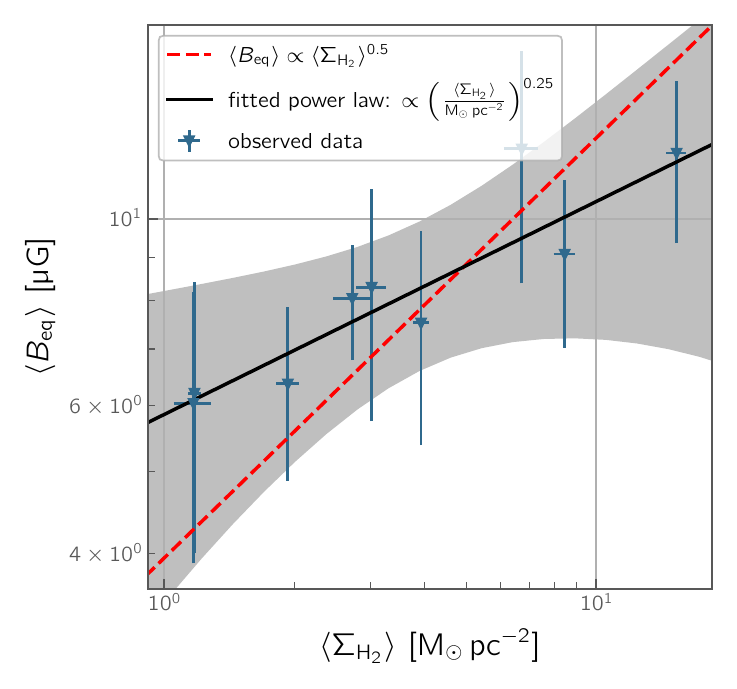}
    \end{subfigure}
    \begin{subfigure}[t]{0.02\textwidth}
        \textbf{(d)}    
    \end{subfigure}
    \begin{subfigure}[t]{0.47\linewidth}
        \includegraphics[width=1.0\linewidth,valign=t]{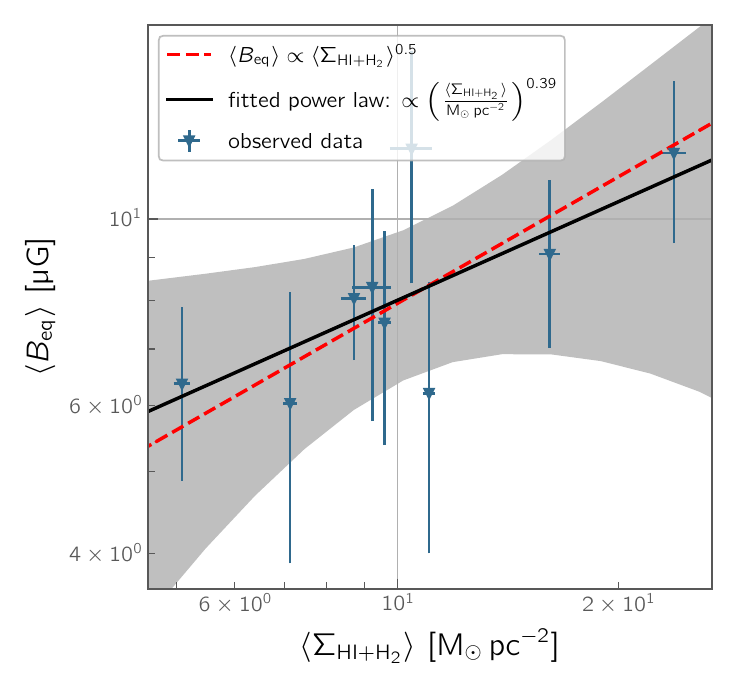}
    \end{subfigure}

		\caption{Global relations between the mean magnetic field strength and the mean star-formation rate surface density (a), the mean atomic gas mass surface density (b), mean molecular gas mass surface density (c), and mean combined atomic and molecular gas mass surface density (d). Best-fitting relations are shown as sold lines and the theoretical expectations as dashed lines.}
		\label{fig:mean_relations}
\end{figure*}%

\begin{figure*}[htp]
\centering
    \begin{subfigure}[t]{0.02\textwidth}
        \textbf{(a)}    
    \end{subfigure}
    \begin{subfigure}[t]{0.47\linewidth}
        \includegraphics[width=1.0\linewidth,valign=t]{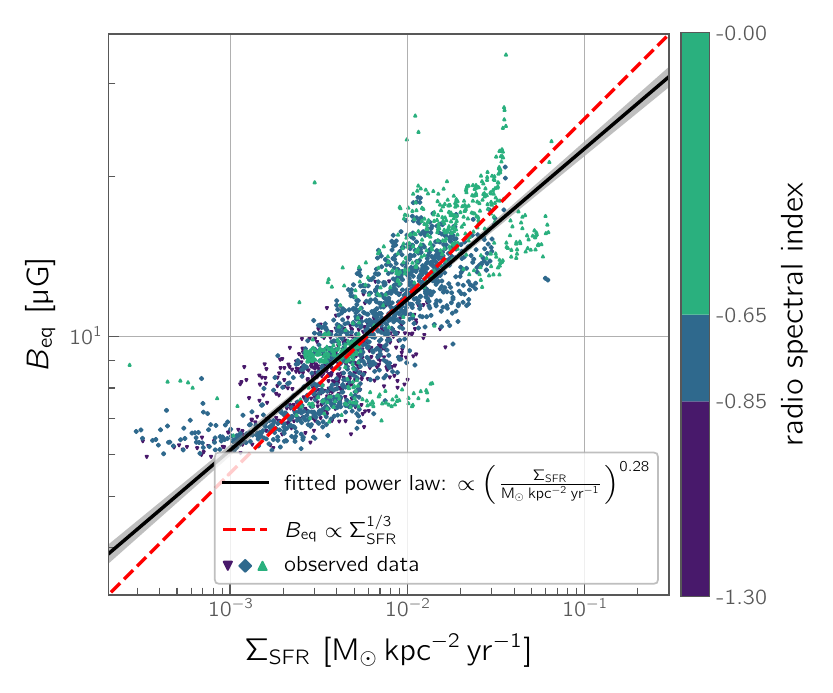}
    \end{subfigure}
    \begin{subfigure}[t]{0.02\textwidth}
        \textbf{(b)}    
    \end{subfigure}
    \begin{subfigure}[t]{0.47\linewidth}
        \includegraphics[width=1.0\linewidth,valign=t]{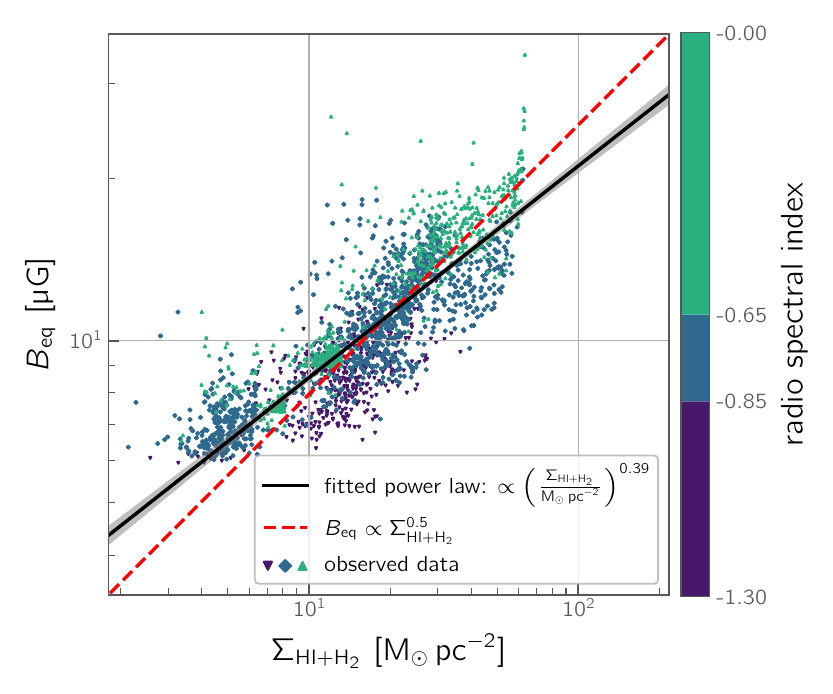}
    \end{subfigure}

		\caption{Spatially resolved relations accounted for cosmic-ray transport between the magnetic field strength and the star-formation rate surface density (a) and the combined atomic and molecular gas mass surface density (b). The \sfrd- and the \mgasd-maps have been smoothed with a Gaussian kernel in order to account for cosmic-ray transport. Data points are coloured according to the radio spectral index and in panel (b) according to the point density. Best-fitting relations are shown as sold lines and the theoretical expectations as dashed lines. }
		\label{fig:smooth_relations}
\end{figure*}%

\begin{figure}[htp]
		\includegraphics[width=\linewidth]{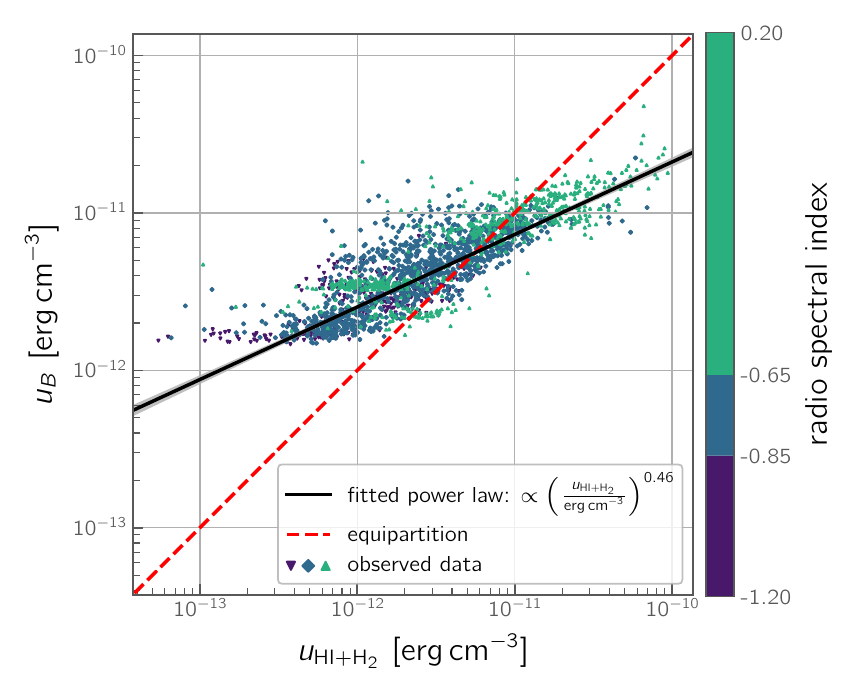}
		\caption{Relation between the magnetic energy density and the total kinetic energy density energy density. The best-fitting relation is shown as solid line, whereas energy equipartition is shown as dashed line. Data points are coloured according to their radio spectral index, highlighting that these relations are universal for all galaxies.}
		\label{fig:energy_relation}
\end{figure}%

}

\begin{table*}
	\centering
	\caption{Correlations with the magnetic field strength.}
	\label{tab:correlations}
	\begin{tabular}{cc ccc cc} 
		\hline
                $y$                                & $x$                                   & $a$                   & $b$               & $c$                                 & $r$     & $\sigma$ \\ \hline
                $B_{\rm eq} / \upmu$G                  & $\Sigma_{\rm SFR}$                      & $7.41 \pm 0.15~\mu$G  & $0.182\pm 0.004$  & $10^{-3}~\rm M_\sun\,yr^{-1}\,kpc^{-2}$ & 0.69    & 0.10~dex \\
                $\langle B_{\rm eq}/\upmu\rm G\rangle$ & $\langle\Sigma_{\rm SFR}\rangle$        & $4.74 \pm 2.14~\mu$G  & $0.22\pm 0.10$    & $10^{-3}~\rm M_\sun\,yr^{-1}\,kpc^{-2}$ & 0.82    & 0.06~dex \\
                $B_{\rm eq}/\upmu\rm G$                & $\lbrace\Sigma_{\rm SFR}\rbrace$        & $6.10 \pm 0.18~\mu$G  & $0.284\pm 0.006$  & $10^{-3}~\rm M_\sun\,yr^{-1}\,kpc^{-2}$ & 0.83    & 0.06~dex \\
                $B_{\rm eq} / \upmu$G                  & $\Sigma_{\rm H\,{\sc i}}$                 & $6.87 \pm 0.04~\mu$G  & $0.163\pm 0.003$  & $\rm M_\sun\,pc^{-2}$                 & 0.36    & 0.14~dex \\
                $\langle B_{\rm eq}/\upmu\rm G\rangle$ & $\langle\Sigma_{\rm H\,{\sc i}}\rangle$   & $6.96 \pm 2.46~\mu$G  & $0.09\pm 0.27$    & $\rm M_\sun\,pc^{-2}$                 & $-0.14$ & 0.12~dex \\
                $B_{\rm eq} / \upmu$G                  & $\Sigma_{\rm H_2}$                      & $7.49 \pm 0.06~\mu$G  & $0.183\pm 0.003$  & $\rm M_\sun\,pc^{-2}$                 & 0.84    & 0.08~dex \\
                $\langle B_{\rm eq}/\upmu\rm G\rangle$ & $\langle\Sigma_{\rm H_2}\rangle$        & $5.85 \pm 0.96~\mu$G  & $0.25\pm 0.10$    & $\rm M_\sun\,pc^{-2}$                 & 0.91    & 0.05~dex \\
                $B_{\rm eq} / \upmu$G                  & $\Sigma_{\rm H\,{\sc i}+H_2}$              & $4.33 \pm 0.07~\mu$G  & $0.309\pm 0.006$  & $\rm M_\sun\,pc^{-2}$                 & 0.80    & 0.08~dex \\
                $\langle B_{\rm eq}/\upmu\rm G\rangle$ & $\langle\Sigma_{\rm H\,{\sc i}+H_2}\rangle$& $3.27 \pm 1.46~\mu$G  & $0.39\pm 0.19$    & $\rm M_\sun\,pc^{-2}$                 & 0.69    & 0.08~dex \\
                $B_{\rm eq}/\upmu\rm G$                & $\lbrace\Sigma_{\rm H\,{\sc i}+H_2}\rbrace$& $3.45 \pm 0.09~\mu$G  & $0.393\pm 0.009$  & $\rm M_\sun\,pc^{-2}$                 & 0.85    & 0.07~dex \\
                \hline
        \end{tabular}
        \flushleft
            {\small {\bf Notes.} The equation which is fitted is $y = a(x/c)^b$. Also, $r$ is Pearson's correlation coefficient and $\sigma$ is the scatter around the best-fitting relation. $B_{\rm eq}$ is the equipartition magnetic field strength, $\langle B_{\rm eq}\rangle$ is the mean equipartition magnetic field strength. Quantities denoted with `$\lbrace \rbrace$' are obtained with the smoothing experiment to correct for cosmic-ray diffusion.}
\end{table*}

\section{Results}
\label{s:results}

\subsection{Magnetic field strength and scale length}
\label{ss:magnetic_field_strength}



In Table~\ref{tab:data-table}, we list the mean and maximum equipartition magnetic field strength in our galaxy sample along with the assumed inclination angles and path lengths that are needed as input parameters. The corresponding equipartition magnetic field strength maps are presented in Appendix~\ref{as:atlas_of_magnetic_fields_in_galaxies}. The mean magnetic field strength varies between $3.6$ and $12.5~\upmu\rm G$ and the maximum magnetic field strength between $7.5$ and $41.1$~$\upmu\rm G$. Taken as a mean across the sample, the mean magnetic field strength is $7.9\pm 2.0$~$\upmu\rm G$ and the maximum magnetic field strength is $19.8\pm 8.6$~$\upmu\rm G$.


In Appendix~\ref{as:radial_profile}, we present the 1-D radial profiles of the magnetic field strengths in our sample galaxies that have moderate inclination angles ($i\lessapprox 75$~degree). 
We determined the magnetic field radial scale-lengths in our sample galaxies with moderate inclination angles by fitting the radial profiles using the form, 
\begin{equation}
B_{\rm eq}(r) = B_0\,{\rm e}^{-r/r_B},     
\end{equation}
where $B_0$ provides an estimate of the magnetic field strength at the centre of a galaxy, and $r_B$ is the radial scale-length. In Table~\ref{tab:radial}, we present the results of the fitted profiles. We found scale-lengths between $3.1$ and $88.6$~kpc disregarding cases where the error is larger than 50\,\%. In some cases the magnetic field strength is nearly constant with radius ($r_B>100~\rm kpc)$, namely, NGC\,925, 2976, 3077, 3938, 4625, and 5457. These galaxies are all of a very late type (Sc, Sd, Sm or P). The median magnetic field scale-length in our sample is $r_B=10.4\pm 26.5$~kpc or, expressed as ratio of $R_{25}$, the  median scale-length is $r_B=(1.0\pm 1.8)R_{25}$.

\subsection{Magnetic field--star formation relation}
\label{ss:magnetic_field_star_formation_relation}


In Fig.~\ref{fig:relations}a, the combined plot for the $B$--SFR relation is presented. There is a strong correlation with a Pearson correlation coefficient of $r = 0.69$. The slope of the relation is $0.182\pm 0.004$, which is significantly flatter than the observed relation of $b\approx 1/3$ for the global magnetic field strength \citep{Niklas1997,Heesen2014,Tabatabaei2017}. A similar slope is found when we consider only gas poor galaxies as shown in Fig.~\ref{fig:relations_poor}a. In Fig.~\ref{fig:mean_relations}a, we present the corresponding global relation of the mean magnetic field strength as function of the mean star-formation rate surface density. The slope is now higher at $b=0.22\pm 0.10$, although is still well below integrated $B$--SFR measurements. Integrated measurements can be regarded as upper limits for the slopes as higher mass galaxies have steeper radio continuum spectra and thus higher calorimetric fractions of the cosmic-ray electrons \citep{Smith2021}. On the other hand, spatially resolved relations are lower limits due to the effects of cosmic-ray transport, which is particularly important at low frequencies. The effects of cosmic-ray electron transport can be seen in Fig.~\ref{fig:relations}a, where the data points with flatter spectral indices fall predominantly below the dashed line shown for comparison, $B\propto \Sigma_{\rm SFR}^{1/3}$, which is, as well as the integrated relation, in line with a simplifying theory assuming energy equipartition and a constant velocity dispersion \citep{Schleicher2013}.
The \mfsfr relation was already investigated previously in individual galaxies. \citet{Tabatabaei2013} found a slope of $0.14\pm 0.01$, whereas \citet{Chyzy2008} found a slope of $0.18\pm 0.01$. The \mfsfr relation is closely related to the radio continuum--star-formation rate (RC--SFR) relation because for equipartition $I_\nu\propto B^{3.6}$ assuming a non-thermal spectral index of $-0.6$. As the slope of the RC--SFR relation changes between arm and inter-arm region with higher slopes in the arm region \citep{Basu2013}, the \mfsfr relation follows accordingly. The large scatter of the \mfsfr relation is in part a consequence of treating the arm and inter-arm region the same, while a differentiation either using morphology \citep{Dumas2011,Basu2013} or radio spectral index \citep{Heesen2019} may lead to a reduced scatter. But since in this work we would like to concentrate on the relation between magnetic field strength and the gas surface density, we did not investigate these influences any further.

Because supernovae as the probable sources of CREs are rare in outer discs and CRE diffusion from inside is limited to a few kpc, it is well possible that energy equipartition no longer holds in the outer discs. Then the equipartition estimates, derived from the synchrotron intensities that are limited by the low density of CREs, yield underestimates for the field strength. The reason is that any under-equipartition CRE density has to be balanced by a higher field strength to explain the observed synchrotron intensity, so that the field strength has to increase. This can be tested by Faraday rotation measures of background sources located behind outer discs. RM data of intervening galaxies in front of QSOs indicate indeed that the extents of magnetic fields are much larger than those of the star-forming disc \citep[see e.g.][]{Kwang_Seong2016}. The field strengths at low \sfrd in Figs.~\ref{fig:relations}a and \ref{fig:smooth_relations}a are above the expected values. The above correction for a possible non-equipartition in the outer discs would even increase this discrepancy, not remove it.

To circumvent the problematic study of magnetic fields in outer discs in emission, one may hence turn in the future to the usage of RM studies. Those with LOFAR give access to weak magnetic fields as are expected for nearby star-forming galaxies \citep{OSullivan2020}.

\subsection{Magnetic field--gas relation}
\label{ss:magnetic_field_gas_relation}

The spatially resolved magnetic field--gas relation for the atomic gas is shown in Figs.~\ref{fig:relations}b and \ref{fig:relations_poor}b. The scatter of $0.14$~dex is the largest in our investigated magnetic field correlations. Consequently, the correlation coefficient for the full sample $r=0.36$ is fairly small, suggesting that the correlation is not significant. This result does not change for the global \mfhi relation as presented in Fig.~\ref{fig:mean_relations}b, where the correlation is even weaker. We note that the data points seem to cluster around several areas, fitting all these with just a single line and average them for the global relations, might underestimate the complexity of the
distribution. Hence, neither the fitted nor the theoretical line are
accurate descriptions of the data. The H\,{\sc i} surface density rarely exceeds $10~\rm M_\sun\,pc^{-2}$ where H\,{\sc i} `saturates' and the transition from H\,{\sc i} to $\rm H_2$ happens \citep{Leroy2008}.

For the molecular gas, the spatially resolved \mfhtwo relation is shown in Figs.~\ref{fig:relations}c and \ref{fig:relations_poor}c. The scatter is fairly small with 0.08~dex and correlation is significant with $r=0.84$ (both refer to the full sample). Clearly, the magnetic field is much better correlated with the distribution of the molecular gas than with the atomic gas. We also note that the slope is identical to the \mfsfr relation within the uncertainties, but the scatter is slightly smaller. A similar slope is expected when one considers $\Sigma_{\rm SFR}\propto \Sigma_{\rm H\,2}$ \citep{Bigiel2008}. A similar result was also found in the global relation presented in Fig.~\ref{fig:mean_relations}c, where the scatter is reduced further to $0.05$~dex and the slope increases to $b=0.25\pm 0.10$. Again, this slope is in agreement with the global \mfsfr relation.






In Fig.~\ref{fig:relations}d, we show the magnetic field--gas relation for the combined atomic and molecular gas surface densities. We find that the \mfgas relation has the steepest slope with $b=0.309\pm 0.006$ in conjunction with a small scatter of $0.08$~dex indicating a  tight correlation ($r=0.80$) for the full sample with similar results for the gas poor sample (Fig.~\ref{fig:relations_poor}d). For the global relation shown in Fig.~\ref{fig:mean_relations}d, the slope is increased further to $b=0.39\pm 0.19$. We note that for the spatially resolved correlation only pixels were taken into account, where CO emission was detected. The higher slope of the \mfgas in comparison to the \mfhtwo relation may be related to the fact that the lowest mass surface densities are removed as there is always H\,{\sc i} gas even if there is only little $\rm H_2$ gas.

It seems that the relation between $B$ and \mhid has different exponents in regions of different density, i.e.\ a flat relation at low densities but a steeper one at high densities. This is consistent with results from Zeeman splitting in our Milky Way \citep[see fig.~1 in][]{Crutcher2010}. A similar but weaker effect is also seen in the relation between $B$ and \mhtwod that also flattens at low densities. The probable reason is that $B$ is not only related to gas density, but also to the turbulent velocity dispersion, which may not be not constant but varies with gas density \citep{Pakmor2017}. Figure~\ref{fig:relations} may tell us that the variation of turbulent velocity across and between galaxies is larger in H\,{\sc i} gas than in $\rm H_2$ gas, possibly related to the different dynamics of molecular clouds \citep{Sun2018} and of atomic gas \citep{Tamburro2009}.

\subsection{Influence of cosmic-ray transport}
\label{ss:influence_of_cosmic_ray_transport}

In Fig.~\ref{fig:smooth_relations}, we present the spatially resolved \mfsfr and the  \mfgas relations, where we have  accounted for cosmic-ray transport by convolving the \sfrd- and \mgasd-maps with a Gaussian kernel that matched the cosmic-ray transport length in each galaxy. In this section we want to concentrate only on these two most relevant relations with the magnetic field strength. The slope of the \mfsfr relation increases to $b=0.284\pm 0.006$, as expected, and the correlation coefficient increases, too, when compared with the relation of the unconvolved maps. For the \mfgas relation, the slope increases to $b=0.393\pm 0.009$, and the correlation coefficient rises slightly. Hence, we conclude that cosmic-ray transport does affect our results of the magnetic field--gas relation, but not by quite as much as the \mfsfr relation. This is because the molecular gas is only detected where the \sfrd-values are high, whereas cosmic-ray transport affects mostly galaxy outskirts and inter-arm regions. This can be seen by the various slopes of the RC--SFR relation in arm and inter-arm regions, where the arm regions are almost linear \citep[e.g.][]{Dumas2011,Heesen2019} in good agreement with the global relation. In contrast, the \sfrd-maps cover the full extent of the galaxies including the outskirts.

\subsection{Kinetic energy density}

In Fig.~\ref{fig:energy_relation}, we show a comparison between the magnetic and the total kinetic energy density of the gas. The best-fitting relation is:

\begin{equation}
	u_{B} = (2.53 \pm 0.62) \times 10^{-12}~{\rm erg\, cm^{-3}} \left( \frac{u_{\rm HI + H_2}}{10^{-12}~\rm erg\, cm^{-3}} \right)^{0.461 \pm 0.009},
	\label{eq:b_kinetic_energy}
\end{equation}
with $r = 0.78$ and a scatter of 0.20 dex. The slope is with $0.50$ smaller than what would be expected if energy equipartition with the gas were to hold everywhere. While on average the magnetic field is in equipartition, at the low energy densities we find an excess of magnetic energy density, whereas at the high kinetic energy densities, the magnetic field is fairly weak. As stated in Section~\ref{ss:influence_of_cosmic_ray_transport}, we can rule out cosmic-ray transport as the sole reason. It is possible that the gas velocity dispersion changes across the galaxy, possibly being correlated with \sfrd \citep{Tamburro2009}. A positive correlation with \sfrd as observed in H\,{\sc i} \citep{Tamburro2009} as well as in CO meaning $\rm H_2$ \citep{Colombo2014} would mean that we underestimate the kinetic energy density in areas of high \sfrd, hence compounding the domination of kinetic over magnetic energy density.

\section{Discussion}
\label{s:discussion}

\subsection{What regulates magnetic fields in galaxies?}

If the magnetic field is frozen into the gas, we would expect a correlation with the gas density as:
\begin{equation}
    B \propto \rho^\kappa,
    \label{eq:b_rho}
\end{equation}
where $\kappa$ depends on physical properties of the medium and on geometry. For isotropic compression in dense media with supersonic turbulence, $\kappa=2/3$, as observed from Zeeman splitting data \citep{Crutcher2010}, whereas for linear compression such as in a shock $\kappa=1$ for the field component parallel to the shock front and $\kappa=0$ for the component perpendicular to the shock front. We note that magnetic field--gas relation  in multi-phase simulations is very complex and cannot be easily described by a single power-law relationship consistent with these simple gas compressions. In particular, the relation depends on the Alfv\'enic Mach number, where $\kappa=2/3$ is found in high Mach number
regions of the cold ISM dominated by molecular hydrogen \citep{seta2022}. However, as such regions would likely occupy a small fraction of the ISM and H\,{\sc i} is mostly sub- to transsonic so that flux freezing does not hold.

On the other hand, if the magnetic energy density is in equipartition with the kinetic energy density of the gas, we expect:
\begin{equation}
    \frac{B^2}{8\pi} = \frac{f}{2}\,\rho\, \varv_{\rm t}^2 ,
    \label{eq:b_ukin}
\end{equation}
where $\varv_t$ is the turbulent velocity. A constant $\varv_{\rm t}$ leads to $\kappa=1/2$, so that $B\propto \rho^{0.5}$. At the moderate densities of the warm ISM, $\varv_{\rm t}$ is trans- to subsonic \citep{Burkhart2012} and similar to the sound velocity $\varv_{\rm s}$. If gas pressure were constant, $\varv_{\rm t} \propto \varv_{\rm s} \, \rho^{0.5}$, so that $B$ would not vary with density. Realistically, pressure is not constant in the ISM \citep{DeAvillez2005}.

The fraction $f$ of magnetic energy density to kinetic energy density is mostly found to be close to unity \citep{Beck2007,Gent2013,Beck2015a}. A 100\,\% efficient small-scale dynamo would, in this picture, correspond to $f=1$. In some regions $f$ is even larger than unity, which cannot be solely explained by the small-scale dynamo and another source of energy sustaining the dynamo action is needed \citep{Beck2015}.  For a constant velocity dispersion, we obtain $B\propto \rho^{0.5}$. If one now assumes a Kennicutt-Schmidt relation between the gas and the SFR surface density, $\Sigma_{\rm SFR}\propto \Sigma_{\rm gas}^N$, we can infer assuming $N=1.5$ \citep{Schleicher2013}:
\begin{equation}
    B \propto \Sigma_{\rm SFR}^{1/3}.
    \label{eq:b_sfr}
\end{equation}
For molecular dominated gas, we have $N=1$ \citep{Leroy2008}, which would result in a steeper \mfsfr relation of $B \propto \Sigma_{\rm SFR}^{1/2}$. This was supported by the numerical simulations by \citet{Steinwandel2020} that yielded slopes between $0.3$ and $0.45$. Equations \eqref{eq:b_rho}--\eqref{eq:b_sfr} are hence the basic theoretical relations we can test with our observations.

We find the best correlations with the smallest scatter and the highest correlation coefficient in the \mfhtwo correlation, when considering only the spatially resolved correlations. However, the dependence of the magnetic field on gas density is only weak with $B\propto \Sigma_{\rm H_2}^{0.183}$. The relation between magnetic field and total gas density, which has the same small scatter but a much larger slope of $B\propto \Sigma_{\rm H\,I+H_2}^{0.309}$, appears more significant. This slope increases further when we either consider global relations or the spatially resolved relation $B\propto \Sigma_{\rm H\,I+H_2}^{0.393\pm 0.009}$, where the cosmic-ray diffusion is accounted for. This slope is already fairly close to the slope as expected for the saturated dynamo. The remaining difference may stem from a decreasing velocity dispersion in regions with larger gas densities. What appears to be rather certain is that pure compression cannot account for the observed $B$--$\rho$ behaviour as the $\kappa$ is too low.

The \mfsfr correlation of $B\propto \Sigma_{\rm SFR}^{0.182}$ agrees with expectation for a saturated small-scale dynamo, where the slope of the relation is slightly too low but increases to $0.284$ when cosmic-ray transport is accounted for. The magnetic energy density is in approximate equipartition with the kinetic energy density further supporting the saturation of the small-scale dynamo. However, locally we find deviations from equipartition, in particular in areas of high kinetic energy densities where the magnetic field is subdominant. In contrast, the magnetic field dominates in areas of low kinetic energy densities. This is in agreement with radial profiles of the energy densities, which show that the magnetic field dominates in the galaxy outskirts \citep{Beck2007,Beck2015a}. Large-scale gravitational instabilities as a source of turbulence are thought to be efficient only in high-density gas \citep{Brucy2020}. A more probable source of turbulence in the outer discs of galaxies is the magneto-rotational instability \citep[MRI;][]{Gressel2013}. Other possibilties are a hypothetical change of efficiency of the large-scale dynamo that operates more effectively in the galaxy outskirts  and the influence of the thermal gas pressure on the magnetic field \citep{Basu2013}.

\subsection{Limitations of the equipartition approach}

When we apply the equipartition estimate for the magnetic field strength (Section~\ref{ss:equipartition_magnetic_field_strength}), we `trim' the radio spectral index to the range of $-1.1\leq \alpha \leq -0.6$. This is in our opinion the best option. The equipartition formula yields unphysically large $B$ values for $\alpha > -0.6$ (very flat) and also for $\alpha <  -1.1$ (very steep). The first case indicates thermal absorption or ionization losses, the second strong synchrotron losses. In both cases the proton-to-electron ratio $K_0$ is larger than the canonical value of $K_0=100$, without knowing the correct value. We performed a few experiments showing that trimming the spectral index gives more reliable $B$ estimates than trying to correct the value of $K_0$. These experiments are presented in Appendix~\ref{as:equipartition_method}. 
Weak and mild synchrotron losses as we have at low radio frequencies do not affect the equipartition estimate of the total field.

Another limitation is the assumption of a constant magnetic field strength along the line-of-sight, which is obviously a simplification and a more complex model would be necessary. The mean magnetic field is systematically lower in the edge-on galaxies as we are seeing the radio halo, where the field strength is weaker. In face-on galaxies, we assume a pathlength of $1.4$~kpc (Section~\ref{ss:equipartition_magnetic_field_strength}), which can only be a crude approximation. Observations of edge-on galaxies show that reality is more complicated. According to \citet{Krause2018}, the radio emission in many edge-on galaxies has two components, a (thin) disc and a halo. The radio disc is a result of the strong magnetic fields in the star-forming regions, while the radio halo mainly reflects the distribution of the CRe leaving the disc. The intensity and vertical extent of the radio disc varies locally, following the tracers of star formation. The radio halo is more homogeneous. \citet{Krause2018} found average values for the halo scale heights of $1.1\pm 0.3$~kpc at 6000~MHz and $1.4\pm 0.7$~kpc at 1500~MHz. Thermal emission was not subtracted, so that the scale heights of pure synchrotron emission are expected to be somewhat larger. The scale heights of the (thin) disc have large uncertainties due to the limited resolution.

In the case of mildly inclined galaxies, a superposition of emission from both components along the line of sight is observed, so that it is hard to estimate their relative contributions. The ratio of disc-to-halo emission has been so far only poorly studied. \citet{Stein2020} found ratios of order unity with the ratio rising with the \sfrd-values, so that some galaxies have relatively high ratios of disc-to-halo emission. In summary, our assumed pathlength for the face-on galaxies is a good approximation for galaxies with strong star formation in the disc, whereas for weakly star-forming galaxies an average between disc and halo path lengths would be more appropriate.

Finally there is the question whether equipartition is applicable at all in galaxies.  Global state-of-the-art MHD models of star-forming galaxies including cosmic rays were investigated recently in a series of papers: \citet{Werhahn2021}; \citet{Werhahn2021b} and \citet{Pfrommer2022}. These simulations  show that equipartition (measured on scales of 10 kpc) is reached in Milky Way-mass galaxies but fails for small galaxies \citep[see fig.~1 in][]{Pfrommer2022}. From similar models, \citet{Buck2020} show radially averaged profiles of magnetic and cosmic-ray energy densities (their fig.~4, panels 2 and 4) that match quite well, although there is a slight suppression of the cosmic-ray energy density compared to the magnetic energy density in galaxy outskirts. If this is indeed the case in our galaxies, this would lead to an overestimation of the magnetic field strength from our equipartition estimates. This could then explain our high magnetic energy densities compared with the kinetic energy densities in these areas. While \citet{Seta2019} show that  equipartition does not hold on very small scales, these scales are not relevant for our work as they are unresolved. They suggest that equipartition could be tested directly in the local ISM, from solar system-corrected cosmic ray data and {\it Voyager} magnetic fields data from outside the solar system. Nevertheless, equipartition does hold to the best of our knowledge on the scale relevant here. Lastly, $\gamma$-ray observations also confirm the validity of energy equipartition for surface star formation rates of $\Sigma_{\rm SFR} \lessapprox 100~\rm M_\sun\,yr^{-1}\,kpc^{-2}$ \citep{Yoast-Hull2016}.

\section{Conclusions}
\label{s:conclusions}

We used the LOFAR data from the LoTSS data release 2 at 144 MHz \citep{Shimwell2022} to calculate magnetic field strengths in a sample of 39 nearby galaxies \citep{Heesen2022} assuming energy equipartition between cosmic rays and the magnetic field. We used radio spectral index maps in order to assess the cosmic-ray electron spectrum. In agreement with former studies, we found that our sample of galaxies shows magnetic fields with strengths of up to 40~$\upmu$G, but with mean values of around 10~$\upmu$G. For a subset of 9 galaxies, we studied the relation between magnetic fields, star formation, and gas density. These are our main conclusions:

\begin{itemize}
    \item The magnetic field strength has the tightest and steepest correlation with the total gas surface density, where we find $B\propto \Sigma_{\rm H\,I+H_2}^{0.309\pm 0.006}$ with a scatter of $0.08$~dex.
    \item Similarly, we find tight correlations of the magnetic field strength with both the SFR and molecular gas surface densities, albeit with smaller slopes of $B\propto \Sigma_{\rm SFR}^{0.182\pm 0.004}$ and $B\propto \Sigma_{\rm H_2}^{0.183\pm 0.003}$, respectively.
    \item We find no significant correlation between magnetic field strength and atomic mass surface density \mhid. The scatter around the best--fitting relation is the largest with $0.14$~dex.
    \item When we account for cosmic-ray transport, smoothing the \sfrd- and \mgasd-maps with a Gaussian kernel, we find steeper slopes of $B\propto \Sigma_{\rm SFR}^{0.284\pm 0.006}$ and $B\propto \Sigma_{\rm H\,I+H_2}^{0.393\pm 0.009}$. Both relations can be explained by a saturated small-scale dynamo, where the magnetic energy density is in equipartition with the kinetic energy density.
    \item Comparing the energy densities, we find that while on average they do agree, locally the magnetic field is sub-dominant in regions of high kinetic energy density, whereas the opposite is the case for regions with a low kinetic energy density.
    \item Either the efficiency of the small-scale dynamo is lower in the inner parts of the disc, the magnetic field strength is boosted by the magneto-rotational instability, or we need to take other components such as the thermal gas into account.
\end{itemize}

The most surprising aspect of our work is that the magnetic field depends most on the total gas mass surface density and not quite so much on the SFR surface density. This suggests a $B$--$\rho$ coupling, where the slope is in agreement with equipartition between magnetic and kinetic energy density. Our angular resolution corresponds to $0.3$--$1.2$~kpc at the distances of our galaxies. A related correlation between CO and radio continuum was found in a sample of galaxies at sub-kpc scales \citep{Murgia2005,Paladino2006,Paladino2008}; the same correlation between CO and radio continuum luminosity was confirmed again in NGC~5194/5 at the 40-pc cloud scale  \citep{Schinnerer2013}.  A $B$--$\rho$ relation has also wider implications for the integrated radio continuum--far-infrared relation of galaxies \citep{Niklas1997} and the radio continuum--star-formation rate relation \citep{Heesen2014}. New insights may be gained in the future from expanding the LoTSS-DR2 nearby galaxies sample to the full LoTSS sky coverage and new CO observations at high angular resolution as provided with the PHANGS--ALMA survey \citep{Leroy2021}.

\begin{acknowledgement}
We thank the anonymous referee for an insightful review, which helped to improve the quality of the manuscript. We also thank Amit Seta for a fruitful discussion about this work. This paper is based (in part) on data obtained with the International LOFAR Telescope (ILT). LOFAR \citep{vanHaarlem2013} is the Low Frequency Array designed and constructed by ASTRON. It has observing, data processing, and data storage facilities in several countries, that are owned by various parties (each with their own funding sources), and that are collectively operated by the ILT foundation under a joint scientific policy. The ILT resources have benefitted from the following recent major funding sources: CNRS-INSU, Observatoire de Paris and Universit\'e d'Orl\'eans, France; BMBF, MIWF-NRW, MPG, Germany; Science Foundation Ireland (SFI), Department of Business, Enterprise and Innovation (DBEI), Ireland; NWO, The Netherlands; The Science and Technology Facilities Council, UK; Ministry of Science and Higher Education, Poland.

This work made use of the {\sc SciPy} project {\tt https://scipy.org}. M.B. acknowledges funding by the Deutsche Forschungsgemeinschaft (DFG, German Research Foundation) under Germany’s Excellence Strategy – EXC 2121 ‘Quantum Universe’ – 390833306.

The J\"ulich LOFAR Long Term Archive and the German LOFAR network are both coordinated and operated by the J\"ulich Supercomputing Centre (JSC), and computing resources on the supercomputer JUWELS at JSC were provided by the Gauss Centre for Supercomputing e.V. (grant CHTB00) through the John von Neumann Institute for Computing (NIC).

MS acknowledges funding from the German Science Foundation DFG, within the Collaborative Research Center SFB1491 "Cosmic Interacting Matters - From Source to Signal".
\end{acknowledgement}
\bibliography{sources.bib}
\bibliographystyle{aa}

\iftoggle{long}{
\appendix

\section{Equipartition method}
\label{as:equipartition_method}


The equipartition estimate of total magnetic field strengths by \citet[][BK05]{Beck2005} is the standard method to estimate magnetic field strengths in star-forming galaxies. Energy equipartition between total cosmic rays and total magnetic fields is thought to be valid on scales larger than the cosmic-ray electron (CRe) propagation length \citep{Seta2019}. A ratio $K_0$ of number densities of cosmic-ray protons (CRp) to the total number of cosmic rays needs to be assumed to extrapolate the CRe spectrum (traced by radio synchrotron emission) to the spectrum of CRp that dominate the total cosmic-ray energy. $K_0=100$ is the standard value for the GeV range that is relevant for GHz emission. $K_0 \approx 100$ is expected from the theory of diffusive shock acceleration \citep{Bell1978} and has also been measured in the local ISM \citep[e.g.][]{Cummings2016}.

Concerns may arise whether the BK05 method is applicable to outer discs and haloes of spiral galaxies, where CRe have lost a significant fraction of their energy. As energy losses of CRp are much weaker, the p/e ratio $K_0$ of number densities becomes larger than the standard value of 100 and varies with frequency. However, considering this complication is not needed because the BK05 method is quite robust against the effects of energy losses of CRe, as shown by the examples below.

\begin{itemize}
    \item[(a)] As the standard case, we assume a star-forming galaxy with inclination $i=30^\circ$, path length $l=1.0$ kpc/$\cos i$, 15\arcsec\ beam, spectral index $\alpha=-0.7$, flux density $S_\nu =1.0~\rm mJy\,beam^{-1}$ at 1.4~GHz, corresponding to $0.286~\rm mJy\,beam^{-1}$ at 8.35 GHz, degree of polarization $p=10\,\%$, and $K_0=100$. Application of the {\sc bfeld} tool (written by M.~Krause) yields a total field of $B_\mathrm{eq} = 13.4~\upmu$G that is regarded to be the ‘correct’ one.
    \item[(b)] Weak synchrotron loss at the higher frequency:  $\alpha=-0.9$ and flux density $S_\nu =1.0~\rm mJy\,beam^{-1}$ at 1.4~GHz (no loss) gives $0.2~\rm mJy\,beam^{-1}$ at 8.35~GHz (loss by a factor of about 1.4 compared to the standard case) and $B_\mathrm{eq} = 13.5~\upmu$G, very close to the correct one.
    \item[(c)] Strong synchrotron loss at the higher frequency: $\alpha=-1.1$ and flux density $S_\nu=1.0~\rm mJy\,beam^{-1}$ at 1.4~GHz (no loss) gives $0.14~\rm mJy\,beam^{-1}$ at 8.35~GHz (loss by a factor of about 2 compared to the standard case). When naively assigning the same low spectral index to CRp, we get $B_\mathrm{eq} = 14.3~\upmu$G, still close to the correct value. The reason is that the integral over the steep CRp energy spectrum, extrapolated from the steep CRe energy spectrum using a constant $K_0$, remains almost constant, even though the real CRp spectrum is flatter.
    \item[(d)] Strong spectral steepening at the higher frequency due to an energy cutoff, e.g. by time-dependent injection:  $\alpha=-1.5$ and flux density $S_\nu =1.0~\rm mJy\,beam^{-1}$ at 1.4~GHz (no loss) gives $0.07~\rm mJy\,beam^{-1}$ at 8.35~GHz (loss by a factor of about 4 compared to the standard case). When assigning the same low spectral index to CRp, we get $B_\mathrm{eq} = 16.4~\upmu$G, a significant overestimate compared to the correct value.    
    \item[(e)] Ionization loss or free-free absorption: $\alpha=-0.55$ and flux density $S_\nu=1.0~\rm mJy/,beam^{-1}$ at 1.4 GHz (no loss) gives $3.55~\rm mJy\,beam^{-1}$ at 0.1~GHz (loss by a factor of about 1.8 compared to the standard case). When assigning the same high spectral index to CRp, we get $B_\mathrm{eq} = 16.7~\upmu$G, again an overestimate.
\end{itemize}

\begin{figure}[htp]
\centering
	\includegraphics[width=0.8\linewidth]{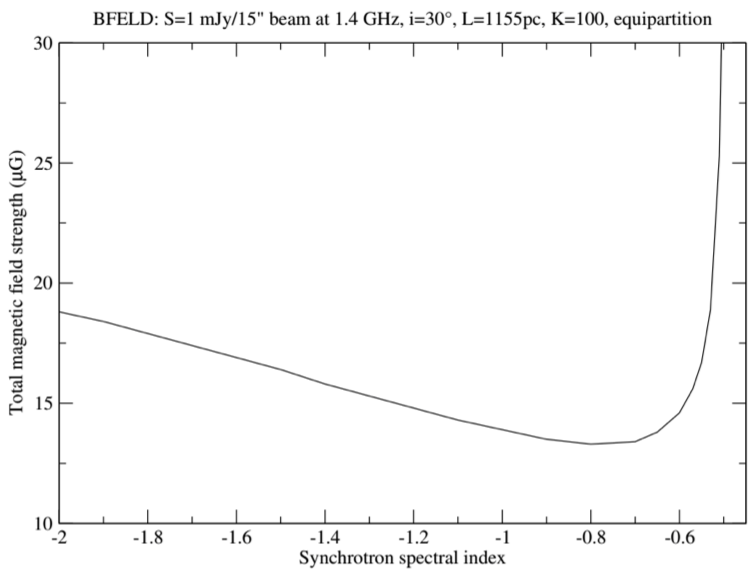}
	\caption{Equipartition estimates of the total magnetic field strength as a function of synchrotron spectral index, assuming a fixed flux density at 1.4 GHz and a constant value of the proton-to-electron ratio of $K_0=100$.}
\label{fig:equipartition}
\end{figure}

Fig.~\ref{fig:equipartition} shows the magnetic field strength as a function of a wide range of spectral indices. The overestimates are large for very steep and very flat spectra. 

We finish this section with our conclusions and some recommendations how to calculate equipartition magnetic field strengths:


\begin{enumerate}
    \item Synchrotron loss of CRe hardly affects the equipartition estimate of the total field. The error is are smaller than the uncertainties of the other input parameters, $K_0$ and $l$. The reason is that the BK05 method extrapolates and integrates the spectrum of CRp over all energies (with a break at the proton rest mass), so that no particles are `lost'.
    \item For very steep CRe spectra the low-energy part of the extrapolated CRp spectrum is overestimated, and so is the field strength. The standard BK05 formula gives approximately correct values only for synchrotron spectra with spectral indices of $\alpha > -1.1$, while steeper spectra lead to significant overestimates of the field strength. Hence, regions with $\alpha < -1.1$ should be clipped.
    \item The BK05 method should only be applied for synchrotron spectra with $\alpha < -0.6$. Flatter spectra, when assumed to apply also to CRp, lead to massive overestimates of the field strength because the high-energy part of the CRp spectrum dominates. The energy integral over the CRp spectrum even diverges for $\alpha \ge -0.5$. Hence, regions with $\alpha > -0.6$ should be clipped.
    \item Energy losses of CRe increase the p/e ratio $K_0$, e.g. to values of 140 (case b), 200 (case c), and 400 (case d). However, using such high values for the equipartition estimate would lead to even larger overestimates of the field strengths. Hence, $K_0=100$ should be used in all cases.
\end{enumerate}

\section{Atlas of magnetic fields in galaxies}
\label{as:atlas_of_magnetic_fields_in_galaxies}

In this Appendix, we present the atlas of magnetic fields. For each galaxy, we show the equipartition magnetic field strength as a map made from LOFAR 144-MHz data with 6~arcsec angular resolution. The first panel shows a grey-scale representation, where the extent of the 6-arcsec emission to the 3$\sigma$ contour line is shown as a blue ellipse \citep[data taken from][]{Heesen2022}. The second panel shows an overlay of the magnetic field strength as contour levels onto an optical image. These \textit{rgb} images are from  the Sloan Digital Sky Survey data release 16 \citep[SDSS;][]{Ahumada2020} or from the Digitized Sky Survey \citep[DSS2;][]{Lasker1996} for NGC~891 and 925, which we got from the \texttt{hips2fits}\footnote{\href{https://alasky.u-strasbg.fr/hips-image-services/hips2fits}{https://alasky.u-strasbg.fr/hips-image-services/hips2fits}} service. In case of NGC~891 and 925, DSS2 because the galaxy is outside the SDSS footprint. These maps are presented in Figs.~\ref{fig:n855}--\ref{fig:n7331}.


\begin{figure*}
	\centering
    \begin{subfigure}[t]{0.02\textwidth}
        \textbf{(a)}    
    \end{subfigure}
    \begin{subfigure}[t]{0.47\linewidth}
        \includegraphics[width=1.0\linewidth,valign=t]{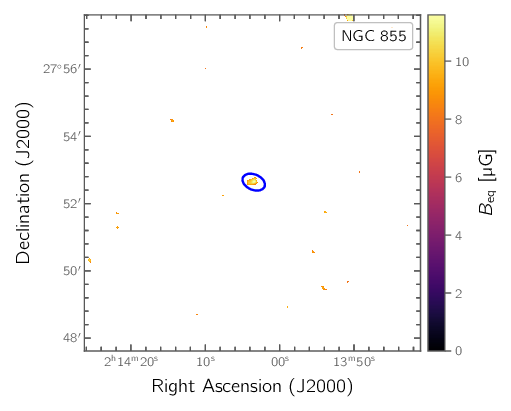}
    \end{subfigure}
    \begin{subfigure}[t]{0.02\textwidth}
        \textbf{(b)}    
    \end{subfigure}
    \begin{subfigure}[t]{0.47\linewidth}
        \includegraphics[width=1.0\linewidth,valign=t]{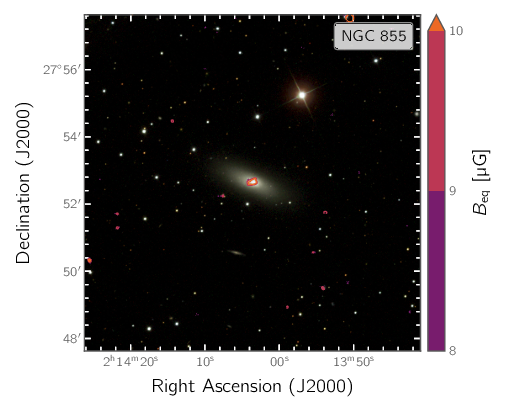}
    \end{subfigure}
    \caption{NGC~855. \map}
    \label{fig:n855}
\end{figure*}
\addcontentsline{toc}{subsection}{NGC 855}

\begin{figure*}
	\centering
    \begin{subfigure}[t]{0.02\textwidth}
        \textbf{(a)}    
    \end{subfigure}
    \begin{subfigure}[t]{0.47\linewidth}
        \includegraphics[width=1.0\linewidth,valign=t]{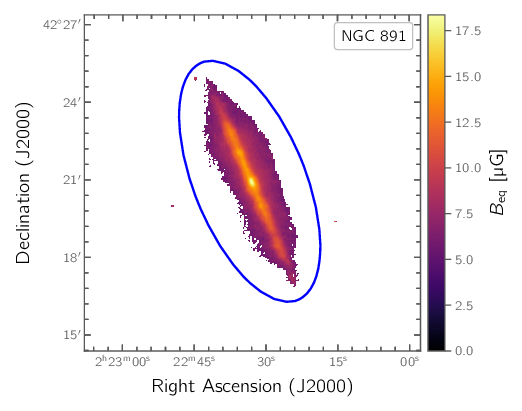}
    \end{subfigure}
    \begin{subfigure}[t]{0.02\textwidth}
        \textbf{(b)}    
    \end{subfigure}
    \begin{subfigure}[t]{0.47\linewidth}
        \includegraphics[width=1.0\linewidth,valign=t]{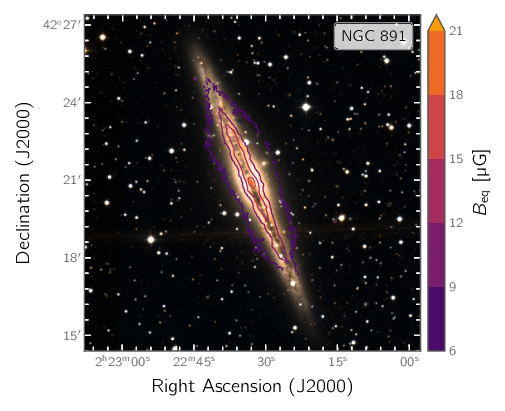}
    \end{subfigure}
    \caption{NGC~891. \map}
    \label{fig:n891}
\end{figure*}
\addcontentsline{toc}{subsection}{NGC 891}

\begin{figure*}
	\centering
    \begin{subfigure}[t]{0.02\textwidth}
        \textbf{(a)}    
    \end{subfigure}
    \begin{subfigure}[t]{0.47\linewidth}
        \includegraphics[width=1.0\linewidth,valign=t]{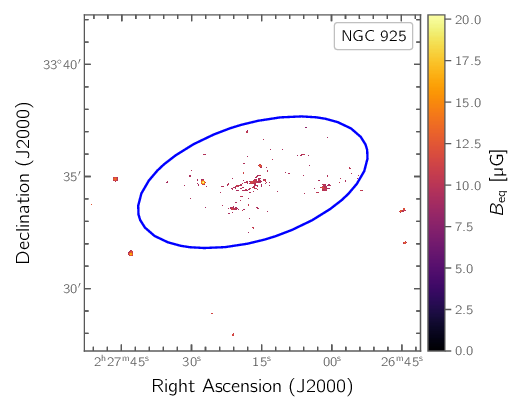}
    \end{subfigure}
    \begin{subfigure}[t]{0.02\textwidth}
        \textbf{(b)}    
    \end{subfigure}
    \begin{subfigure}[t]{0.47\linewidth}
        \includegraphics[width=1.0\linewidth,valign=t]{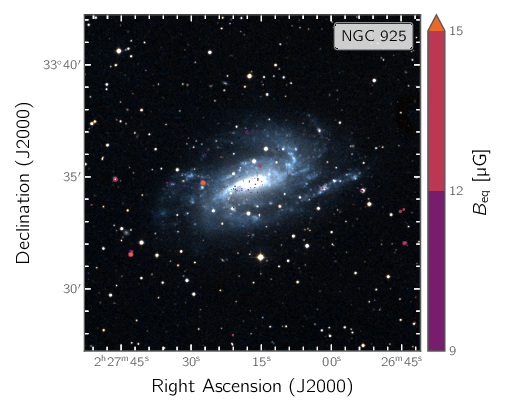}
    \end{subfigure}
    \caption{NGC~925. \map}
    \label{fig:n925}
\end{figure*}
\addcontentsline{toc}{subsection}{NGC 925}

\begin{figure*}
	\centering
    \begin{subfigure}[t]{0.02\textwidth}
        \textbf{(a)}    
    \end{subfigure}
    \begin{subfigure}[t]{0.47\linewidth}
        \includegraphics[width=1.0\linewidth,valign=t]{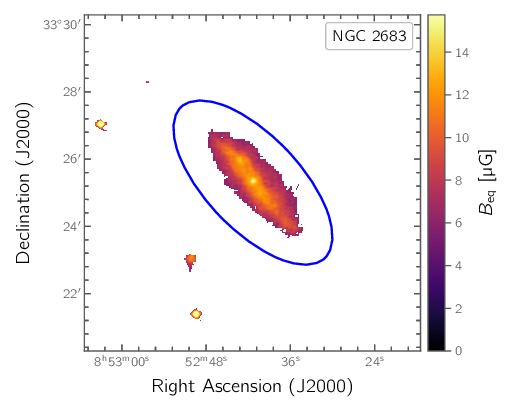}
    \end{subfigure}
    \begin{subfigure}[t]{0.02\textwidth}
        \textbf{(b)}    
    \end{subfigure}
    \begin{subfigure}[t]{0.47\linewidth}
        \includegraphics[width=1.0\linewidth,valign=t]{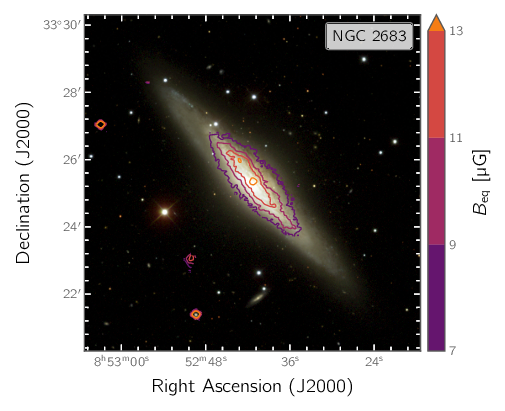}
    \end{subfigure}
    \caption{NGC~2683. \map}
    \label{fig:n2683}
\end{figure*}
\addcontentsline{toc}{subsection}{NGC 2683}

\begin{figure*}
	\centering
    \begin{subfigure}[t]{0.02\textwidth}
        \textbf{(a)}    
    \end{subfigure}
    \begin{subfigure}[t]{0.47\linewidth}
        \includegraphics[width=1.0\linewidth,valign=t]{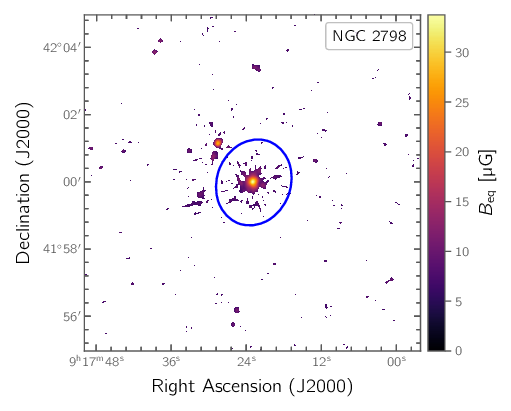}
    \end{subfigure}
    \begin{subfigure}[t]{0.02\textwidth}
        \textbf{(b)}    
    \end{subfigure}
    \begin{subfigure}[t]{0.47\linewidth}
        \includegraphics[width=1.0\linewidth,valign=t]{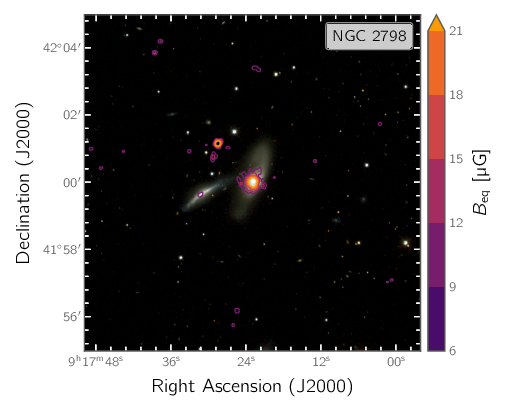}
    \end{subfigure}
    \caption{NGC~2798. \map}
    \label{fig:n2798}
\end{figure*}
\addcontentsline{toc}{subsection}{NGC 2798}

\begin{figure*}
	\centering
    \begin{subfigure}[t]{0.02\textwidth}
        \textbf{(a)}    
    \end{subfigure}
    \begin{subfigure}[t]{0.47\linewidth}
        \includegraphics[width=1.0\linewidth,valign=t]{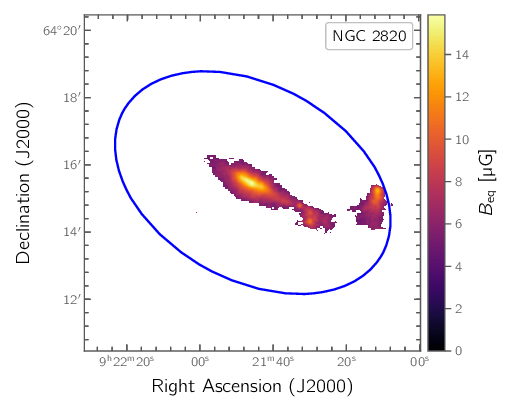}
    \end{subfigure}
    \begin{subfigure}[t]{0.02\textwidth}
        \textbf{(b)}    
    \end{subfigure}
    \begin{subfigure}[t]{0.47\linewidth}
        \includegraphics[width=1.0\linewidth,valign=t]{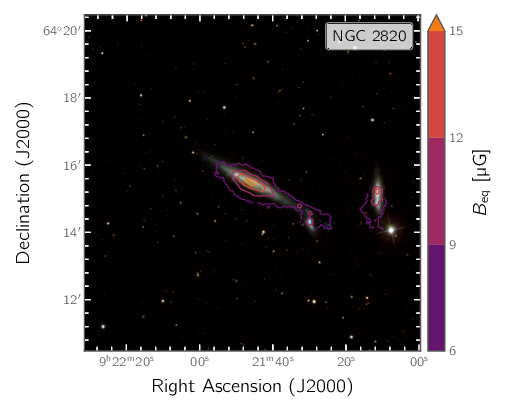}
    \end{subfigure}
    \caption{NGC~2820. \map}
    \label{fig:n2820}
\end{figure*}
\addcontentsline{toc}{subsection}{NGC 2820}

\begin{figure*}
	\centering
    \begin{subfigure}[t]{0.02\textwidth}
        \textbf{(a)}    
    \end{subfigure}
    \begin{subfigure}[t]{0.47\linewidth}
        \includegraphics[width=1.0\linewidth,valign=t]{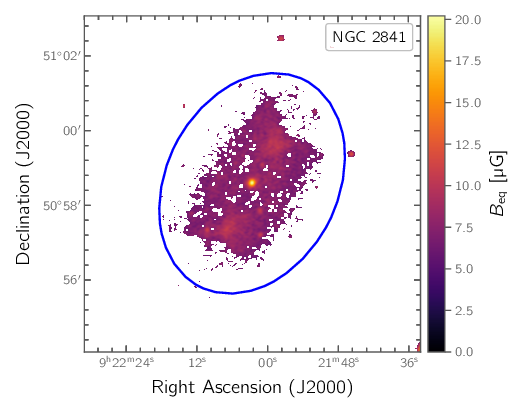}
    \end{subfigure}
    \begin{subfigure}[t]{0.02\textwidth}
        \textbf{(b)}    
    \end{subfigure}
    \begin{subfigure}[t]{0.47\linewidth}
        \includegraphics[width=1.0\linewidth,valign=t]{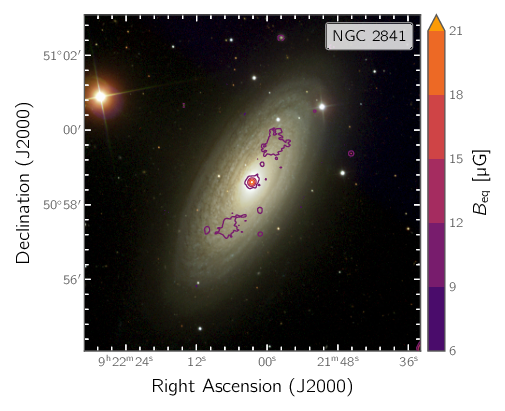}
    \end{subfigure}
    \caption{NGC~2841. \map}
    \label{fig:n2841}
\end{figure*}
\addcontentsline{toc}{subsection}{NGC 2841}

\begin{figure*}
	\centering
    \begin{subfigure}[t]{0.02\textwidth}
        \textbf{(a)}    
    \end{subfigure}
    \begin{subfigure}[t]{0.47\linewidth}
        \includegraphics[width=1.0\linewidth,valign=t]{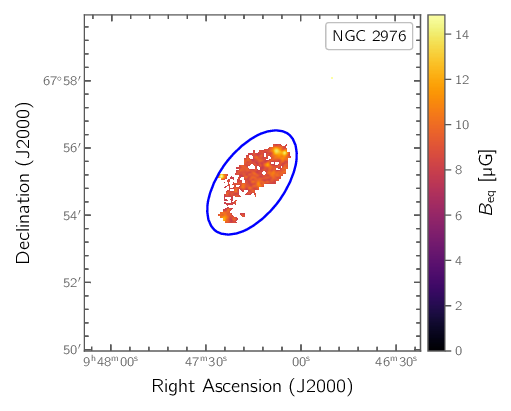}
    \end{subfigure}
    \begin{subfigure}[t]{0.02\textwidth}
        \textbf{(b)}    
    \end{subfigure}
    \begin{subfigure}[t]{0.47\linewidth}
        \includegraphics[width=1.0\linewidth,valign=t]{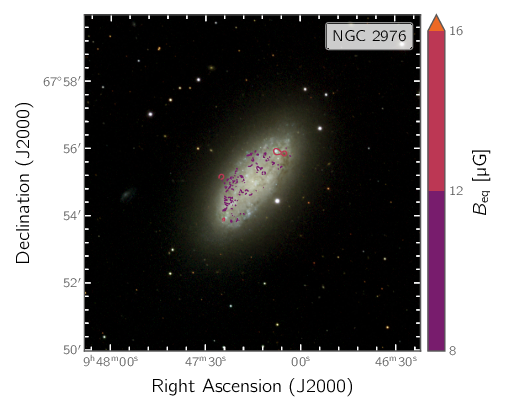}
    \end{subfigure}
    \caption{NGC~2976. \map}
    \label{fig:n2976}
\end{figure*}
\addcontentsline{toc}{subsection}{NGC 2976}

\begin{figure*}
	\centering
    \begin{subfigure}[t]{0.02\textwidth}
        \textbf{(a)}    
    \end{subfigure}
    \begin{subfigure}[t]{0.47\linewidth}
        \includegraphics[width=1.0\linewidth,valign=t]{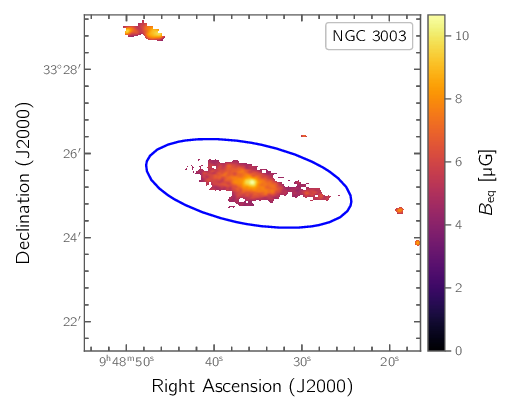}
    \end{subfigure}
    \begin{subfigure}[t]{0.02\textwidth}
        \textbf{(b)}    
    \end{subfigure}
    \begin{subfigure}[t]{0.47\linewidth}
        \includegraphics[width=1.0\linewidth,valign=t]{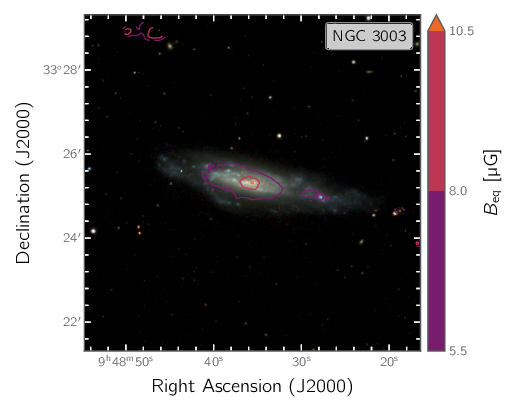}
    \end{subfigure}
    \caption{NGC~3003. \map}
    \label{fig:n3003}
\end{figure*}
\addcontentsline{toc}{subsection}{NGC 3003}

\begin{figure*}
	\centering
    \begin{subfigure}[t]{0.02\textwidth}
        \textbf{(a)}    
    \end{subfigure}
    \begin{subfigure}[t]{0.47\linewidth}
        \includegraphics[width=1.0\linewidth,valign=t]{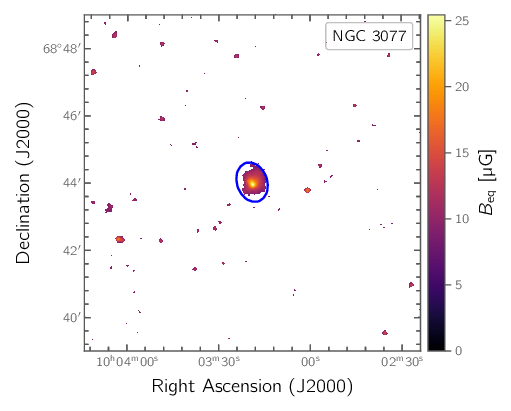}
    \end{subfigure}
    \begin{subfigure}[t]{0.02\textwidth}
        \textbf{(b)}    
    \end{subfigure}
    \begin{subfigure}[t]{0.47\linewidth}
        \includegraphics[width=1.0\linewidth,valign=t]{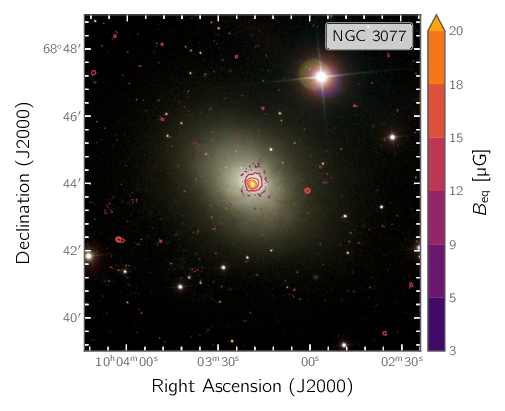}
    \end{subfigure}
    \caption{NGC~3077. \map}
    \label{fig:n3077}
\end{figure*}
\addcontentsline{toc}{subsection}{NGC 3077}

\begin{figure*}
	\centering
    \begin{subfigure}[t]{0.02\textwidth}
        \textbf{(a)}    
    \end{subfigure}
    \begin{subfigure}[t]{0.47\linewidth}
        \includegraphics[width=1.0\linewidth,valign=t]{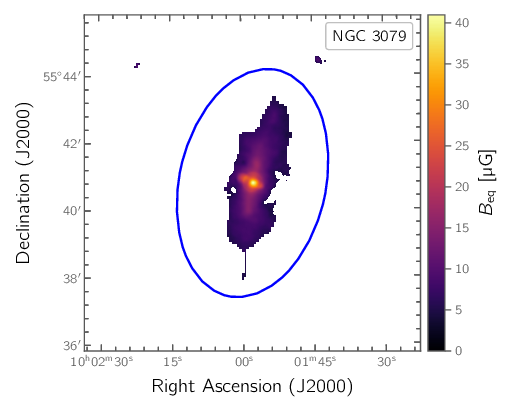}
    \end{subfigure}
    \begin{subfigure}[t]{0.02\textwidth}
        \textbf{(b)}    
    \end{subfigure}
    \begin{subfigure}[t]{0.47\linewidth}
        \includegraphics[width=1.0\linewidth,valign=t]{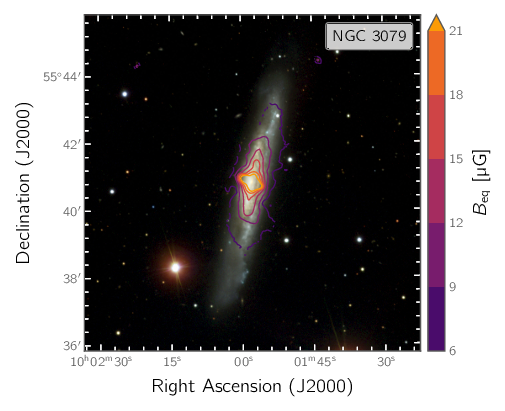}
    \end{subfigure}
    \caption{NGC~3079. \map}
    \label{fig:n3079}
\end{figure*}
\addcontentsline{toc}{subsection}{NGC 3079}

\begin{figure*}
	\centering
    \begin{subfigure}[t]{0.02\textwidth}
        \textbf{(a)}    
    \end{subfigure}
    \begin{subfigure}[t]{0.47\linewidth}
        \includegraphics[width=1.0\linewidth,valign=t]{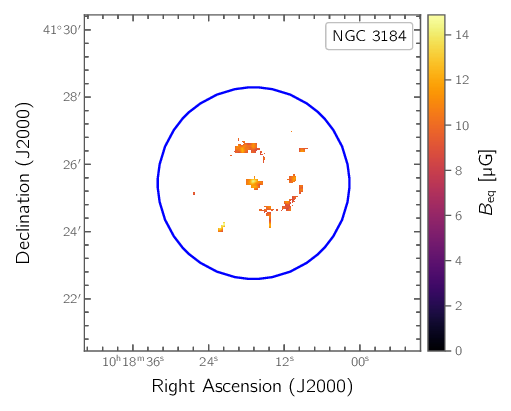}
    \end{subfigure}
    \begin{subfigure}[t]{0.02\textwidth}
        \textbf{(b)}    
    \end{subfigure}
    \begin{subfigure}[t]{0.47\linewidth}
        \includegraphics[width=1.0\linewidth,valign=t]{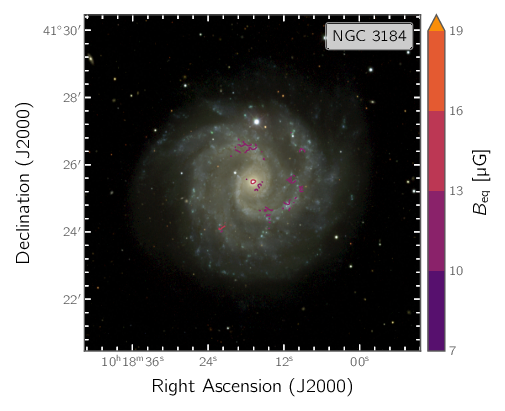}
    \end{subfigure}
    \caption{NGC~3184. \map}
    \label{fig:n3184}
\end{figure*}
\addcontentsline{toc}{subsection}{NGC 3184}

\begin{figure*}
	\centering
    \begin{subfigure}[t]{0.02\textwidth}
        \textbf{(a)}    
    \end{subfigure}
    \begin{subfigure}[t]{0.47\linewidth}
        \includegraphics[width=1.0\linewidth,valign=t]{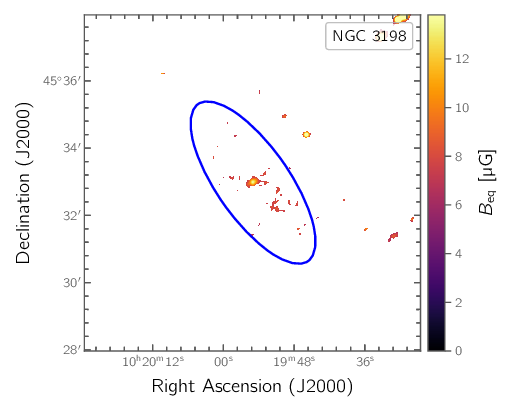}
    \end{subfigure}
    \begin{subfigure}[t]{0.02\textwidth}
        \textbf{(b)}    
    \end{subfigure}
    \begin{subfigure}[t]{0.47\linewidth}
        \includegraphics[width=1.0\linewidth,valign=t]{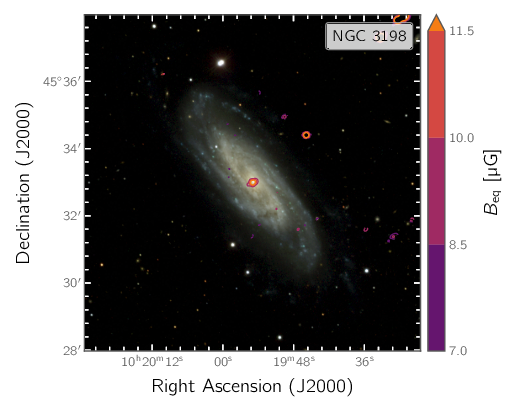}
    \end{subfigure}
    \caption{NGC~3198. \map}
    \label{fig:n3198}
\end{figure*}
\addcontentsline{toc}{subsection}{NGC 3198}

\begin{figure*}
	\centering
    \begin{subfigure}[t]{0.02\textwidth}
        \textbf{(a)}    
    \end{subfigure}
    \begin{subfigure}[t]{0.47\linewidth}
        \includegraphics[width=1.0\linewidth,valign=t]{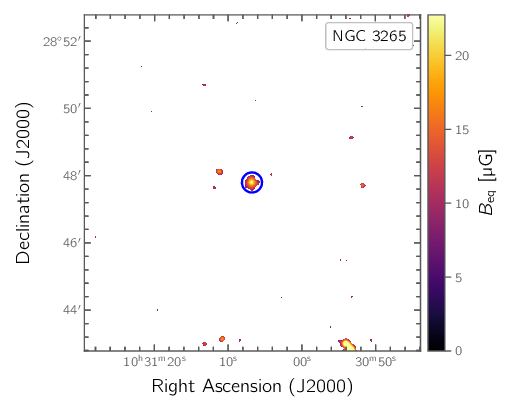}
    \end{subfigure}
    \begin{subfigure}[t]{0.02\textwidth}
        \textbf{(b)}    
    \end{subfigure}
    \begin{subfigure}[t]{0.47\linewidth}
        \includegraphics[width=1.0\linewidth,valign=t]{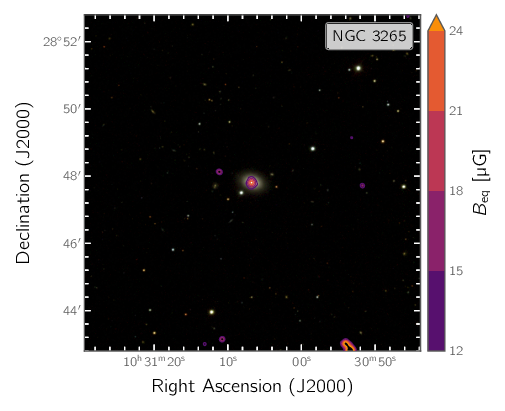}
    \end{subfigure}
    \caption{NGC~3265. \map}
    \label{fig:n3265}
\end{figure*}
\addcontentsline{toc}{subsection}{NGC 3265}

\begin{figure*}
	\centering
    \begin{subfigure}[t]{0.02\textwidth}
        \textbf{(a)}    
    \end{subfigure}
    \begin{subfigure}[t]{0.47\linewidth}
        \includegraphics[width=1.0\linewidth,valign=t]{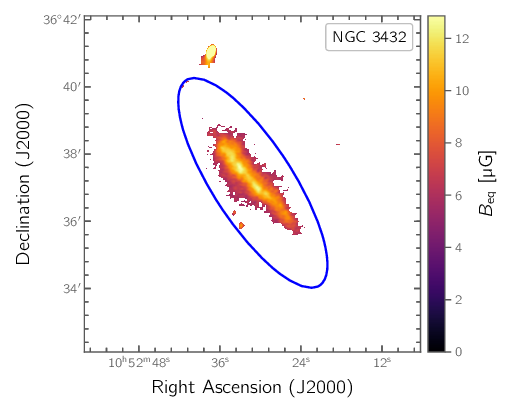}
    \end{subfigure}
    \begin{subfigure}[t]{0.02\textwidth}
        \textbf{(b)}    
    \end{subfigure}
    \begin{subfigure}[t]{0.47\linewidth}
        \includegraphics[width=1.0\linewidth,valign=t]{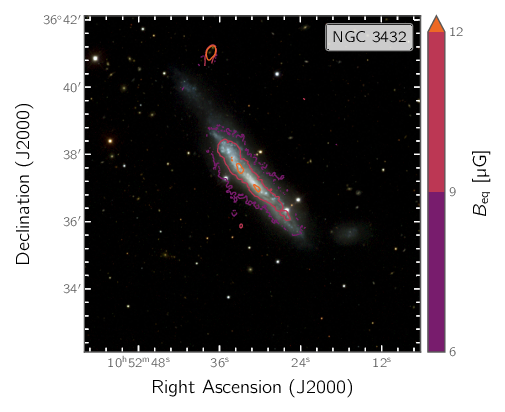}
    \end{subfigure}
    \caption{NGC~3432. \map}
    \label{fig:n3432}
\end{figure*}
\addcontentsline{toc}{subsection}{NGC 3432}

\begin{figure*}
	\centering
    \begin{subfigure}[t]{0.02\textwidth}
        \textbf{(a)}    
    \end{subfigure}
    \begin{subfigure}[t]{0.47\linewidth}
        \includegraphics[width=1.0\linewidth,valign=t]{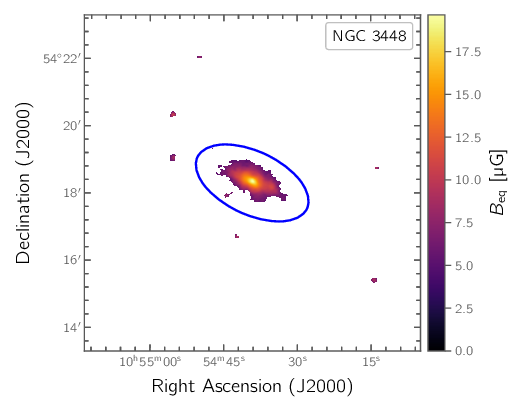}
    \end{subfigure}
    \begin{subfigure}[t]{0.02\textwidth}
        \textbf{(b)}    
    \end{subfigure}
    \begin{subfigure}[t]{0.47\linewidth}
        \includegraphics[width=1.0\linewidth,valign=t]{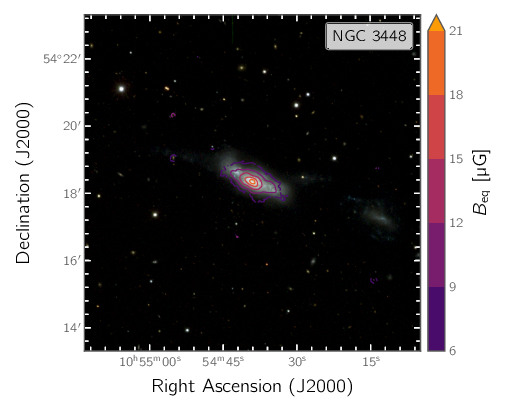}
    \end{subfigure}
    \caption{NGC~3448. \map}
    \label{fig:n3448}
\end{figure*}
\addcontentsline{toc}{subsection}{NGC 3448}

\begin{figure*}
	\centering
    \begin{subfigure}[t]{0.02\textwidth}
        \textbf{(a)}    
    \end{subfigure}
    \begin{subfigure}[t]{0.47\linewidth}
        \includegraphics[width=1.0\linewidth,valign=t]{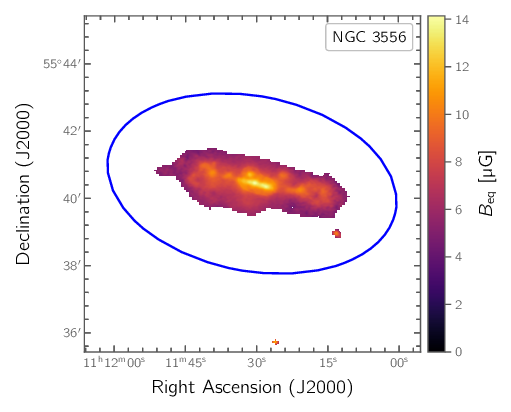}
    \end{subfigure}
    \begin{subfigure}[t]{0.02\textwidth}
        \textbf{(b)}    
    \end{subfigure}
    \begin{subfigure}[t]{0.47\linewidth}
        \includegraphics[width=1.0\linewidth,valign=t]{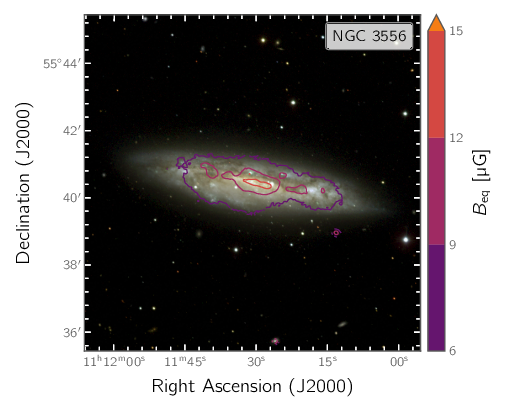}
    \end{subfigure}
    \caption{NGC~3556. \map}
    \label{fig:n3556}
\end{figure*}
\addcontentsline{toc}{subsection}{NGC 3556}

\begin{figure*}
	\centering
    \begin{subfigure}[t]{0.02\textwidth}
        \textbf{(a)}    
    \end{subfigure}
    \begin{subfigure}[t]{0.47\linewidth}
        \includegraphics[width=1.0\linewidth,valign=t]{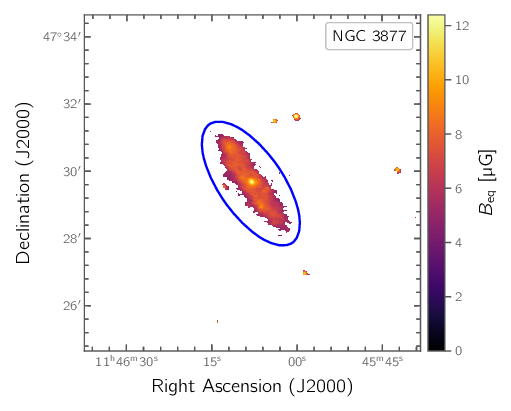}
    \end{subfigure}
    \begin{subfigure}[t]{0.02\textwidth}
        \textbf{(b)}    
    \end{subfigure}
    \begin{subfigure}[t]{0.47\linewidth}
        \includegraphics[width=1.0\linewidth,valign=t]{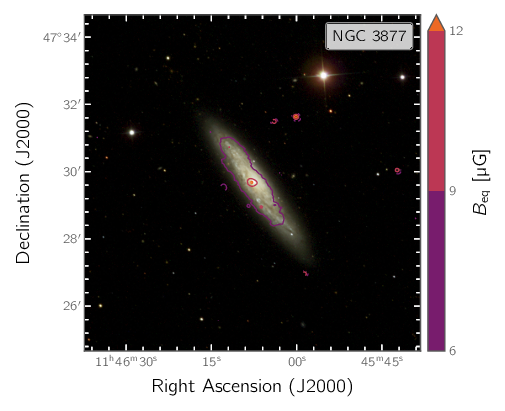}
    \end{subfigure}
    \caption{NGC~3877. \map}
    \label{fig:n3877}
\end{figure*}
\addcontentsline{toc}{subsection}{NGC 3877}

\begin{figure*}
	\centering
    \begin{subfigure}[t]{0.02\textwidth}
        \textbf{(a)}    
    \end{subfigure}
    \begin{subfigure}[t]{0.47\linewidth}
        \includegraphics[width=1.0\linewidth,valign=t]{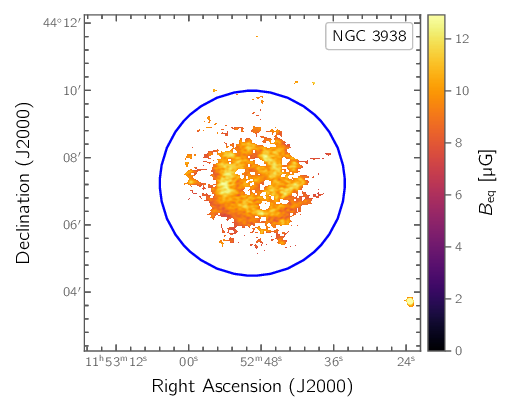}
    \end{subfigure}
    \begin{subfigure}[t]{0.02\textwidth}
        \textbf{(b)}    
    \end{subfigure}
    \begin{subfigure}[t]{0.47\linewidth}
        \includegraphics[width=1.0\linewidth,valign=t]{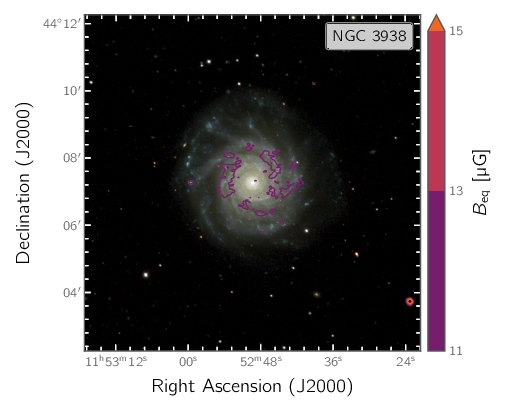}
    \end{subfigure}
    \caption{NGC~3938. \map}
    \label{fig:n3938}
\end{figure*}
\addcontentsline{toc}{subsection}{NGC 3938}

\begin{figure*}
	\centering
    \begin{subfigure}[t]{0.02\textwidth}
        \textbf{(a)}    
    \end{subfigure}
    \begin{subfigure}[t]{0.47\linewidth}
        \includegraphics[width=1.0\linewidth,valign=t]{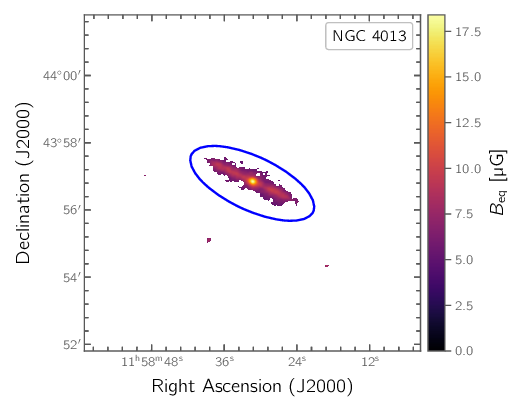}
    \end{subfigure}
    \begin{subfigure}[t]{0.02\textwidth}
        \textbf{(b)}    
    \end{subfigure}
    \begin{subfigure}[t]{0.47\linewidth}
        \includegraphics[width=1.0\linewidth,valign=t]{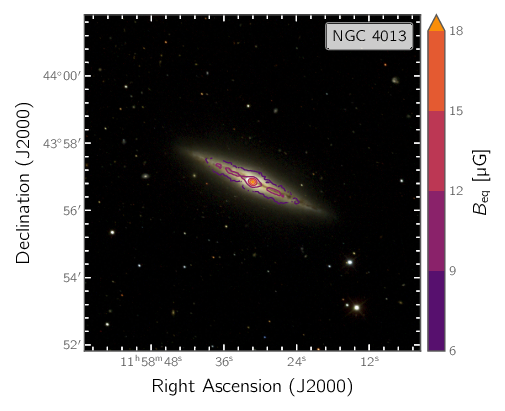}
    \end{subfigure}
    \caption{NGC~4013. \map}
    \label{fig:n4013}
\end{figure*}
\addcontentsline{toc}{subsection}{NGC 4013}

\begin{figure*}
	\centering
    \begin{subfigure}[t]{0.02\textwidth}
        \textbf{(a)}    
    \end{subfigure}
    \begin{subfigure}[t]{0.47\linewidth}
        \includegraphics[width=1.0\linewidth,valign=t]{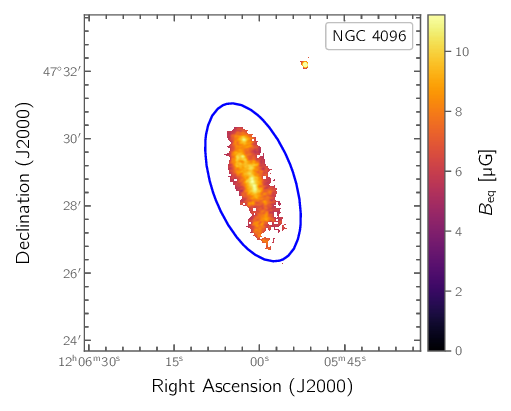}
    \end{subfigure}
    \begin{subfigure}[t]{0.02\textwidth}
        \textbf{(b)}    
    \end{subfigure}
    \begin{subfigure}[t]{0.47\linewidth}
        \includegraphics[width=1.0\linewidth,valign=t]{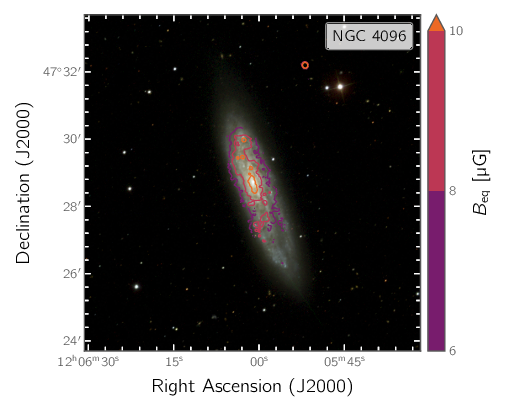}
    \end{subfigure}
    \caption{NGC~4096. \map}
    \label{fig:n4096}
\end{figure*}
\addcontentsline{toc}{subsection}{NGC 4096}

\begin{figure*}
	\centering
    \begin{subfigure}[t]{0.02\textwidth}
        \textbf{(a)}    
    \end{subfigure}
    \begin{subfigure}[t]{0.47\linewidth}
        \includegraphics[width=1.0\linewidth,valign=t]{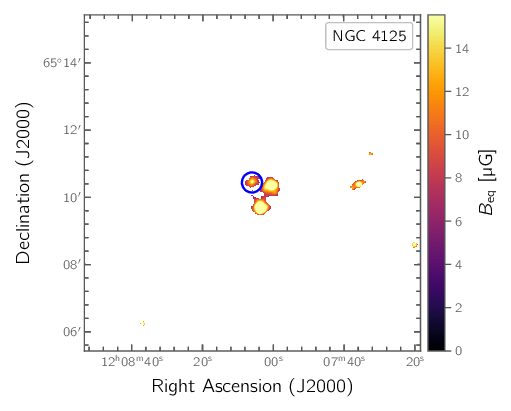}
    \end{subfigure}
    \begin{subfigure}[t]{0.02\textwidth}
        \textbf{(b)}    
    \end{subfigure}
    \begin{subfigure}[t]{0.47\linewidth}
        \includegraphics[width=1.0\linewidth,valign=t]{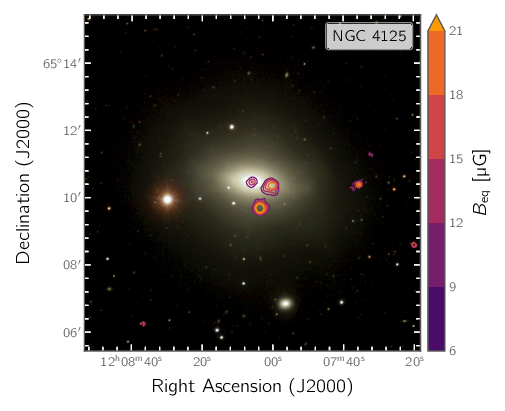}
    \end{subfigure}
    \caption{NGC~4125. \map}
    \label{fig:n4125}
\end{figure*}
\addcontentsline{toc}{subsection}{NGC 4125}

\begin{figure*}
	\centering
    \begin{subfigure}[t]{0.02\textwidth}
        \textbf{(a)}    
    \end{subfigure}
    \begin{subfigure}[t]{0.47\linewidth}
        \includegraphics[width=1.0\linewidth,valign=t]{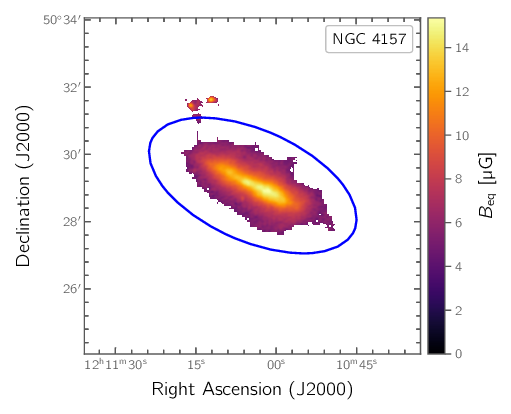}
    \end{subfigure}
    \begin{subfigure}[t]{0.02\textwidth}
        \textbf{(b)}    
    \end{subfigure}
    \begin{subfigure}[t]{0.47\linewidth}
        \includegraphics[width=1.0\linewidth,valign=t]{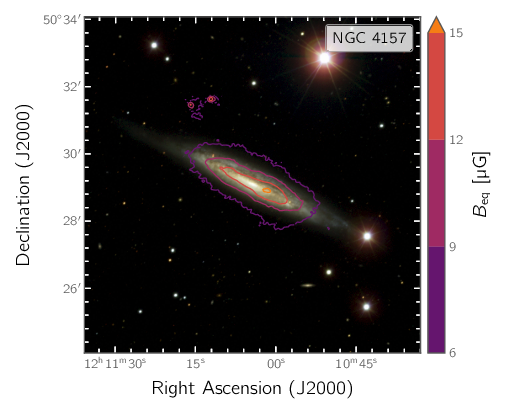}
    \end{subfigure}
    \caption{NGC~4157. \map}
    \label{fig:n4157}
\end{figure*}
\addcontentsline{toc}{subsection}{NGC 4157}

\begin{figure*}
	\centering
    \begin{subfigure}[t]{0.02\textwidth}
        \textbf{(a)}    
    \end{subfigure}
    \begin{subfigure}[t]{0.47\linewidth}
        \includegraphics[width=1.0\linewidth,valign=t]{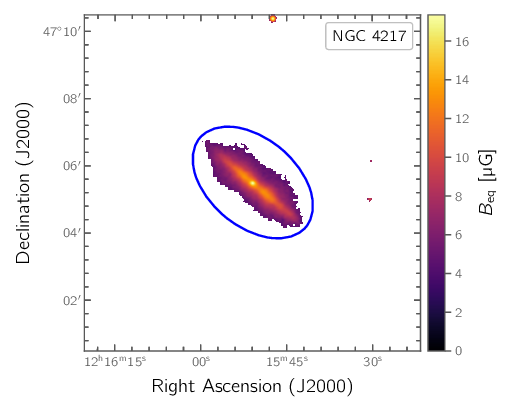}
    \end{subfigure}
    \begin{subfigure}[t]{0.02\textwidth}
        \textbf{(b)}    
    \end{subfigure}
    \begin{subfigure}[t]{0.47\linewidth}
        \includegraphics[width=1.0\linewidth,valign=t]{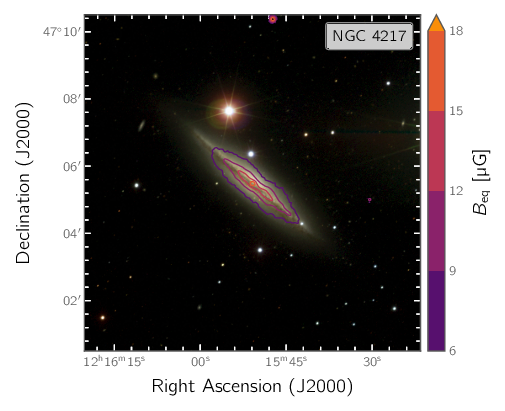}
    \end{subfigure}
    \caption{NGC~4217. \map}
    \label{fig:n4217}
\end{figure*}
\addcontentsline{toc}{subsection}{NGC 4217}

\begin{figure*}
	\centering
    \begin{subfigure}[t]{0.02\textwidth}
        \textbf{(a)}    
    \end{subfigure}
    \begin{subfigure}[t]{0.47\linewidth}
        \includegraphics[width=1.0\linewidth,valign=t]{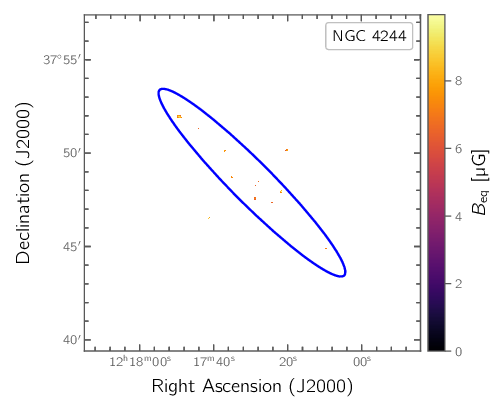}
    \end{subfigure}
    \begin{subfigure}[t]{0.02\textwidth}
        \textbf{(b)}    
    \end{subfigure}
    \begin{subfigure}[t]{0.47\linewidth}
        \includegraphics[width=1.0\linewidth,valign=t]{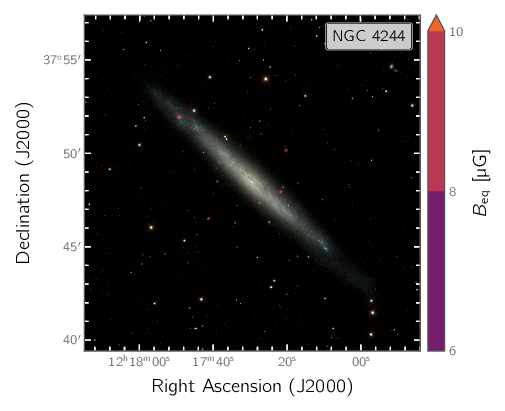}
    \end{subfigure}
    \caption{NGC~4244. \map}
    \label{fig:n4244}
\end{figure*}
\addcontentsline{toc}{subsection}{NGC 4244}

\begin{figure*}
	\centering
    \begin{subfigure}[t]{0.02\textwidth}
        \textbf{(a)}    
    \end{subfigure}
    \begin{subfigure}[t]{0.47\linewidth}
        \includegraphics[width=1.0\linewidth,valign=t]{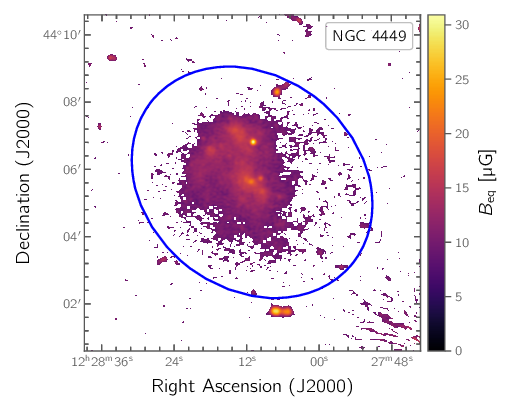}
    \end{subfigure}
    \begin{subfigure}[t]{0.02\textwidth}
        \textbf{(b)}    
    \end{subfigure}
    \begin{subfigure}[t]{0.47\linewidth}
        \includegraphics[width=1.0\linewidth,valign=t]{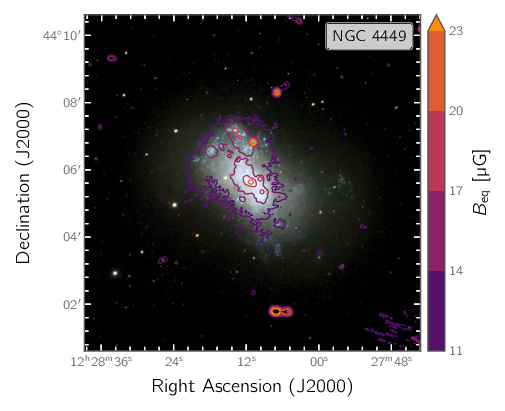}
    \end{subfigure}
    \caption{NGC~4449. \map}
    \label{fig:n4449}
\end{figure*}
\addcontentsline{toc}{subsection}{NGC 4449}

\begin{figure*}
	\centering
    \begin{subfigure}[t]{0.02\textwidth}
        \textbf{(a)}    
    \end{subfigure}
    \begin{subfigure}[t]{0.47\linewidth}
        \includegraphics[width=1.0\linewidth,valign=t]{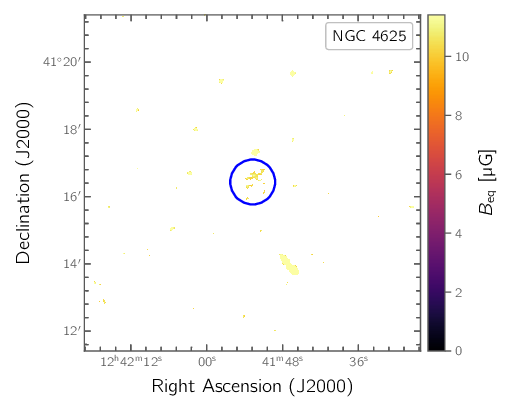}
    \end{subfigure}
    \begin{subfigure}[t]{0.02\textwidth}
        \textbf{(b)}    
    \end{subfigure}
    \begin{subfigure}[t]{0.47\linewidth}
        \includegraphics[width=1.0\linewidth,valign=t]{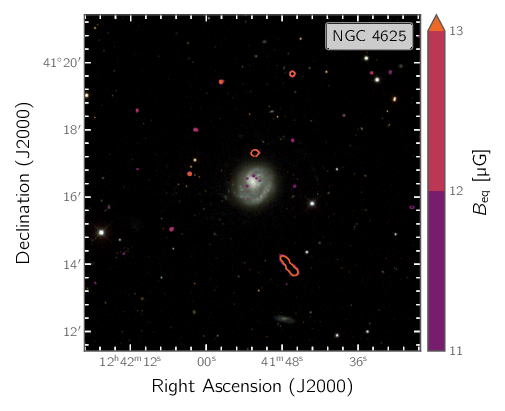}
    \end{subfigure}
    \caption{NGC~4625. \map}
    \label{fig:n4625}
\end{figure*}
\addcontentsline{toc}{subsection}{NGC 4625}

\begin{figure*}
	\centering
    \begin{subfigure}[t]{0.02\textwidth}
        \textbf{(a)}    
    \end{subfigure}
    \begin{subfigure}[t]{0.47\linewidth}
        \includegraphics[width=1.0\linewidth,valign=t]{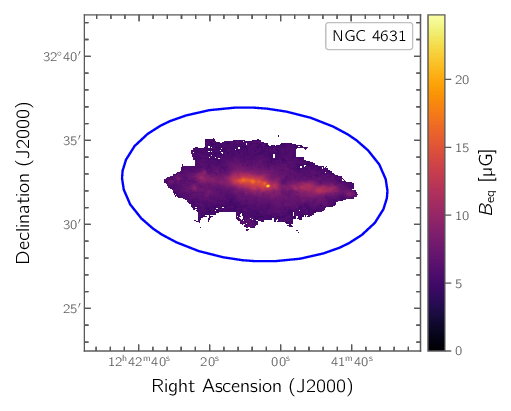}
    \end{subfigure}
    \begin{subfigure}[t]{0.02\textwidth}
        \textbf{(b)}    
    \end{subfigure}
    \begin{subfigure}[t]{0.47\linewidth}
        \includegraphics[width=1.0\linewidth,valign=t]{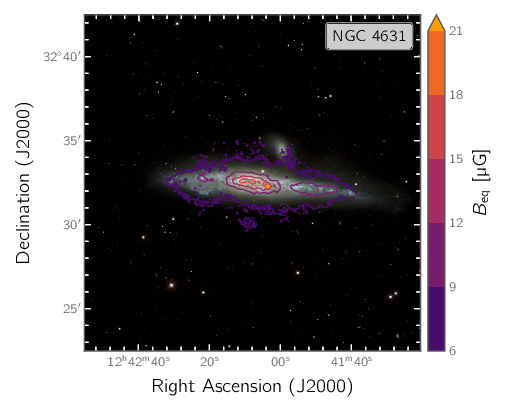}
    \end{subfigure}
    \caption{NGC~4631. \map}
    \label{fig:n4631}
\end{figure*}
\addcontentsline{toc}{subsection}{NGC 4631}

\begin{figure*}
	\centering
    \begin{subfigure}[t]{0.02\textwidth}
        \textbf{(a)}    
    \end{subfigure}
    \begin{subfigure}[t]{0.47\linewidth}
        \includegraphics[width=1.0\linewidth,valign=t]{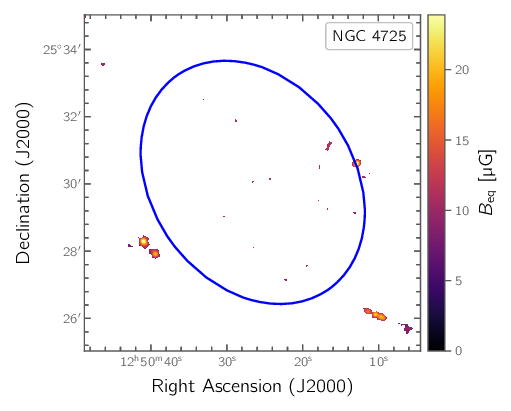}
    \end{subfigure}
    \begin{subfigure}[t]{0.02\textwidth}
        \textbf{(b)}    
    \end{subfigure}
    \begin{subfigure}[t]{0.47\linewidth}
        \includegraphics[width=1.0\linewidth,valign=t]{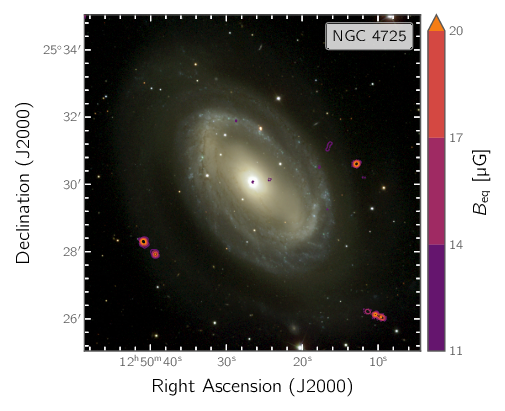}
    \end{subfigure}
    \caption{NGC~4725. \map}
    \label{fig:n4725}
\end{figure*}
\addcontentsline{toc}{subsection}{NGC 4725}

\begin{figure*}
	\centering
    \begin{subfigure}[t]{0.02\textwidth}
        \textbf{(a)}    
    \end{subfigure}
    \begin{subfigure}[t]{0.47\linewidth}
        \includegraphics[width=1.0\linewidth,valign=t]{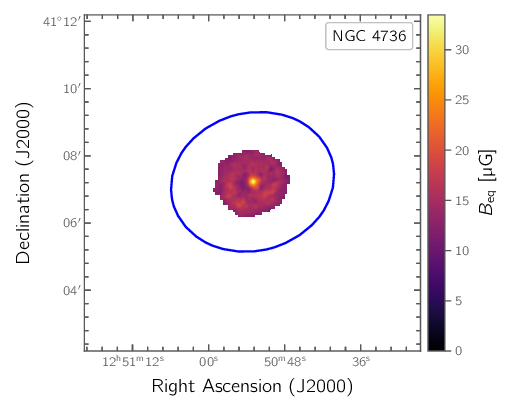}
    \end{subfigure}
    \begin{subfigure}[t]{0.02\textwidth}
        \textbf{(b)}    
    \end{subfigure}
    \begin{subfigure}[t]{0.47\linewidth}
        \includegraphics[width=1.0\linewidth,valign=t]{atlas/n4736_magnetic_non_thermal.pdf.png}
    \end{subfigure}
    \caption{NGC~4736. \map}
    \label{fig:n4736}
\end{figure*}
\addcontentsline{toc}{subsection}{NGC 4736}

\begin{figure*}
	\centering
    \begin{subfigure}[t]{0.02\textwidth}
        \textbf{(a)}    
    \end{subfigure}
    \begin{subfigure}[t]{0.47\linewidth}
        \includegraphics[width=1.0\linewidth,valign=t]{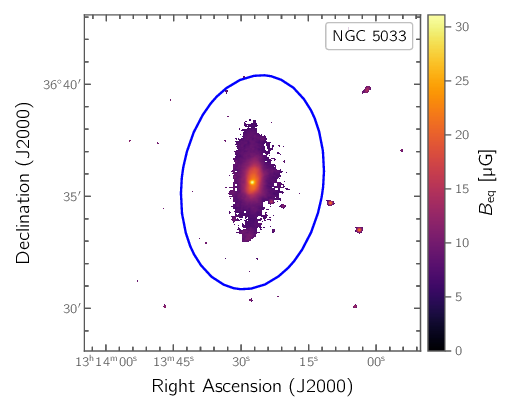}
    \end{subfigure}
    \begin{subfigure}[t]{0.02\textwidth}
        \textbf{(b)}    
    \end{subfigure}
    \begin{subfigure}[t]{0.47\linewidth}
        \includegraphics[width=1.0\linewidth,valign=t]{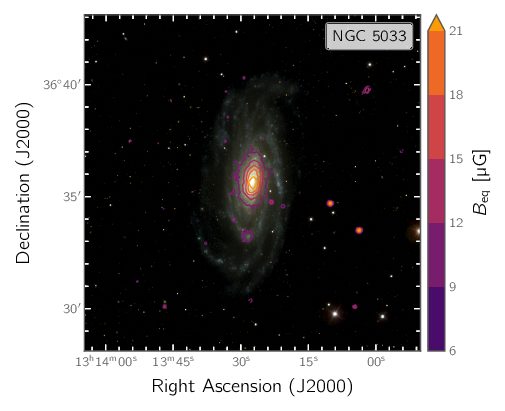}
    \end{subfigure}
    \caption{NGC~5033. \map}
    \label{fig:n5033}
\end{figure*}
\addcontentsline{toc}{subsection}{NGC 5033}

\begin{figure*}
	\centering
    \begin{subfigure}[t]{0.02\textwidth}
        \textbf{(a)}    
    \end{subfigure}
    \begin{subfigure}[t]{0.47\linewidth}
        \includegraphics[width=1.0\linewidth,valign=t]{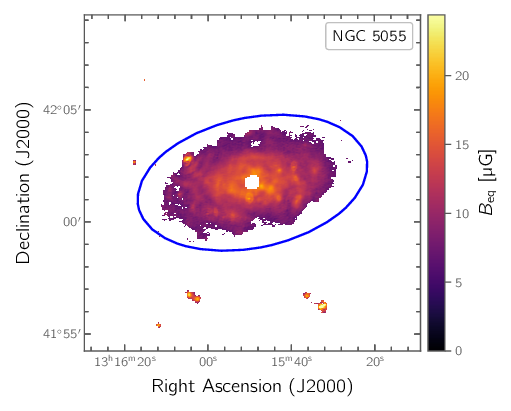}
    \end{subfigure}
    \begin{subfigure}[t]{0.02\textwidth}
        \textbf{(b)}    
    \end{subfigure}
    \begin{subfigure}[t]{0.47\linewidth}
        \includegraphics[width=1.0\linewidth,valign=t]{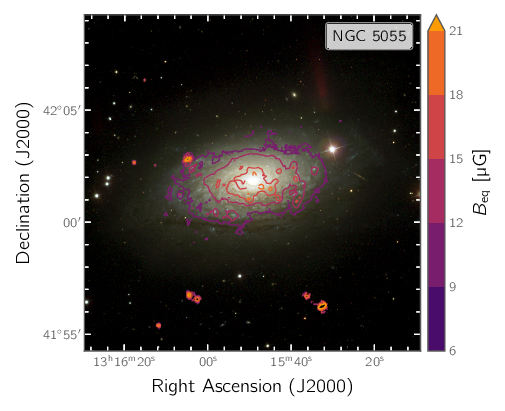}
    \end{subfigure}
    \caption{NGC~5055. \map}
    \label{fig:n5055}
\end{figure*}
\addcontentsline{toc}{subsection}{NGC 5055}

\begin{figure*}
	\centering
    \begin{subfigure}[t]{0.02\textwidth}
        \textbf{(a)}    
    \end{subfigure}
    \begin{subfigure}[t]{0.47\linewidth}
        \includegraphics[width=1.0\linewidth,valign=t]{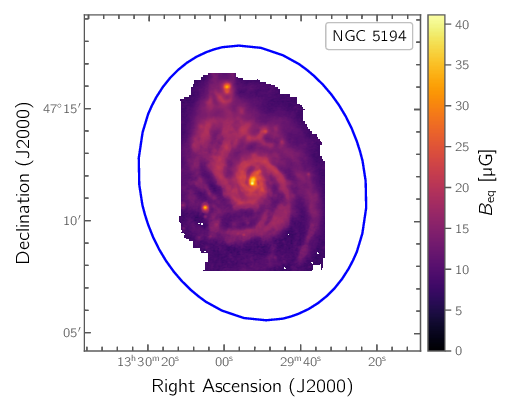}
    \end{subfigure}
    \begin{subfigure}[t]{0.02\textwidth}
        \textbf{(b)}    
    \end{subfigure}
    \begin{subfigure}[t]{0.47\linewidth}
        \includegraphics[width=1.0\linewidth,valign=t]{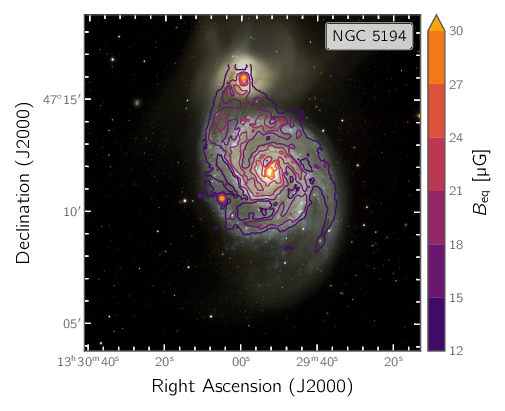}
    \end{subfigure}
    \caption{NGC~5194. \map}
    \label{fig:n5194}
\end{figure*}
\addcontentsline{toc}{subsection}{NGC 5194}

\begin{figure*}
	\centering
    \begin{subfigure}[t]{0.02\textwidth}
        \textbf{(a)}    
    \end{subfigure}
    \begin{subfigure}[t]{0.47\linewidth}
        \includegraphics[width=1.0\linewidth,valign=t]{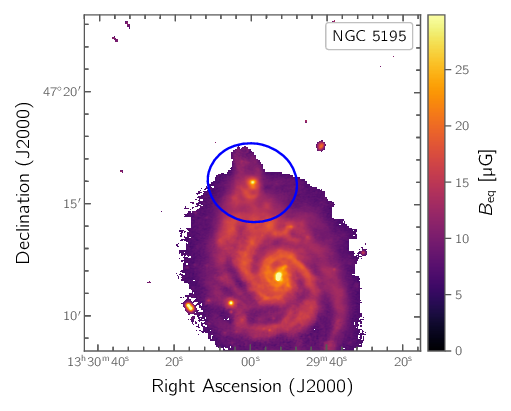}
    \end{subfigure}
    \begin{subfigure}[t]{0.02\textwidth}
        \textbf{(b)}    
    \end{subfigure}
    \begin{subfigure}[t]{0.47\linewidth}
        \includegraphics[width=1.0\linewidth,valign=t]{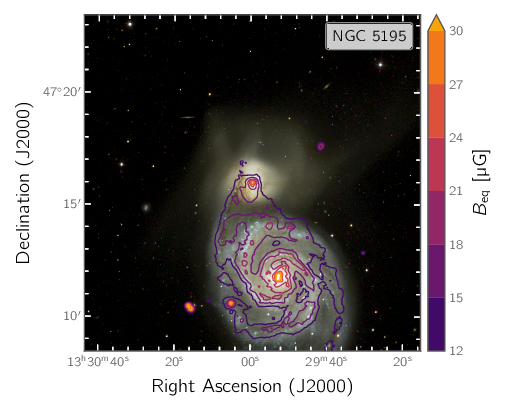}
    \end{subfigure}
    \caption{NGC~5195. \map}
    \label{fig:n5195}
\end{figure*}
\addcontentsline{toc}{subsection}{NGC 5195}

\begin{figure*}
	\centering
    \begin{subfigure}[t]{0.02\textwidth}
        \textbf{(a)}    
    \end{subfigure}
    \begin{subfigure}[t]{0.47\linewidth}
        \includegraphics[width=1.0\linewidth,valign=t]{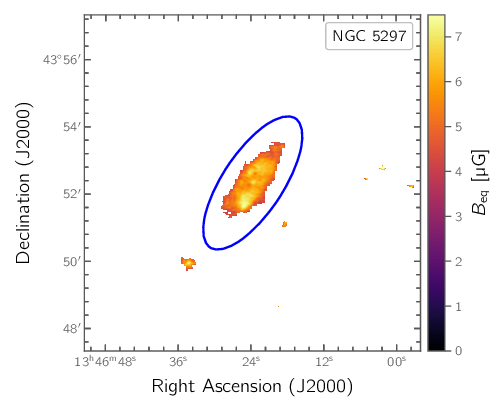}
    \end{subfigure}
    \begin{subfigure}[t]{0.02\textwidth}
        \textbf{(b)}    
    \end{subfigure}
    \begin{subfigure}[t]{0.47\linewidth}
        \includegraphics[width=1.0\linewidth,valign=t]{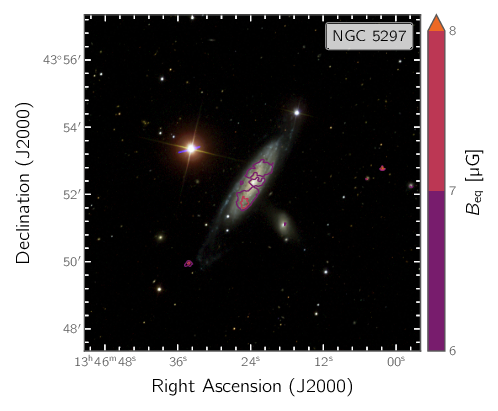}
    \end{subfigure}
    \caption{NGC~5297. \map}
    \label{fig:n5297}
\end{figure*}
\addcontentsline{toc}{subsection}{NGC 5297}

\begin{figure*}
	\centering
    \begin{subfigure}[t]{0.02\textwidth}
        \textbf{(a)}    
    \end{subfigure}
    \begin{subfigure}[t]{0.47\linewidth}
        \includegraphics[width=1.0\linewidth,valign=t]{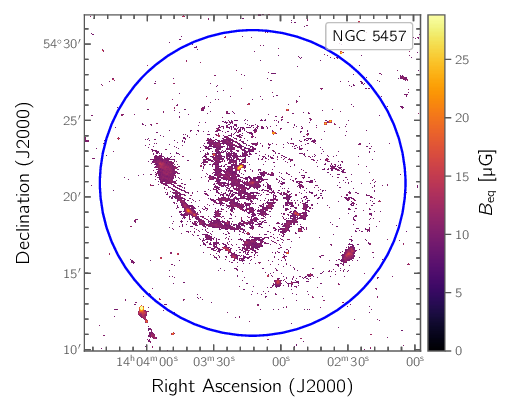}
    \end{subfigure}
    \begin{subfigure}[t]{0.02\textwidth}
        \textbf{(b)}    
    \end{subfigure}
    \begin{subfigure}[t]{0.47\linewidth}
        \includegraphics[width=1.0\linewidth,valign=t]{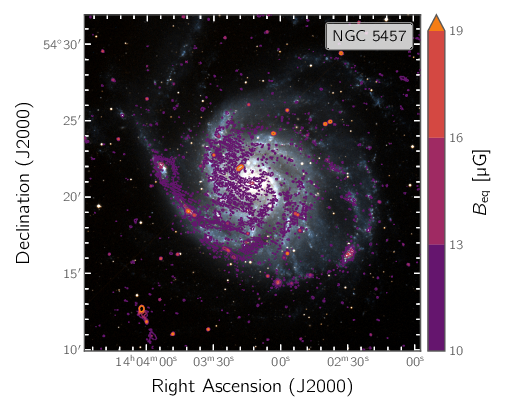}
    \end{subfigure}
    \caption{NGC~5457. \map}
    \label{fig:n5457}
\end{figure*}
\addcontentsline{toc}{subsection}{NGC 5457}

\begin{figure*}
	\centering
    \begin{subfigure}[t]{0.02\textwidth}
        \textbf{(a)}    
    \end{subfigure}
    \begin{subfigure}[t]{0.47\linewidth}
        \includegraphics[width=1.0\linewidth,valign=t]{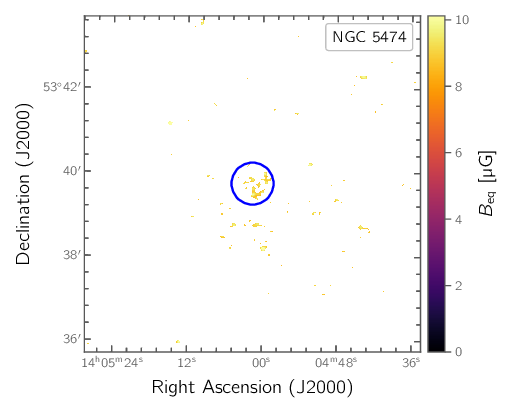}
    \end{subfigure}
    \begin{subfigure}[t]{0.02\textwidth}
        \textbf{(b)}    
    \end{subfigure}
    \begin{subfigure}[t]{0.47\linewidth}
        \includegraphics[width=1.0\linewidth,valign=t]{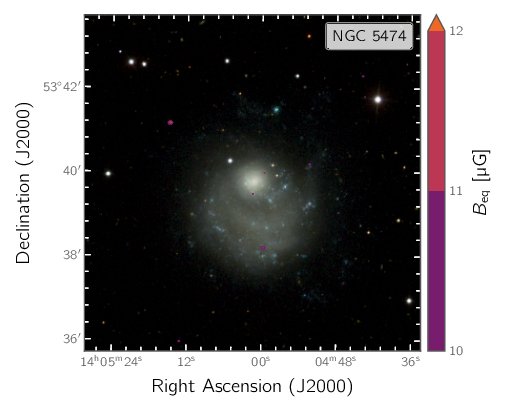}
    \end{subfigure}
    \caption{NGC~5474. \map}
    \label{fig:n5474}
\end{figure*}
\addcontentsline{toc}{subsection}{NGC 5474}

\begin{figure*}
	\centering
    \begin{subfigure}[t]{0.02\textwidth}
        \textbf{(a)}    
    \end{subfigure}
    \begin{subfigure}[t]{0.47\linewidth}
        \includegraphics[width=1.0\linewidth,valign=t]{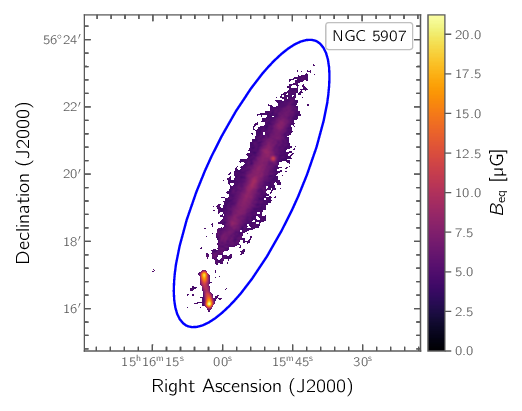}
    \end{subfigure}
    \begin{subfigure}[t]{0.02\textwidth}
        \textbf{(b)}    
    \end{subfigure}
    \begin{subfigure}[t]{0.47\linewidth}
        \includegraphics[width=1.0\linewidth,valign=t]{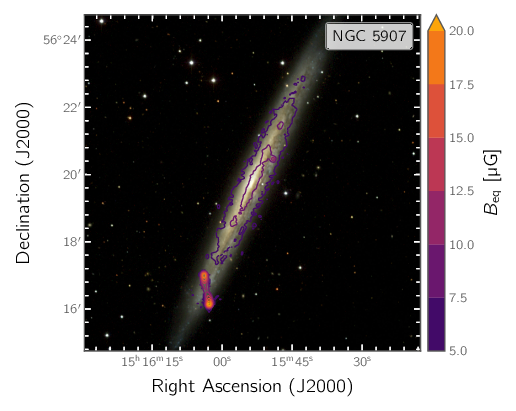}
    \end{subfigure}
    \caption{NGC~5907. \map}
    \label{fig:n5907}
\end{figure*}
\addcontentsline{toc}{subsection}{NGC 5907}

\begin{figure*}
	\centering
    \begin{subfigure}[t]{0.02\textwidth}
        \textbf{(a)}    
    \end{subfigure}
    \begin{subfigure}[t]{0.47\linewidth}
        \includegraphics[width=1.0\linewidth,valign=t]{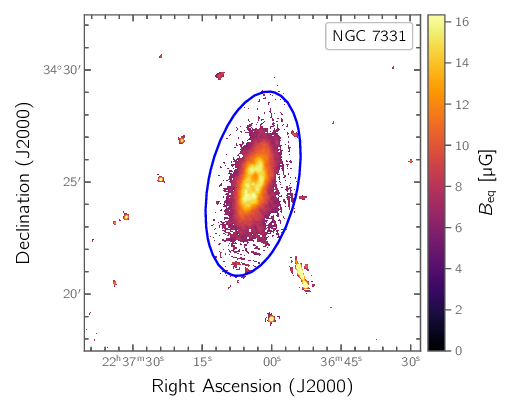}
    \end{subfigure}
    \begin{subfigure}[t]{0.02\textwidth}
        \textbf{(b)}    
    \end{subfigure}
    \begin{subfigure}[t]{0.47\linewidth}
        \includegraphics[width=1.0\linewidth,valign=t]{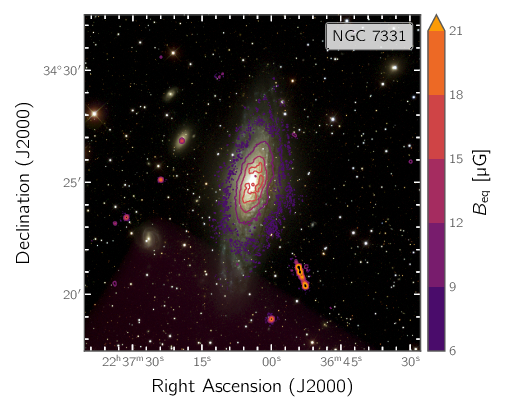}
    \end{subfigure}
    \caption{NGC~7331. \map}
    \label{fig:n7331}
\end{figure*}
\addcontentsline{toc}{subsection}{NGC 7331}

\section{Magnetic field--gas relation in individual galaxies}
\label{as:correlations}

In this Appendix, we present the magnetic field--gas relation in the sub-sample of nine galaxies, where we have analysed the data. We show the individual relations, separated as \mfhi relation for the atomic gas only, as \mfhtwo for the molecular gas only and as \mfgas for the combined atomic and molecular gas. We present for each galaxy an overlay of the equipartition magnetic field strength as contours on the gas mass surface densities and the corresponding relations. The contour levels of the magnetic field strength can read of from the colour bars of the maps. The data points in the individual relations are coloured according to the radio spectral index. Best-fitting relations are shown as solid lines and 1$\sigma$ confidence levels as grey-shaded areas. The theoretical lines for simplified assumptions such as constant velocity dispersion is shown as red dashed lines. The relations are presented in Figs.~\ref{fig:n925_corr}--\ref{fig:n7331_corr}.

\begin{figure*}
	\centering
    \begin{subfigure}[t]{0.02\textwidth}
        \textbf{(a)}    
    \end{subfigure}
    \begin{subfigure}[t]{0.47\linewidth}
        \includegraphics[width=0.9\linewidth,valign=t]{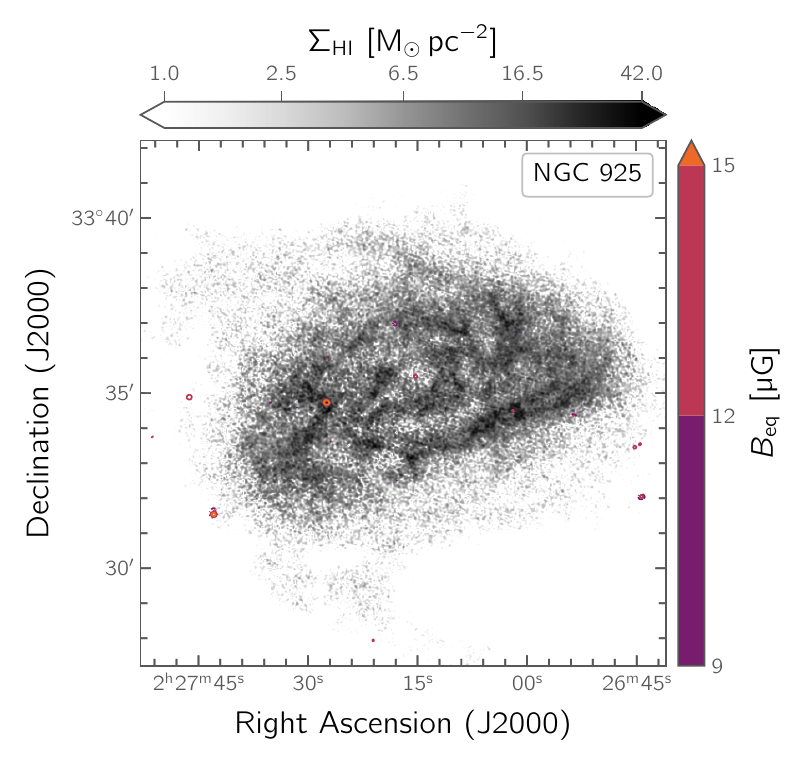}
    \end{subfigure}
    \begin{subfigure}[t]{0.02\textwidth}
        \textbf{(b)}    
    \end{subfigure}
    \begin{subfigure}[t]{0.47\linewidth}
        \vspace{0.7cm}
        \includegraphics[width=0.9\linewidth,valign=t]{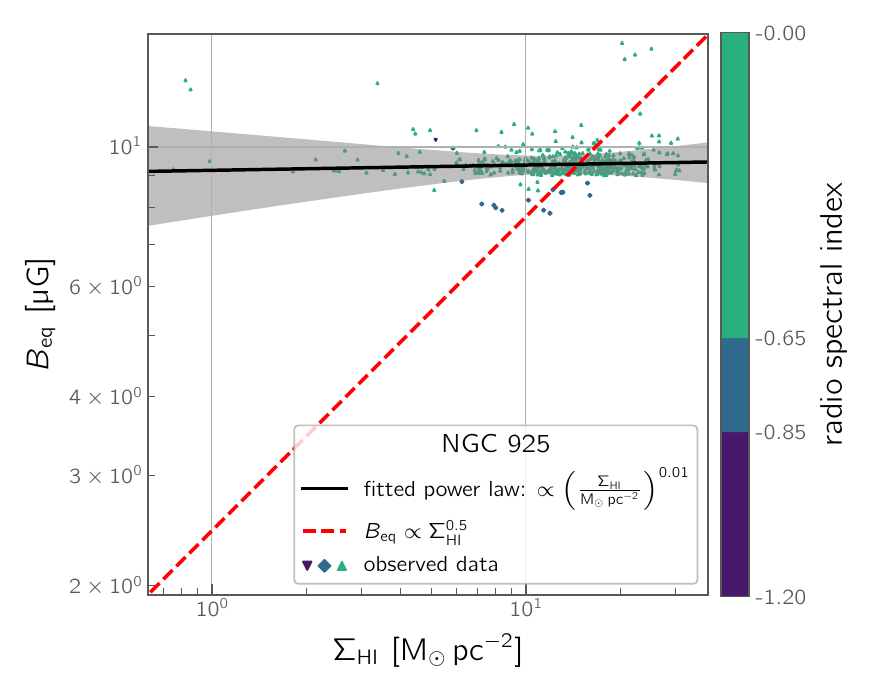}
    \end{subfigure}
    \\
    \begin{subfigure}[t]{0.02\textwidth}
        \textbf{(c)}    
    \end{subfigure}
    \begin{subfigure}[t]{0.47\linewidth}
        \includegraphics[width=0.9\linewidth,valign=t]{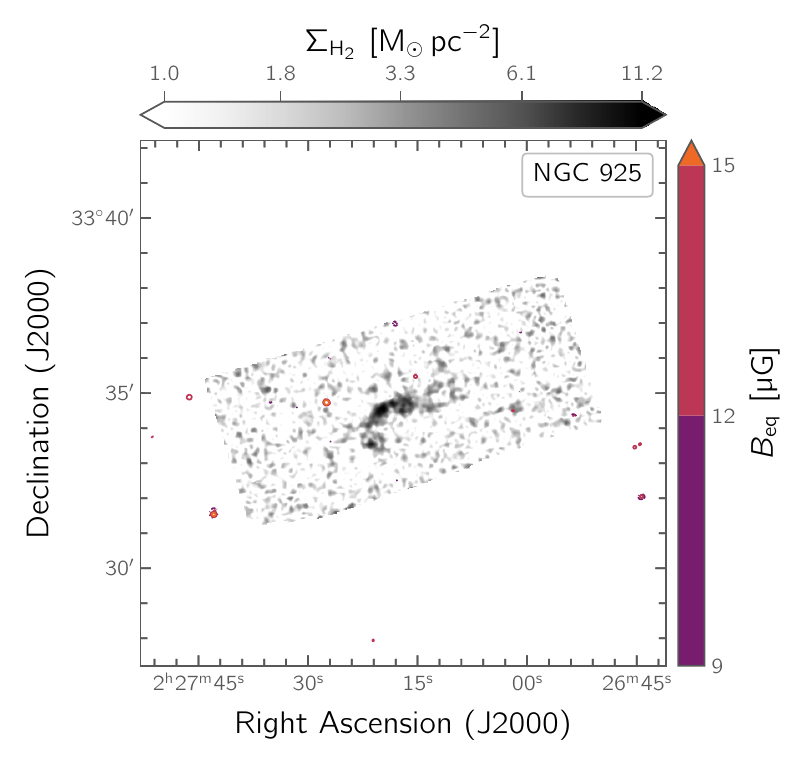}
    \end{subfigure}
    \begin{subfigure}[t]{0.02\textwidth}
        \textbf{(d)}    
    \end{subfigure}
    \begin{subfigure}[t]{0.47\linewidth}
        \vspace{0.7cm}
        \includegraphics[width=0.9\linewidth,valign=t]{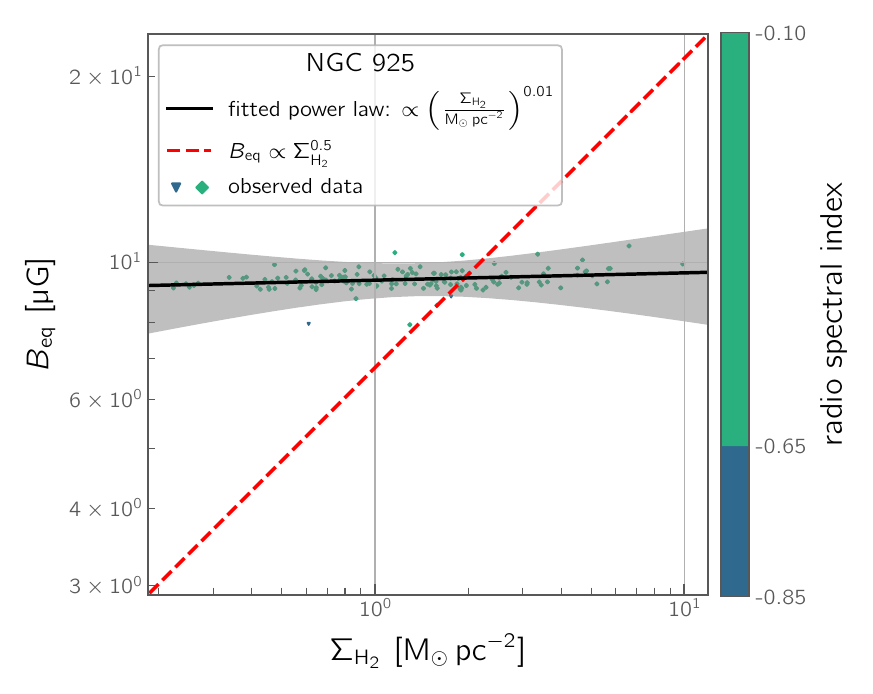}
    \end{subfigure}
    \\
    \begin{subfigure}[t]{0.02\textwidth}
        \textbf{(e)}    
    \end{subfigure}
    \begin{subfigure}[t]{0.47\linewidth}
        \includegraphics[width=0.9\linewidth,valign=t]{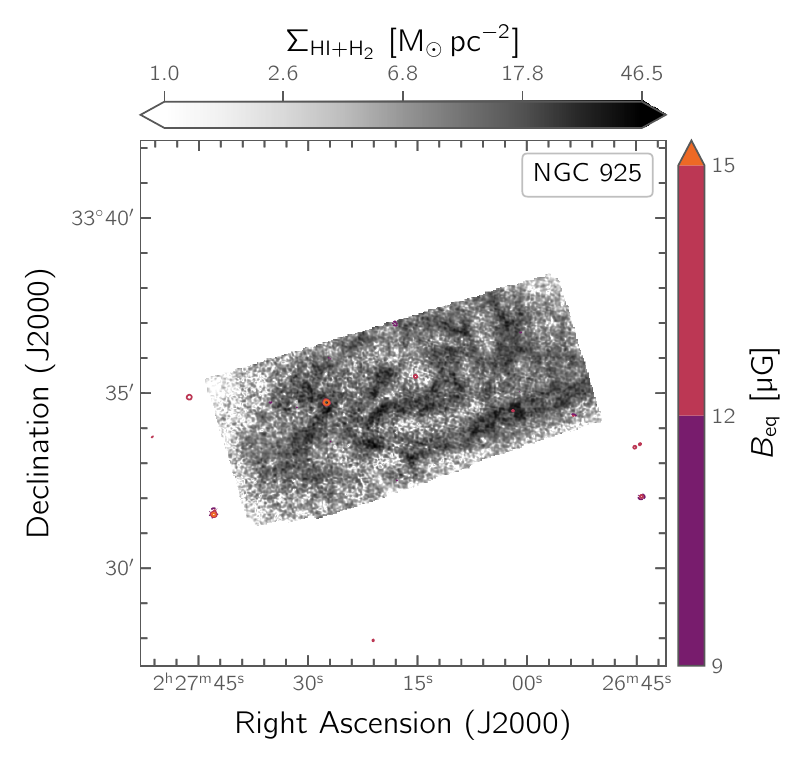}
    \end{subfigure}
    \begin{subfigure}[t]{0.02\textwidth}
        \textbf{(f)}    
    \end{subfigure}
    \begin{subfigure}[t]{0.47\linewidth}
        \vspace{0.7cm}
        \includegraphics[width=0.9\linewidth,valign=t]{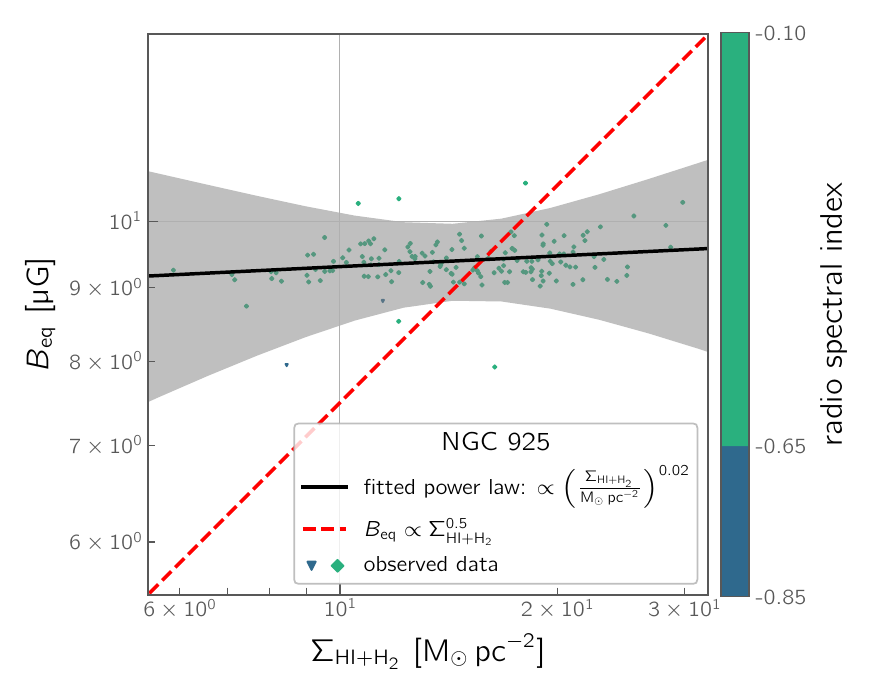}
    \end{subfigure}
    \caption{NGC 925. \corr}
    \label{fig:n925_corr}
    \end{figure*}
\addcontentsline{toc}{subsection}{NGC 925}

\begin{figure*}
	\centering
    \begin{subfigure}[t]{0.02\textwidth}
        \textbf{(a)}    
    \end{subfigure}
    \begin{subfigure}[t]{0.47\linewidth}
        \includegraphics[width=0.9\linewidth,valign=t]{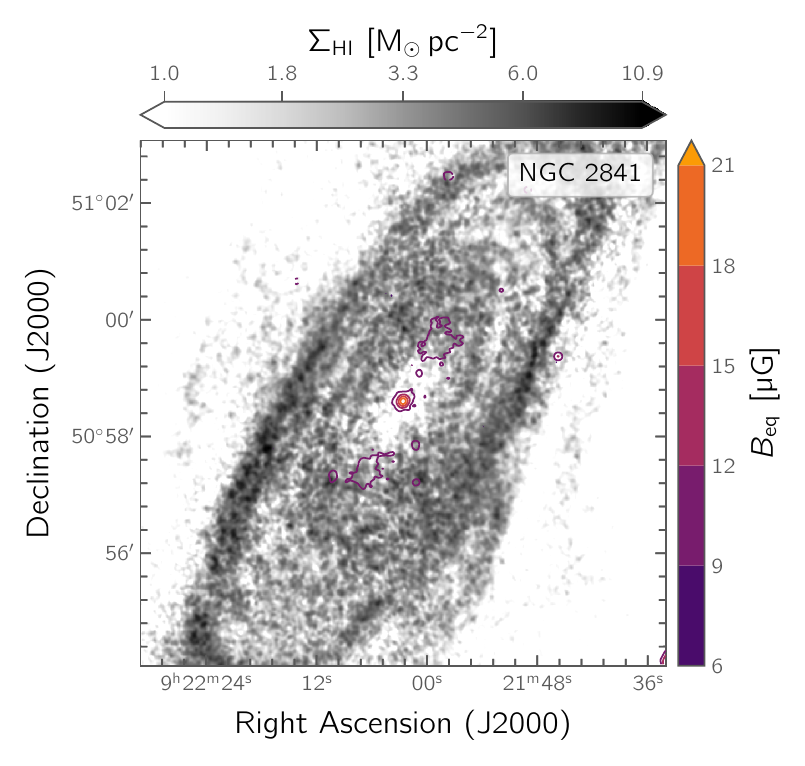}
    \end{subfigure}
    \begin{subfigure}[t]{0.02\textwidth}
        \textbf{(b)}    
    \end{subfigure}
    \begin{subfigure}[t]{0.47\linewidth}
        \vspace{0.7cm}
        \includegraphics[width=0.9\linewidth,valign=t]{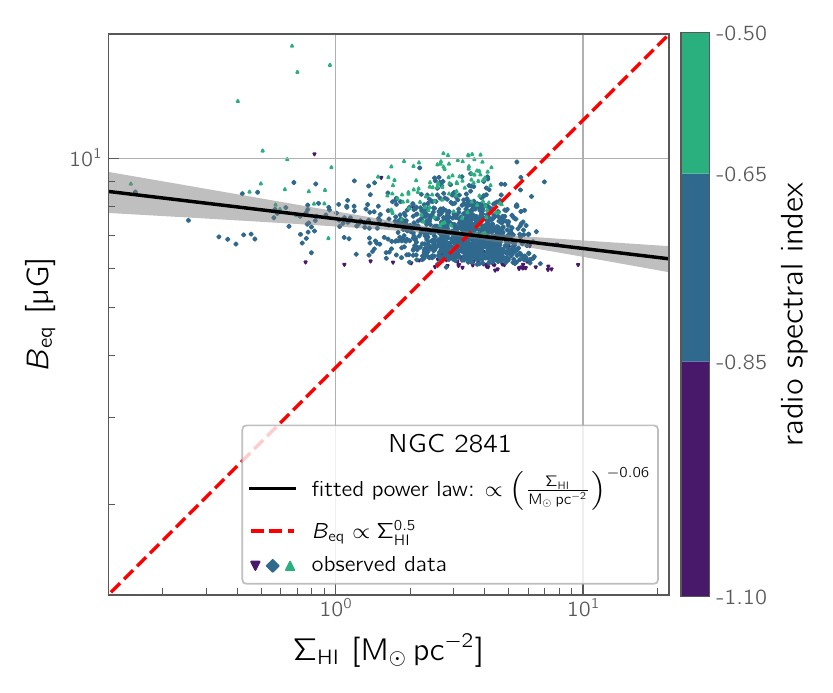}
    \end{subfigure}
    \\
    \begin{subfigure}[t]{0.02\textwidth}
        \textbf{(c)}    
    \end{subfigure}
    \begin{subfigure}[t]{0.47\linewidth}
        \includegraphics[width=0.9\linewidth,valign=t]{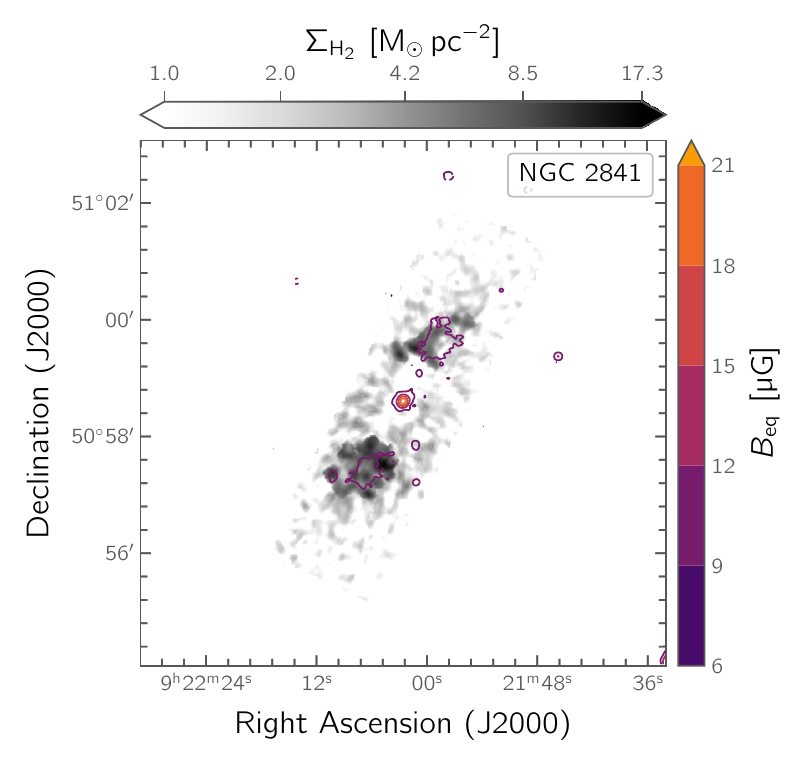}
    \end{subfigure}
    \begin{subfigure}[t]{0.02\textwidth}
        \textbf{(d)}    
    \end{subfigure}
    \begin{subfigure}[t]{0.47\linewidth}
        \vspace{0.7cm}
        \includegraphics[width=0.9\linewidth,valign=t]{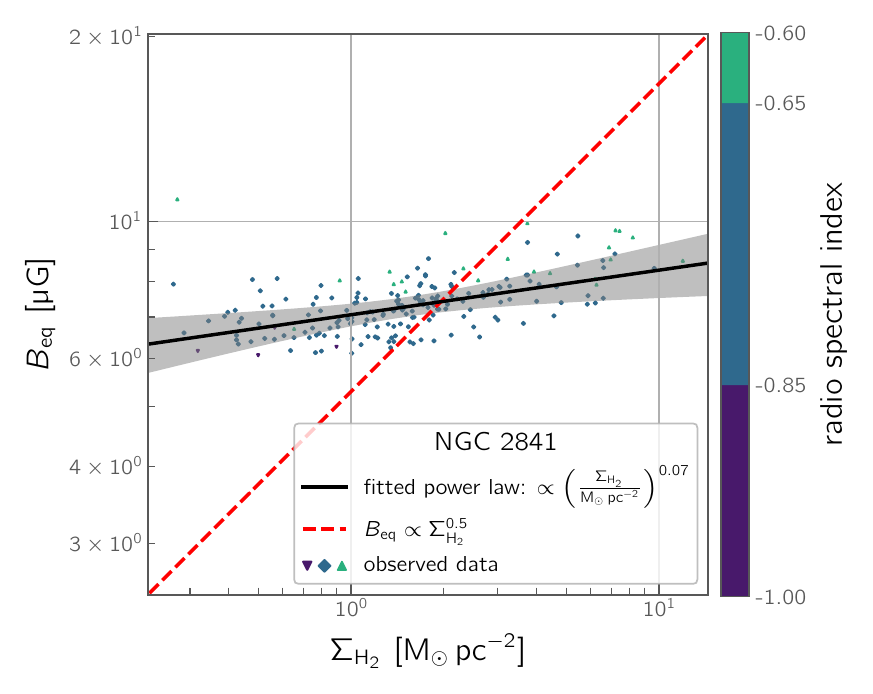}
    \end{subfigure}
    \\
    \begin{subfigure}[t]{0.02\textwidth}
        \textbf{(e)}    
    \end{subfigure}
    \begin{subfigure}[t]{0.47\linewidth}
        \includegraphics[width=0.9\linewidth,valign=t]{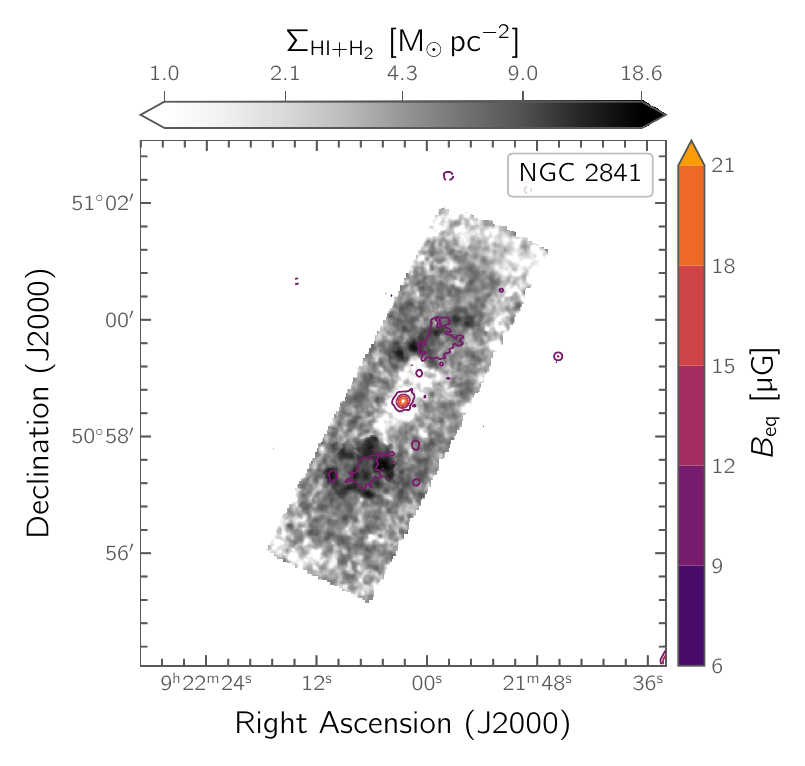}
    \end{subfigure}
    \begin{subfigure}[t]{0.02\textwidth}
        \textbf{(f)}    
    \end{subfigure}
    \begin{subfigure}[t]{0.47\linewidth}
        \vspace{0.7cm}
        \includegraphics[width=0.9\linewidth,valign=t]{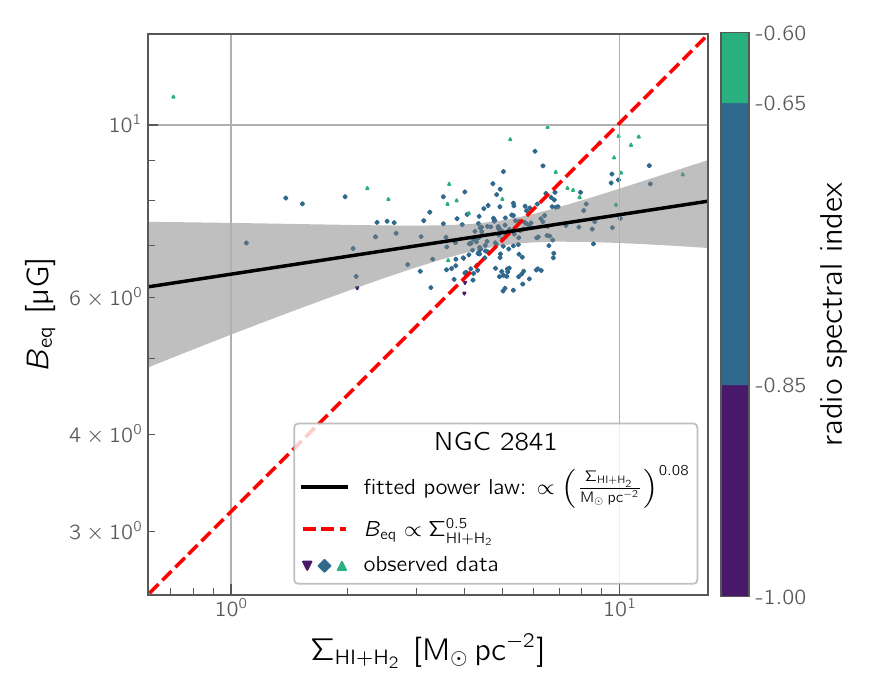}
    \end{subfigure}
    \caption{NGC 2841. \corr}
    \label{fig:n2841_corr}
\end{figure*}
\addcontentsline{toc}{subsection}{NGC 2841}

\begin{figure*}
	\centering
    \begin{subfigure}[t]{0.02\textwidth}
        \textbf{(a)}    
    \end{subfigure}
    \begin{subfigure}[t]{0.47\linewidth}
        \includegraphics[width=0.9\linewidth,valign=t]{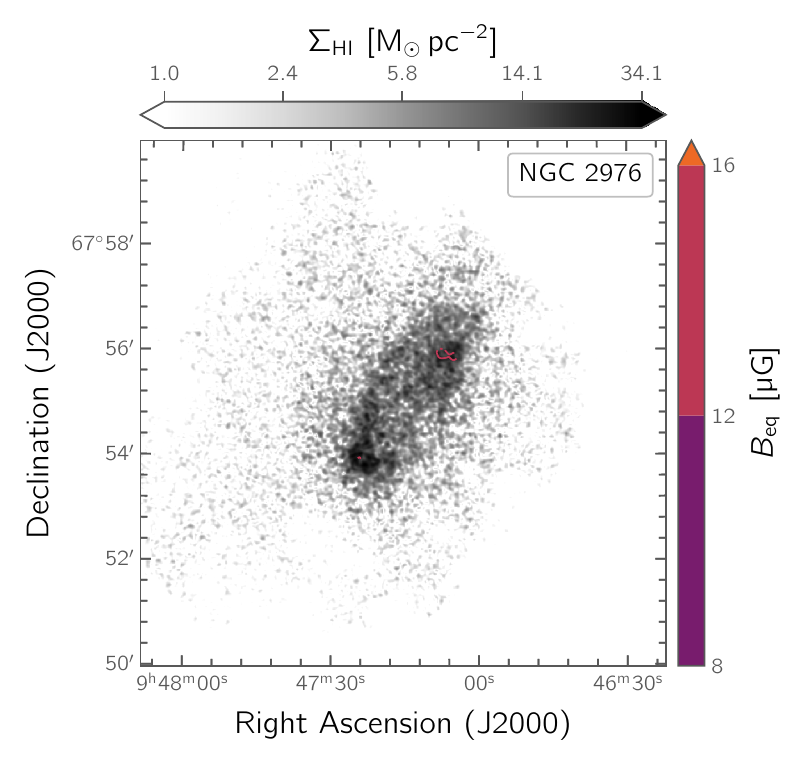}
    \end{subfigure}
    \begin{subfigure}[t]{0.02\textwidth}
        \textbf{(b)}    
    \end{subfigure}
    \begin{subfigure}[t]{0.47\linewidth}
        \vspace{0.7cm}
        \includegraphics[width=0.9\linewidth,valign=t]{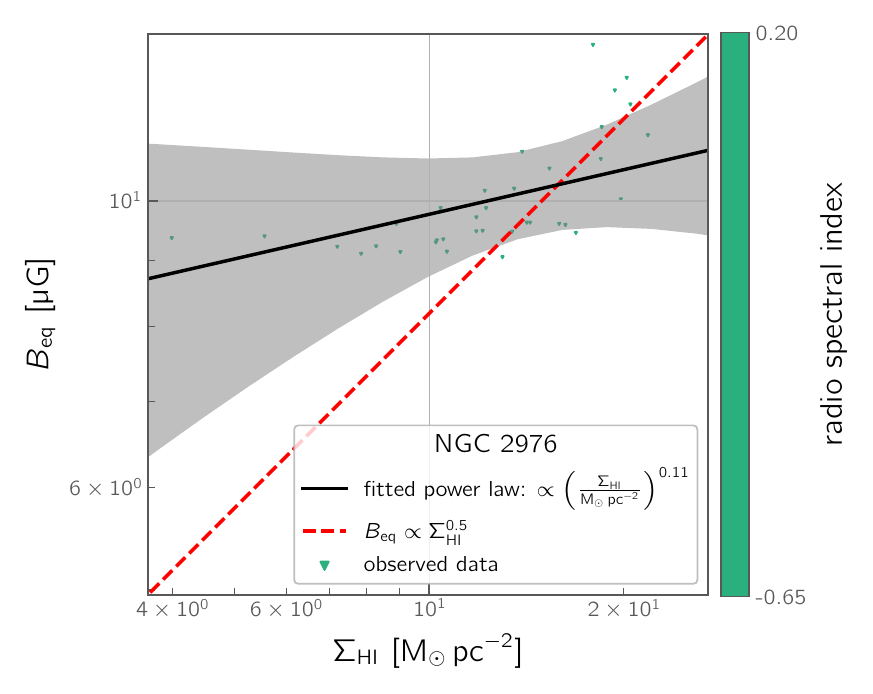}
    \end{subfigure}
    \\
    \begin{subfigure}[t]{0.02\textwidth}
        \textbf{(c)}    
    \end{subfigure}
    \begin{subfigure}[t]{0.47\linewidth}
        \includegraphics[width=0.9\linewidth,valign=t]{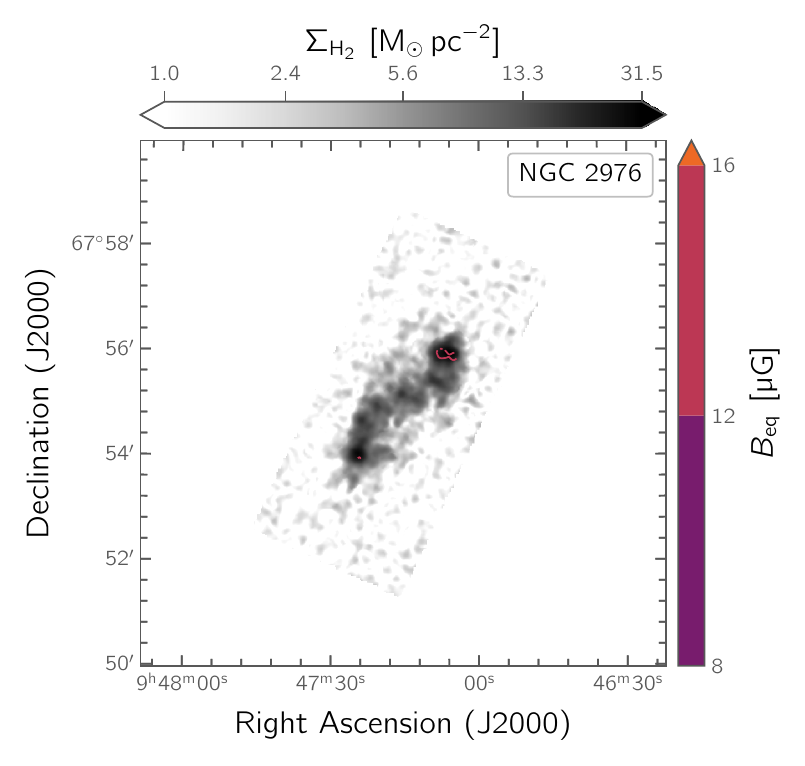}
    \end{subfigure}
    \begin{subfigure}[t]{0.02\textwidth}
        \textbf{(d)}    
    \end{subfigure}
    \begin{subfigure}[t]{0.47\linewidth}
        \vspace{0.7cm}
        \includegraphics[width=0.9\linewidth,valign=t]{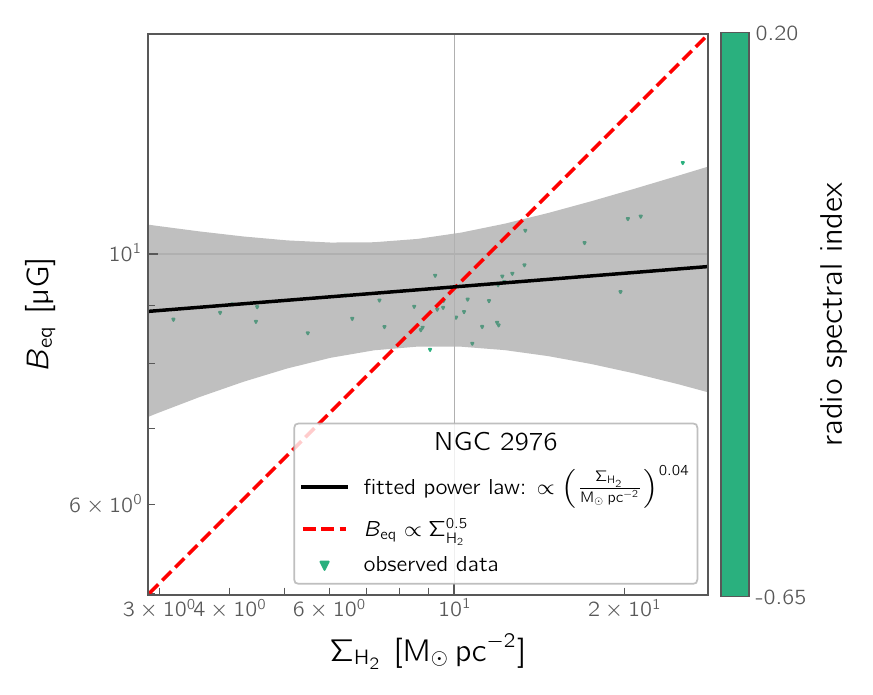}
    \end{subfigure}
    \\
    \begin{subfigure}[t]{0.02\textwidth}
        \textbf{(e)}    
    \end{subfigure}
    \begin{subfigure}[t]{0.47\linewidth}
        \includegraphics[width=0.9\linewidth,valign=t]{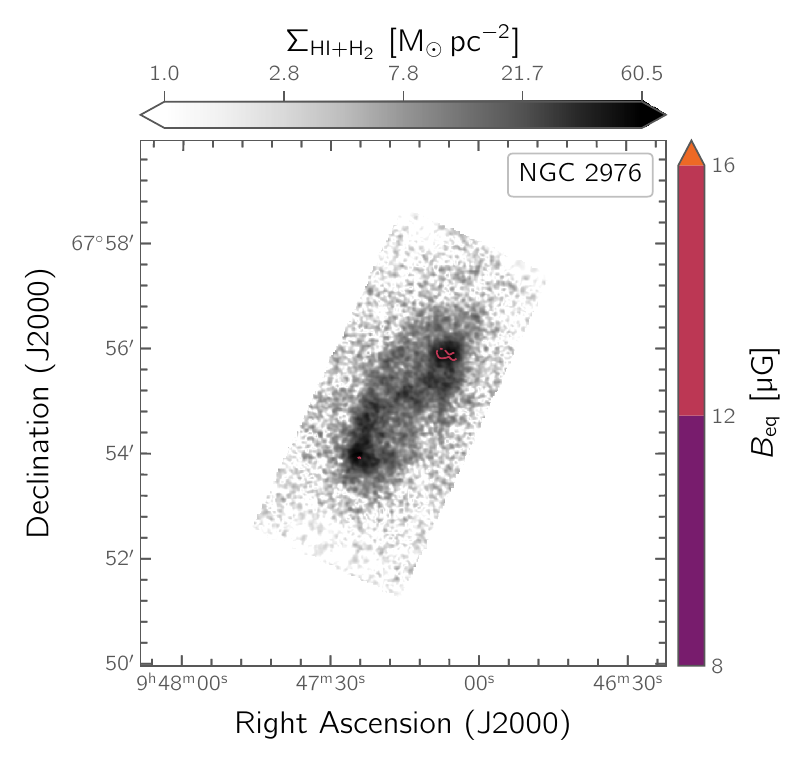}
    \end{subfigure}
    \begin{subfigure}[t]{0.02\textwidth}
        \textbf{(f)}    
    \end{subfigure}
    \begin{subfigure}[t]{0.47\linewidth}
        \vspace{0.7cm}
        \includegraphics[width=0.9\linewidth,valign=t]{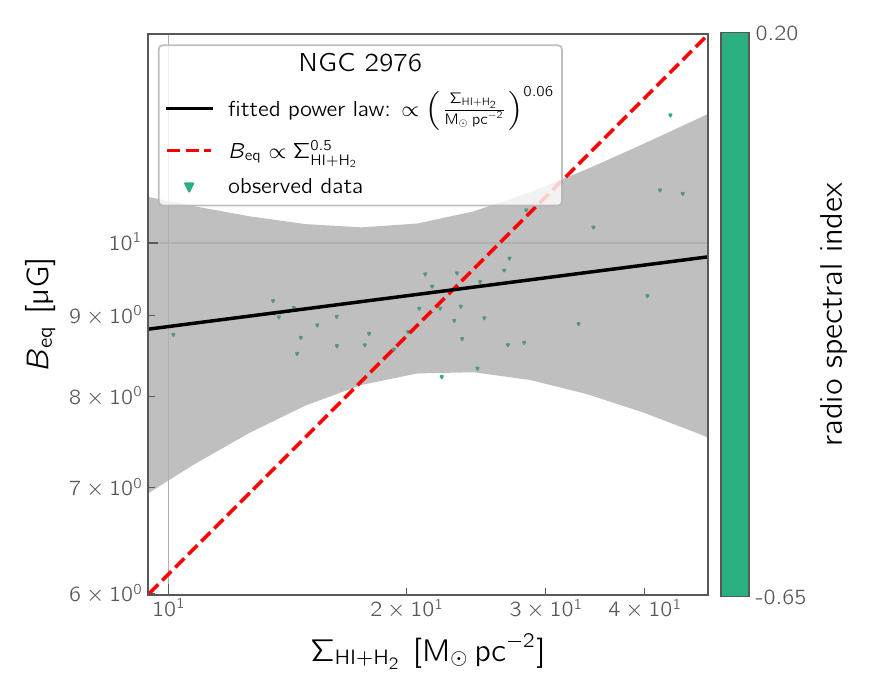}
    \end{subfigure}
    \caption{NGC 2976. \corr}
    \label{fig:n2976_corr}
\end{figure*}
\addcontentsline{toc}{subsection}{NGC 2976}

\begin{figure*}
	\centering
    \begin{subfigure}[t]{0.02\textwidth}
        \textbf{(a)}    
    \end{subfigure}
    \begin{subfigure}[t]{0.47\linewidth}
        \includegraphics[width=0.9\linewidth,valign=t]{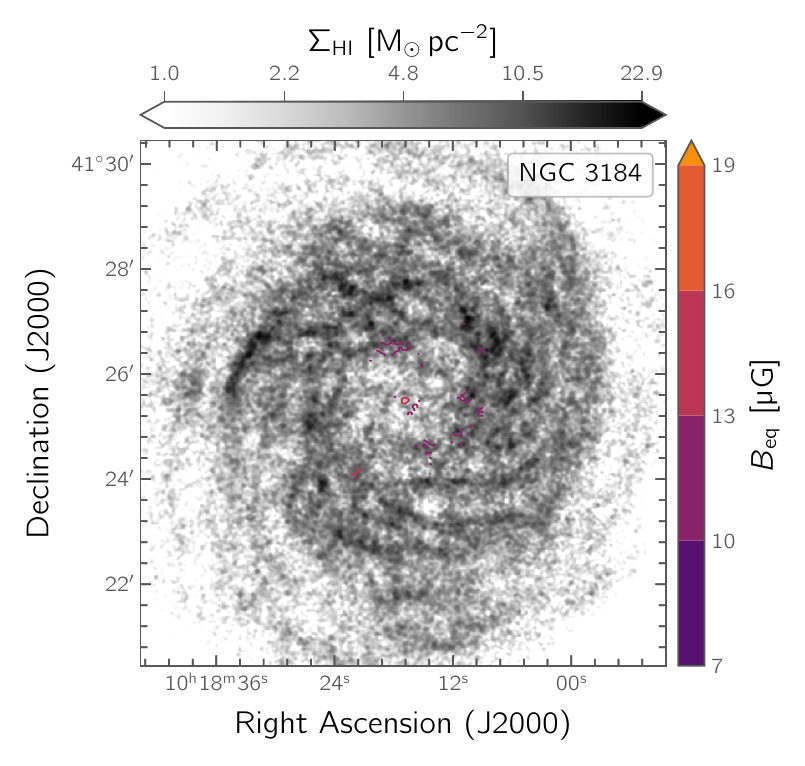}
    \end{subfigure}
    \begin{subfigure}[t]{0.02\textwidth}
        \textbf{(b)}    
    \end{subfigure}
    \begin{subfigure}[t]{0.47\linewidth}
        \vspace{0.7cm}
        \includegraphics[width=0.9\linewidth,valign=t]{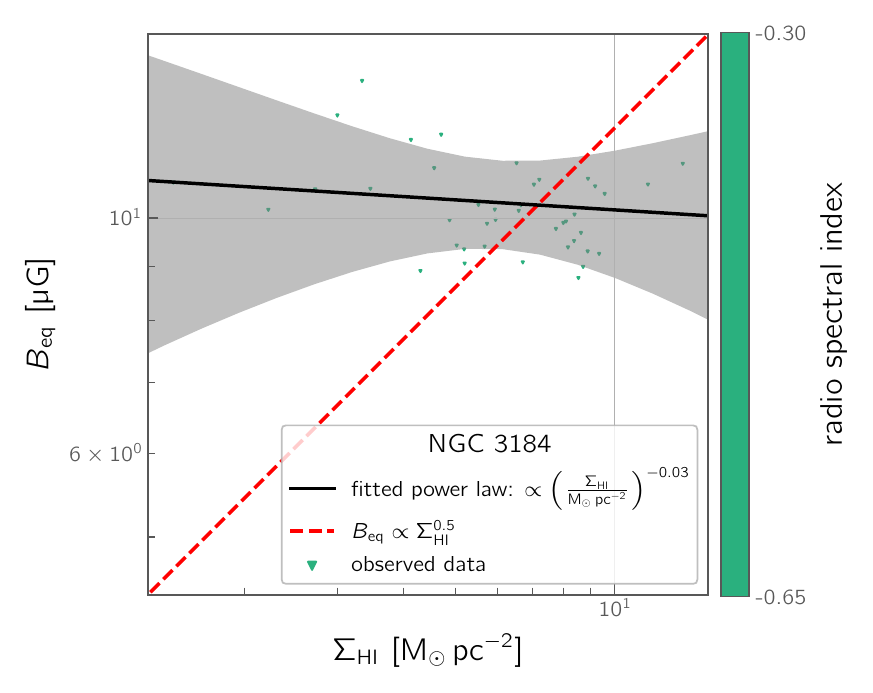}
    \end{subfigure}
    \\
    \begin{subfigure}[t]{0.02\textwidth}
        \textbf{(c)}    
    \end{subfigure}
    \begin{subfigure}[t]{0.47\linewidth}
        \includegraphics[width=0.9\linewidth,valign=t]{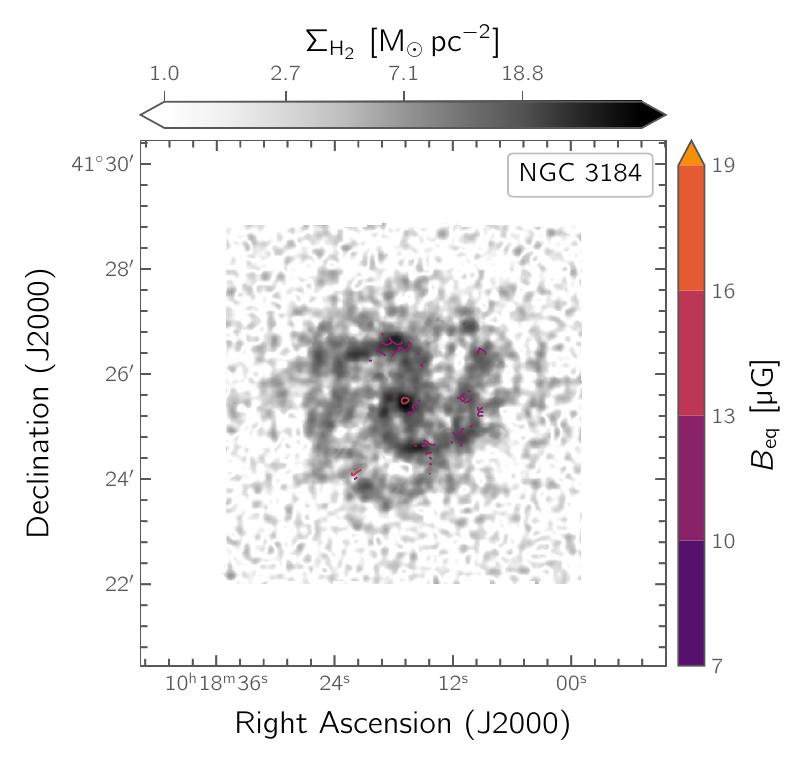}
    \end{subfigure}
    \begin{subfigure}[t]{0.02\textwidth}
        \textbf{(d)}    
    \end{subfigure}
    \begin{subfigure}[t]{0.47\linewidth}
        \vspace{0.7cm}
        \includegraphics[width=0.9\linewidth,valign=t]{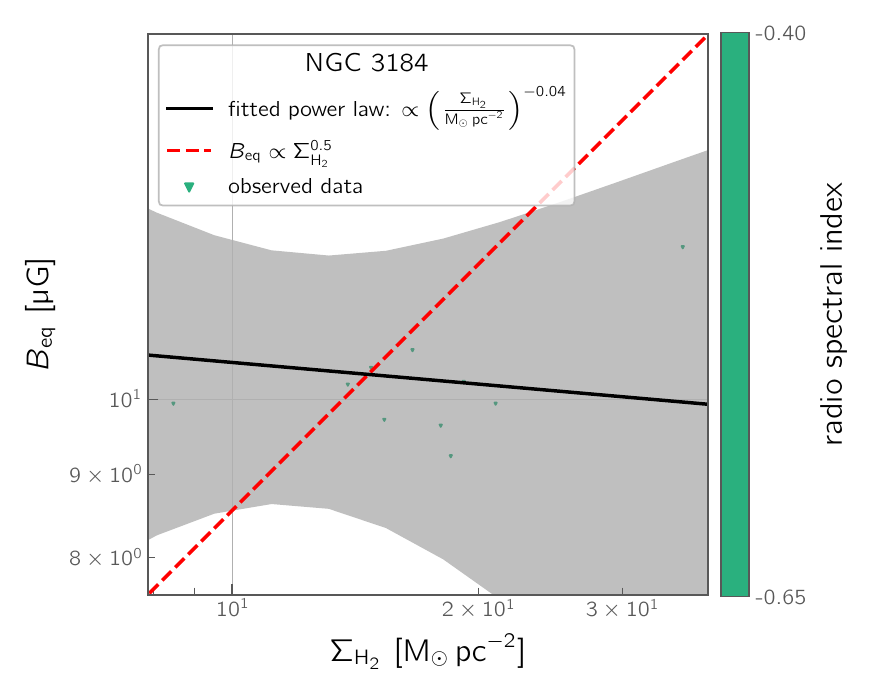}
    \end{subfigure}
    \\
    \begin{subfigure}[t]{0.02\textwidth}
        \textbf{(e)}    
    \end{subfigure}
    \begin{subfigure}[t]{0.47\linewidth}
        \includegraphics[width=0.9\linewidth,valign=t]{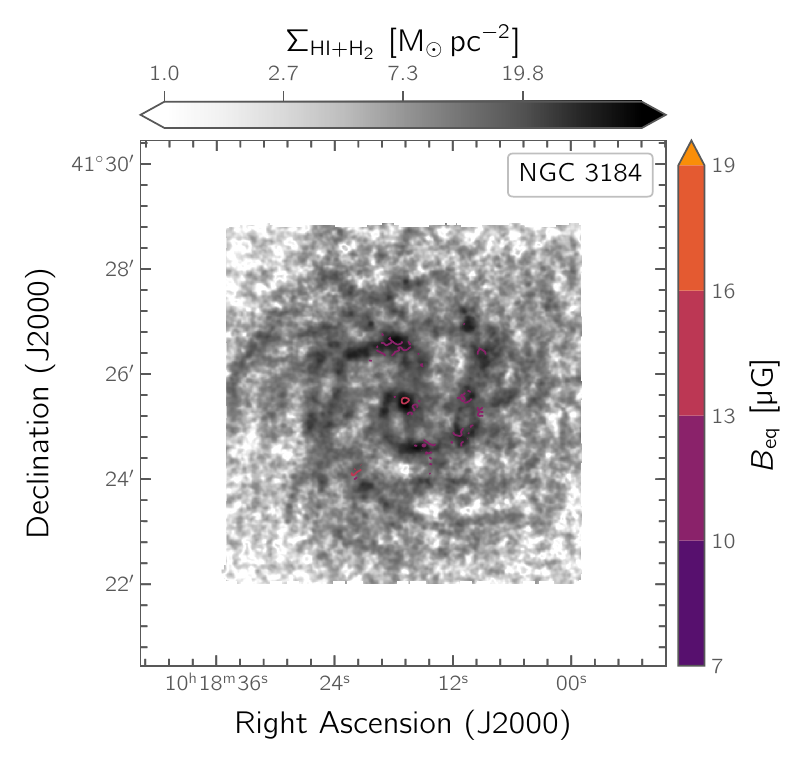}
    \end{subfigure}
    \begin{subfigure}[t]{0.02\textwidth}
        \textbf{(f)}    
    \end{subfigure}
    \begin{subfigure}[t]{0.47\linewidth}
        \vspace{0.7cm}
        \includegraphics[width=0.9\linewidth,valign=t]{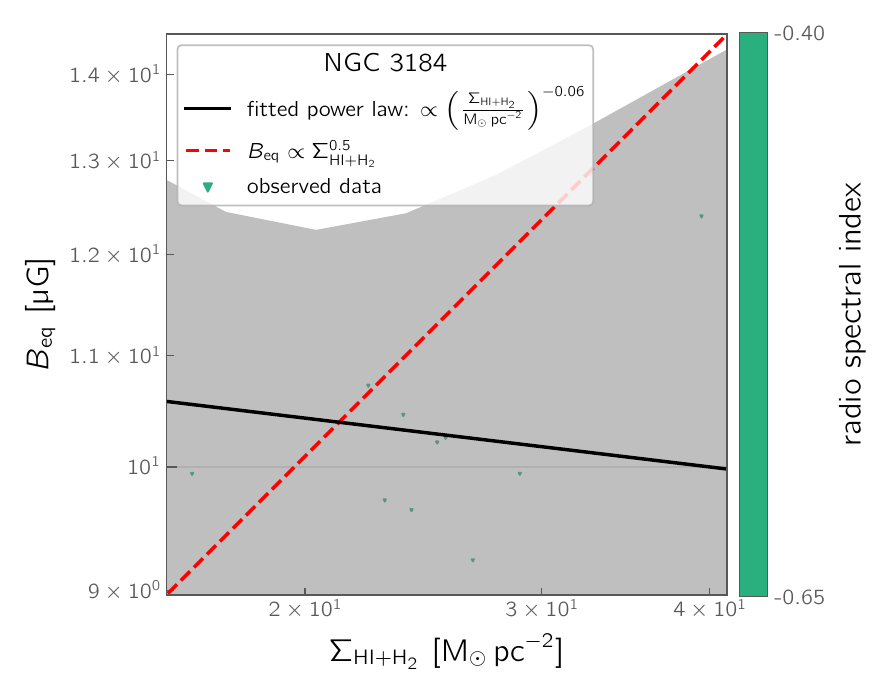}
    \end{subfigure}
    \caption{NGC 3184. \corr}
    \label{fig:n3184_corr}
\end{figure*}
\addcontentsline{toc}{subsection}{NGC 3184}

\begin{figure*}
	\centering
    \begin{subfigure}[t]{0.02\textwidth}
        \textbf{(a)}    
    \end{subfigure}
    \begin{subfigure}[t]{0.47\linewidth}
        \includegraphics[width=0.9\linewidth,valign=t]{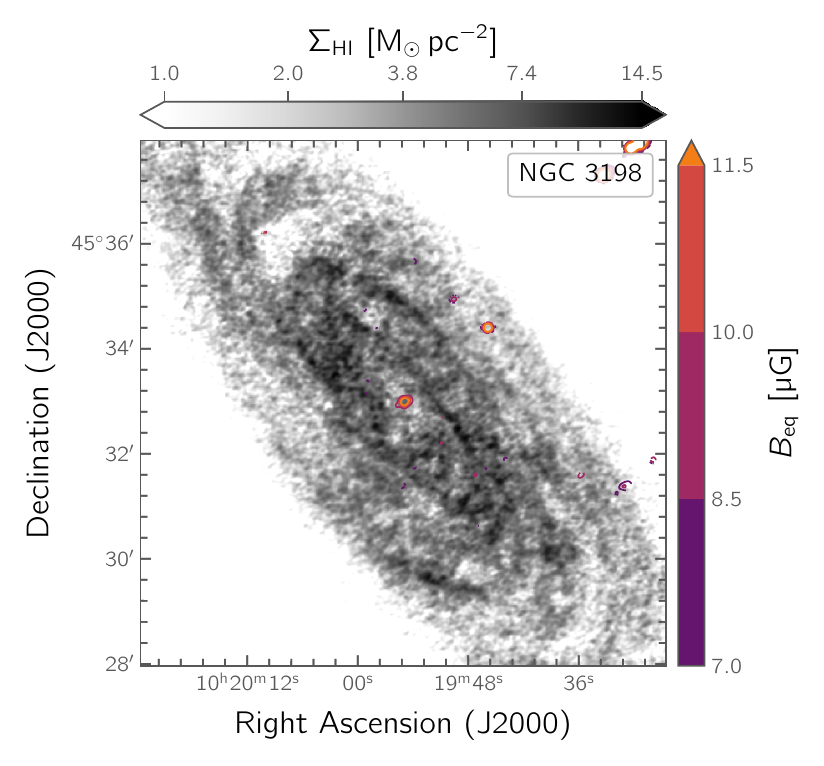}
    \end{subfigure}
    \begin{subfigure}[t]{0.02\textwidth}
        \textbf{(b)}    
    \end{subfigure}
    \begin{subfigure}[t]{0.47\linewidth}
        \vspace{0.7cm}
        \includegraphics[width=0.9\linewidth,valign=t]{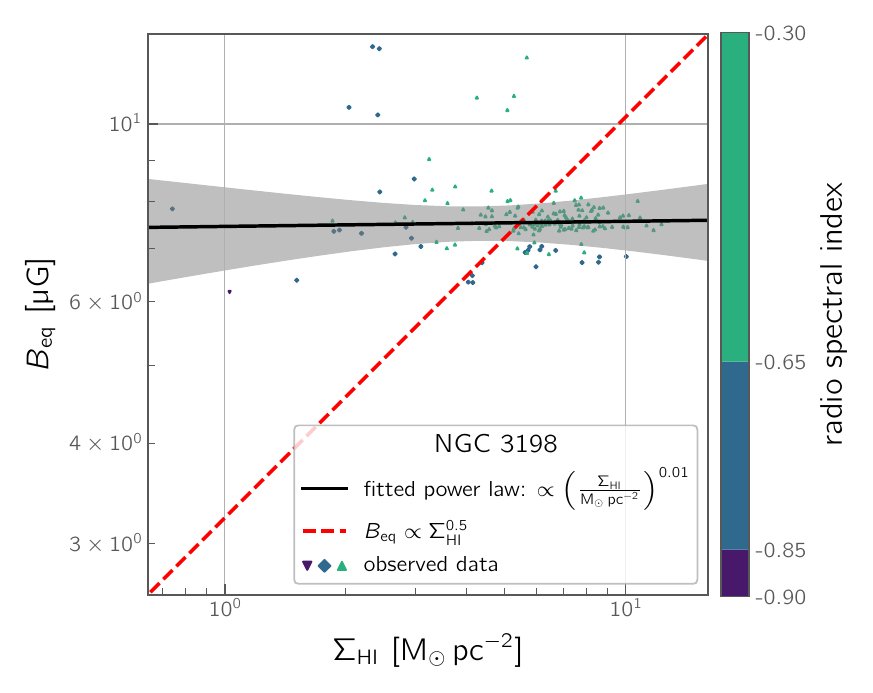}
    \end{subfigure}
    \\
    \begin{subfigure}[t]{0.02\textwidth}
        \textbf{(c)}    
    \end{subfigure}
    \begin{subfigure}[t]{0.47\linewidth}
        \includegraphics[width=0.9\linewidth,valign=t]{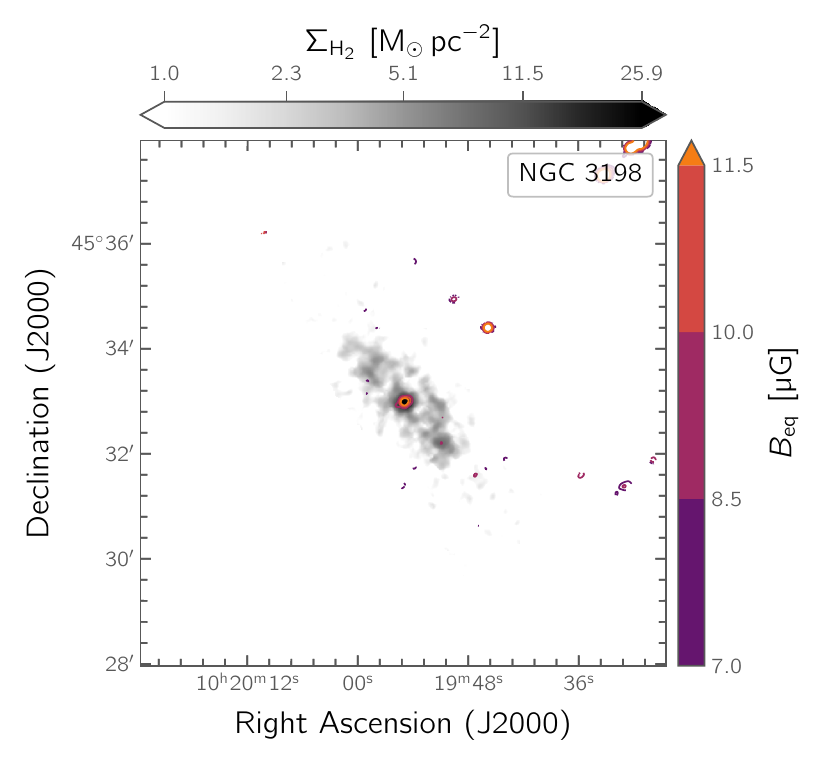}
    \end{subfigure}
    \begin{subfigure}[t]{0.02\textwidth}
        \textbf{(d)}    
    \end{subfigure}
    \begin{subfigure}[t]{0.47\linewidth}
        \vspace{0.7cm}
        \includegraphics[width=0.9\linewidth,valign=t]{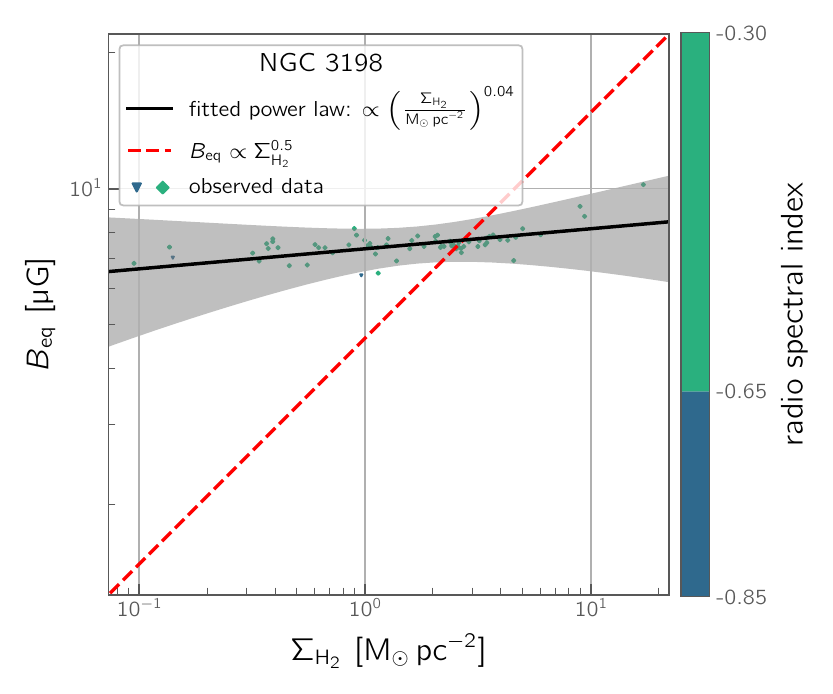}
    \end{subfigure}
    \\
    \begin{subfigure}[t]{0.02\textwidth}
        \textbf{(e)}    
    \end{subfigure}
    \begin{subfigure}[t]{0.47\linewidth}
        \includegraphics[width=0.9\linewidth,valign=t]{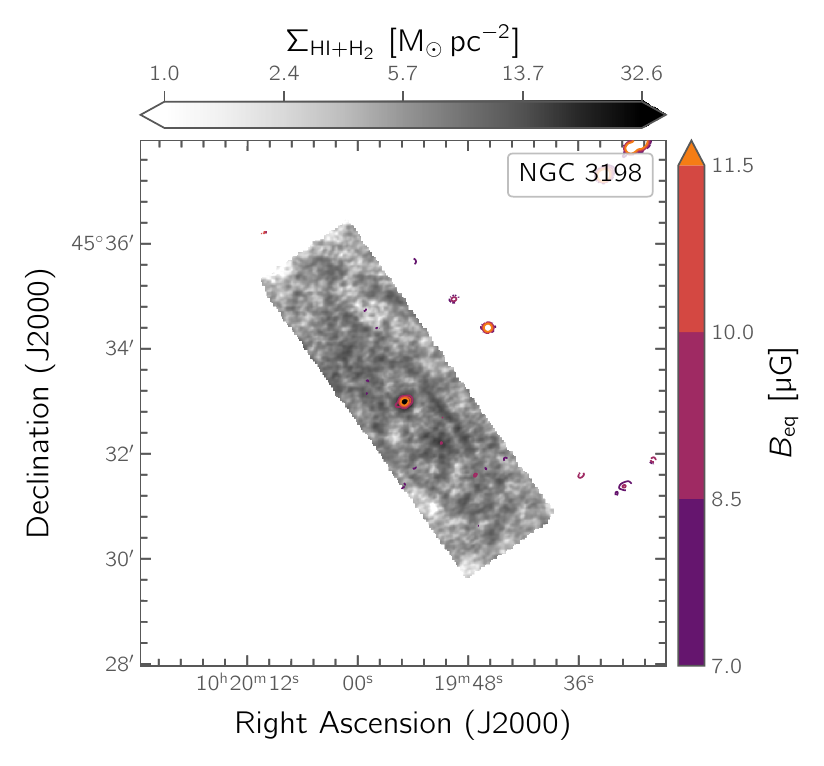}
    \end{subfigure}
    \begin{subfigure}[t]{0.02\textwidth}
        \textbf{(f)}    
    \end{subfigure}
    \begin{subfigure}[t]{0.47\linewidth}
        \vspace{0.7cm}
        \includegraphics[width=0.9\linewidth,valign=t]{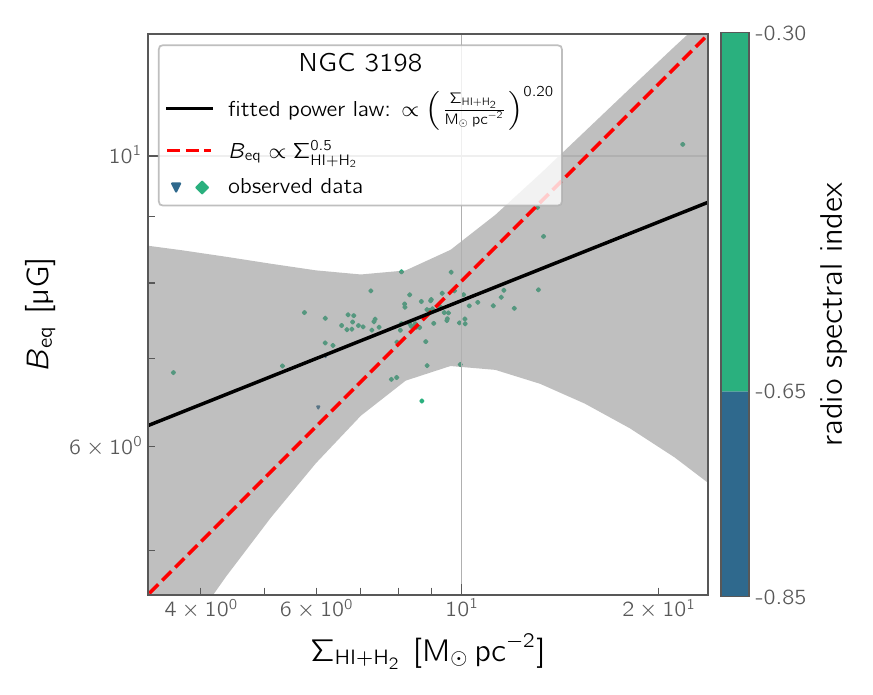}
    \end{subfigure}
    \caption{NGC 3198. \corr}
    \label{fig:n3198_corr}
\end{figure*}
\addcontentsline{toc}{subsection}{NGC 3198}

\begin{figure*}
	\centering
    \begin{subfigure}[t]{0.02\textwidth}
        \textbf{(a)}    
    \end{subfigure}
    \begin{subfigure}[t]{0.47\linewidth}
        \includegraphics[width=0.9\linewidth,valign=t]{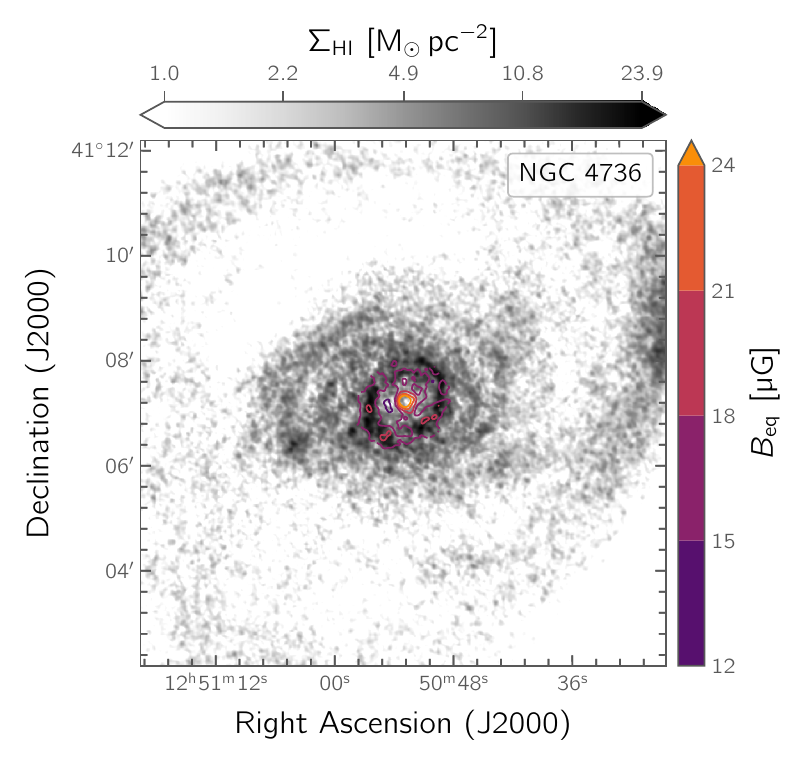}
    \end{subfigure}
    \begin{subfigure}[t]{0.02\textwidth}
        \textbf{(b)}    
    \end{subfigure}
    \begin{subfigure}[t]{0.47\linewidth}
        \vspace{0.7cm}
        \includegraphics[width=0.9\linewidth,valign=t]{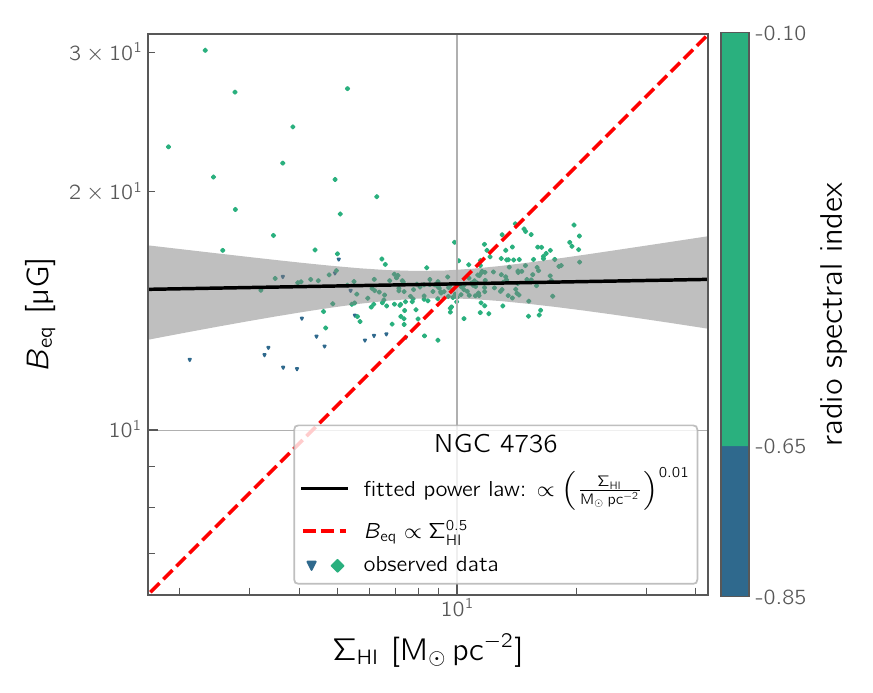}
    \end{subfigure}
    \\
    \begin{subfigure}[t]{0.02\textwidth}
        \textbf{(c)}    
    \end{subfigure}
    \begin{subfigure}[t]{0.47\linewidth}
        \includegraphics[width=0.9\linewidth,valign=t]{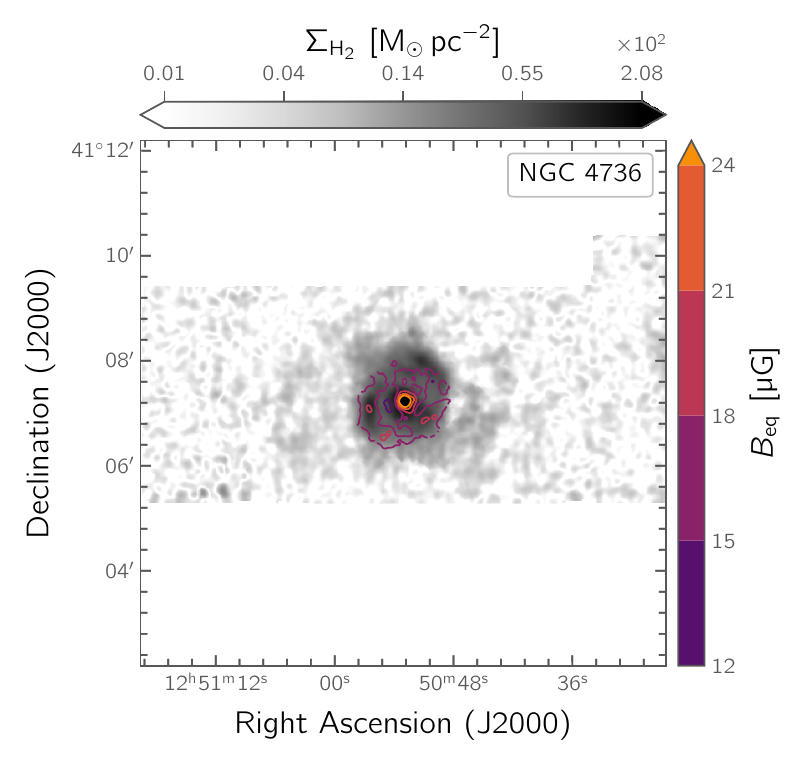}
    \end{subfigure}
    \begin{subfigure}[t]{0.02\textwidth}
        \textbf{(d)}    
    \end{subfigure}
    \begin{subfigure}[t]{0.47\linewidth}
        \vspace{0.7cm}
        \includegraphics[width=0.9\linewidth,valign=t]{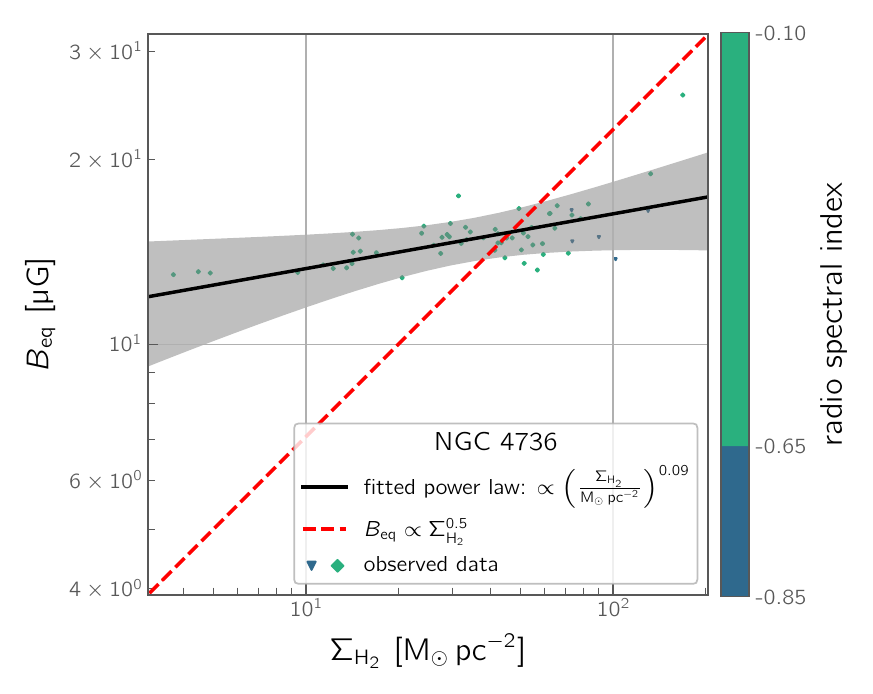}
    \end{subfigure}
    \\
    \begin{subfigure}[t]{0.02\textwidth}
        \textbf{(e)}    
    \end{subfigure}
    \begin{subfigure}[t]{0.47\linewidth}
        \includegraphics[width=0.9\linewidth,valign=t]{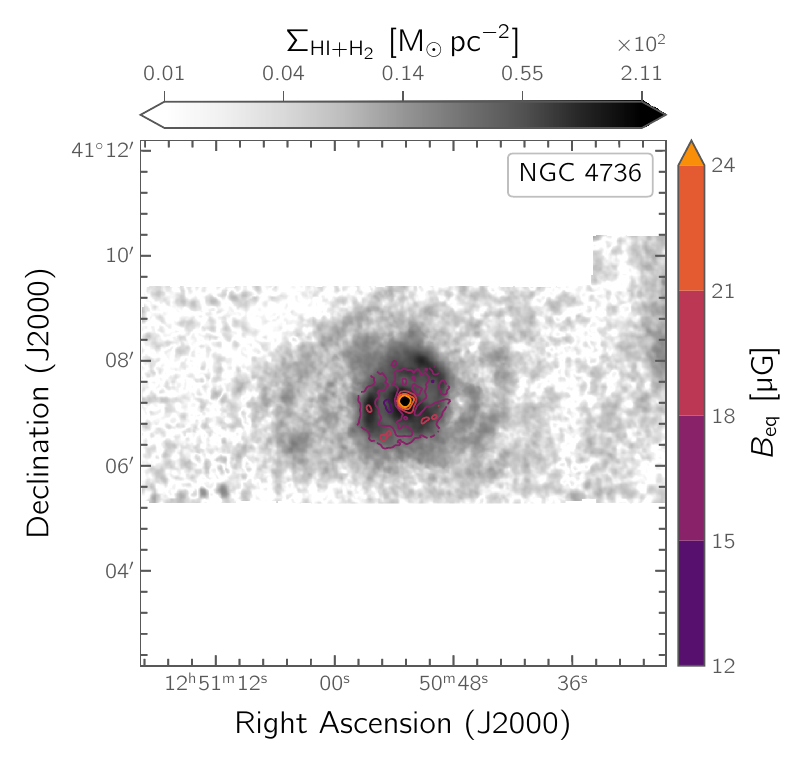}
    \end{subfigure}
    \begin{subfigure}[t]{0.02\textwidth}
        \textbf{(f)}    
    \end{subfigure}
    \begin{subfigure}[t]{0.47\linewidth}
        \vspace{0.7cm}
        \includegraphics[width=0.9\linewidth,valign=t]{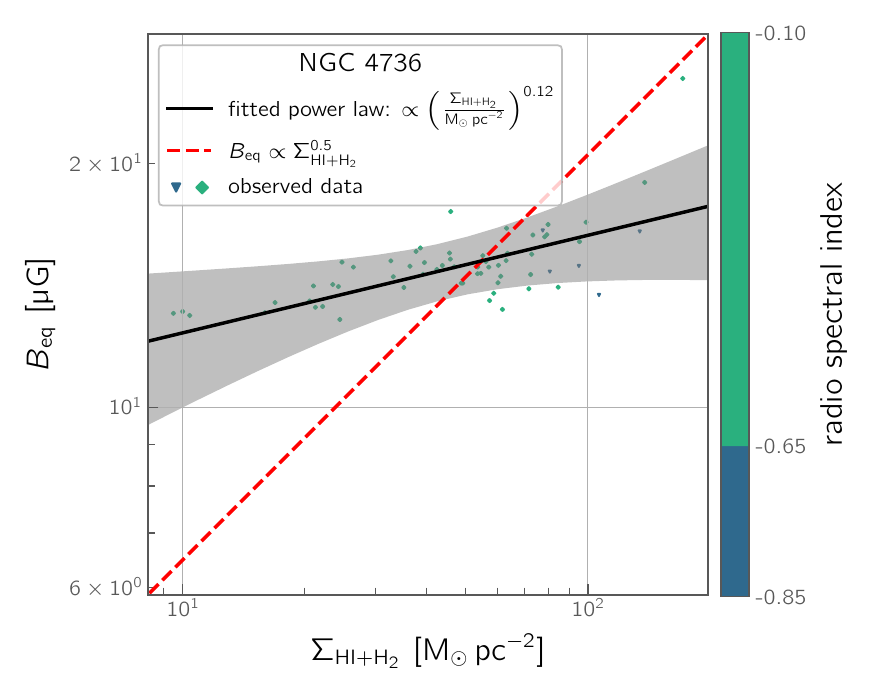}
    \end{subfigure}
    \caption{NGC 4736. \corr}
    \label{fig:n4736_corr}
\end{figure*}
\addcontentsline{toc}{subsection}{NGC 4736}

\begin{figure*}
	\centering
    \begin{subfigure}[t]{0.02\textwidth}
        \textbf{(a)}    
    \end{subfigure}
    \begin{subfigure}[t]{0.47\linewidth}
        \includegraphics[width=0.9\linewidth,valign=t]{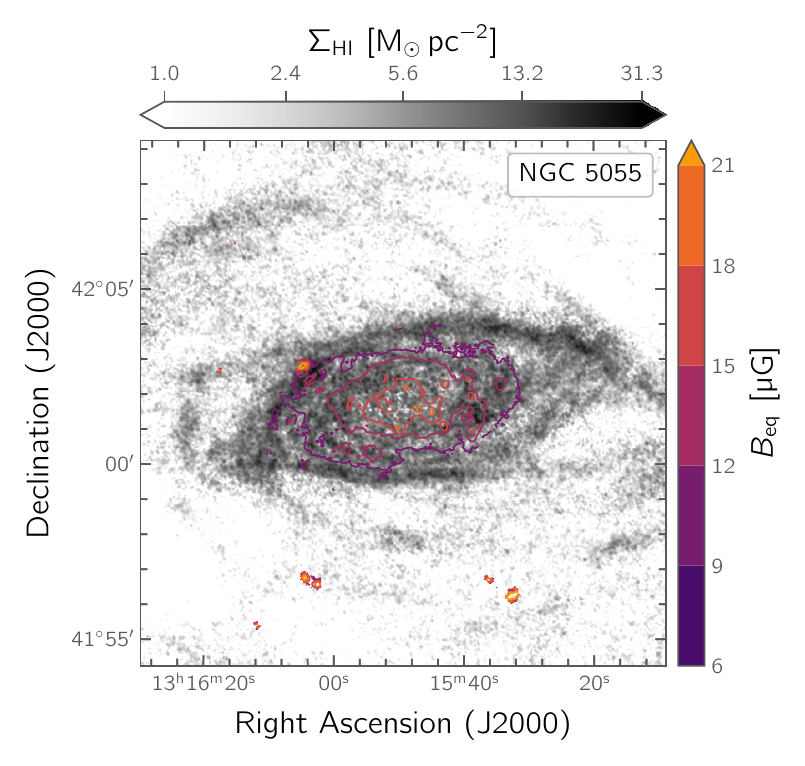}
    \end{subfigure}
    \begin{subfigure}[t]{0.02\textwidth}
        \textbf{(b)}    
    \end{subfigure}
    \begin{subfigure}[t]{0.47\linewidth}
        \vspace{0.7cm}
        \includegraphics[width=0.9\linewidth,valign=t]{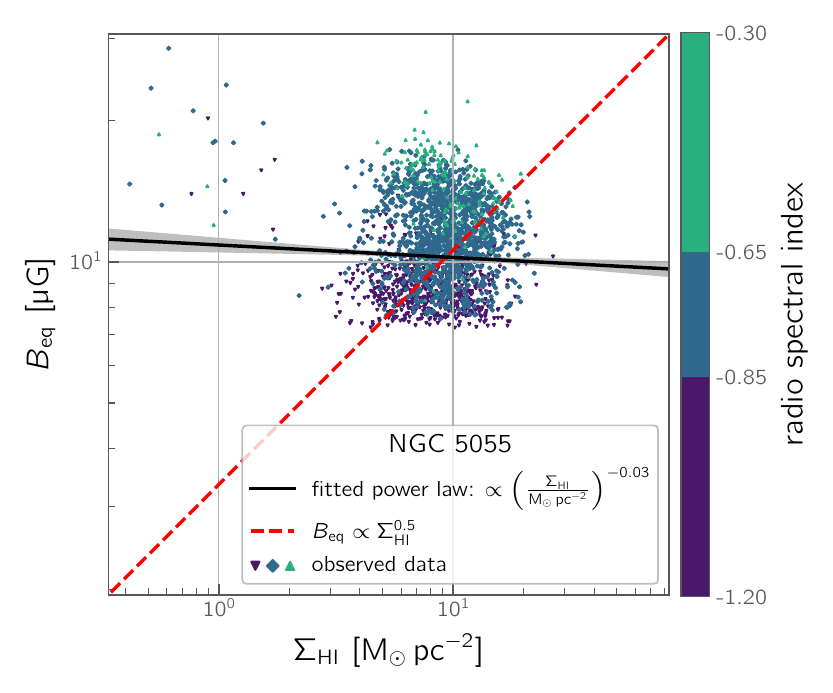}
    \end{subfigure}
    \\
    \begin{subfigure}[t]{0.02\textwidth}
        \textbf{(c)}    
    \end{subfigure}
    \begin{subfigure}[t]{0.47\linewidth}
        \includegraphics[width=0.9\linewidth,valign=t]{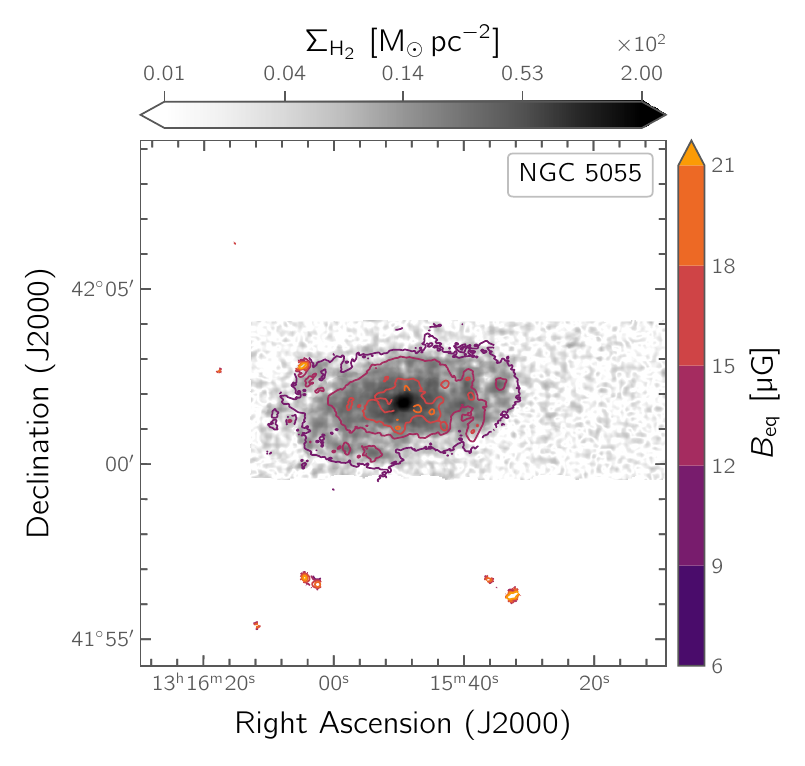}
    \end{subfigure}
    \begin{subfigure}[t]{0.02\textwidth}
        \textbf{(d)}    
    \end{subfigure}
    \begin{subfigure}[t]{0.47\linewidth}
        \vspace{0.7cm}
        \includegraphics[width=0.9\linewidth,valign=t]{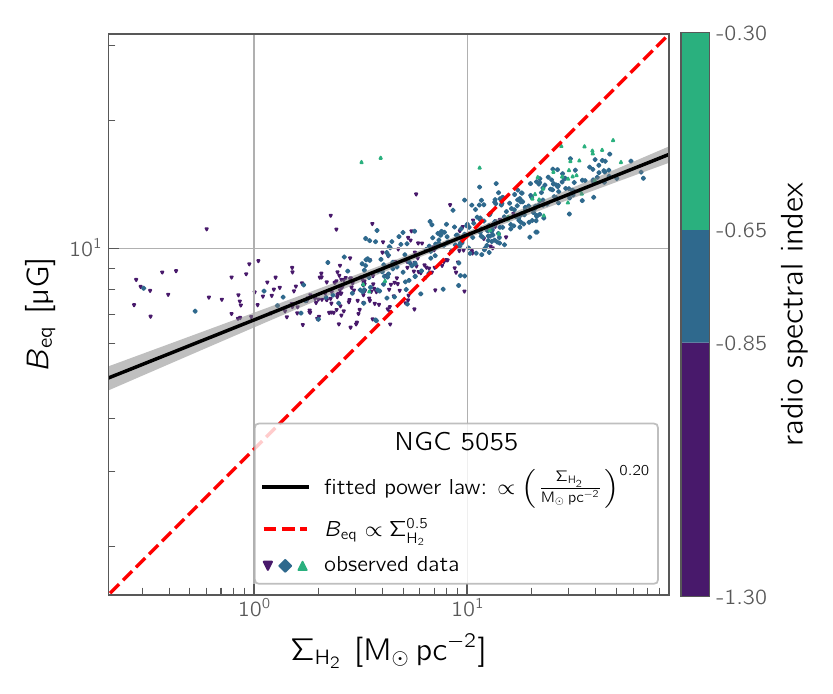}
    \end{subfigure}
    \\
    \begin{subfigure}[t]{0.02\textwidth}
        \textbf{(e)}    
    \end{subfigure}
    \begin{subfigure}[t]{0.47\linewidth}
        \includegraphics[width=0.9\linewidth,valign=t]{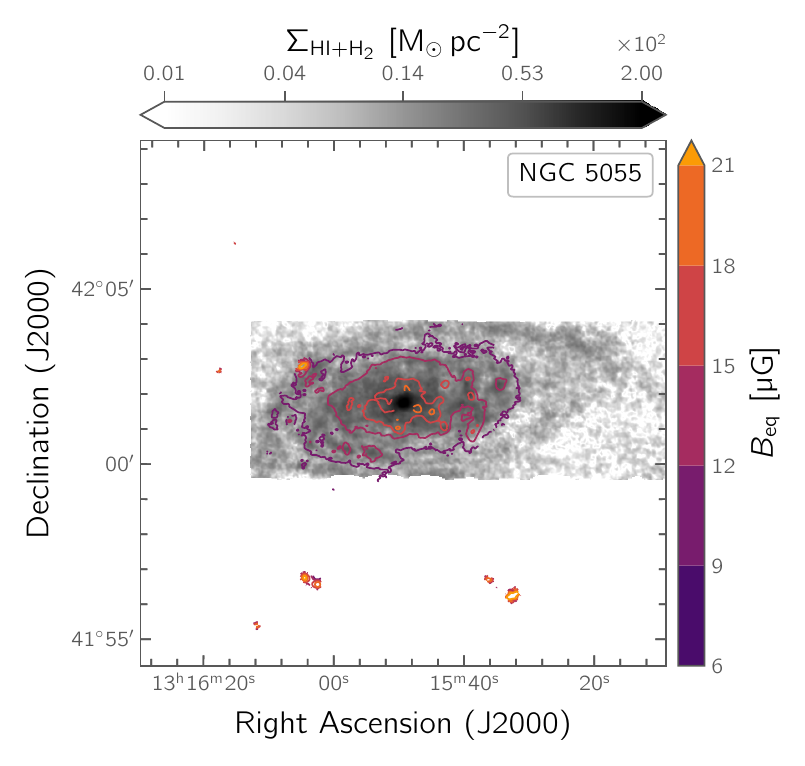}
    \end{subfigure}
    \begin{subfigure}[t]{0.02\textwidth}
        \textbf{(f)}    
    \end{subfigure}
    \begin{subfigure}[t]{0.47\linewidth}
        \vspace{0.7cm}
        \includegraphics[width=0.9\linewidth,valign=t]{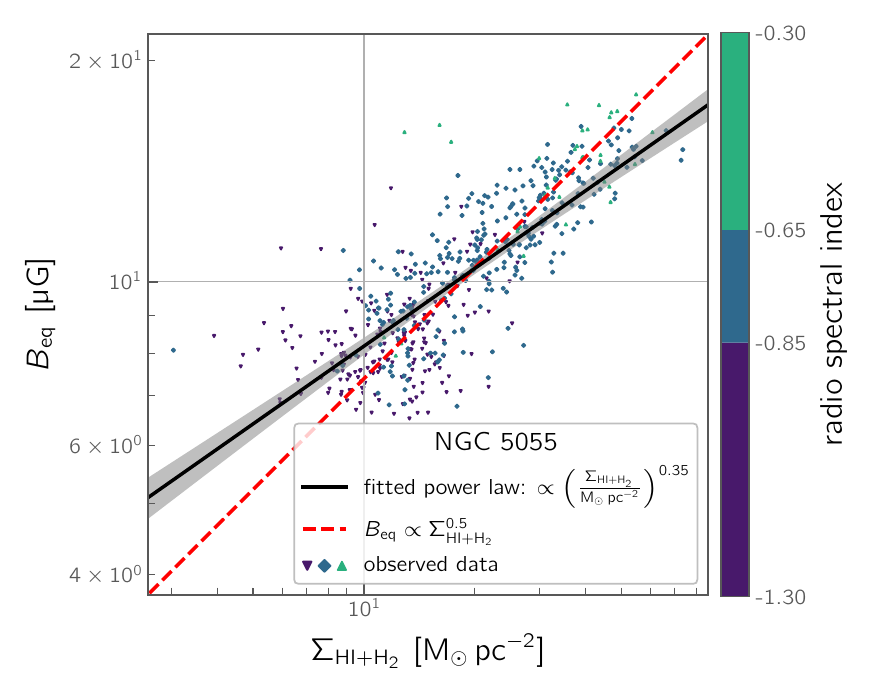}
    \end{subfigure}
    \caption{NGC 5055. \corr}
    \label{fig:n5055_corr}
\end{figure*}
\addcontentsline{toc}{subsection}{NGC 5055}

\begin{figure*}
	\centering
    \begin{subfigure}[t]{0.02\textwidth}
        \textbf{(a)}    
    \end{subfigure}
    \begin{subfigure}[t]{0.47\linewidth}
        \includegraphics[width=0.9\linewidth,valign=t]{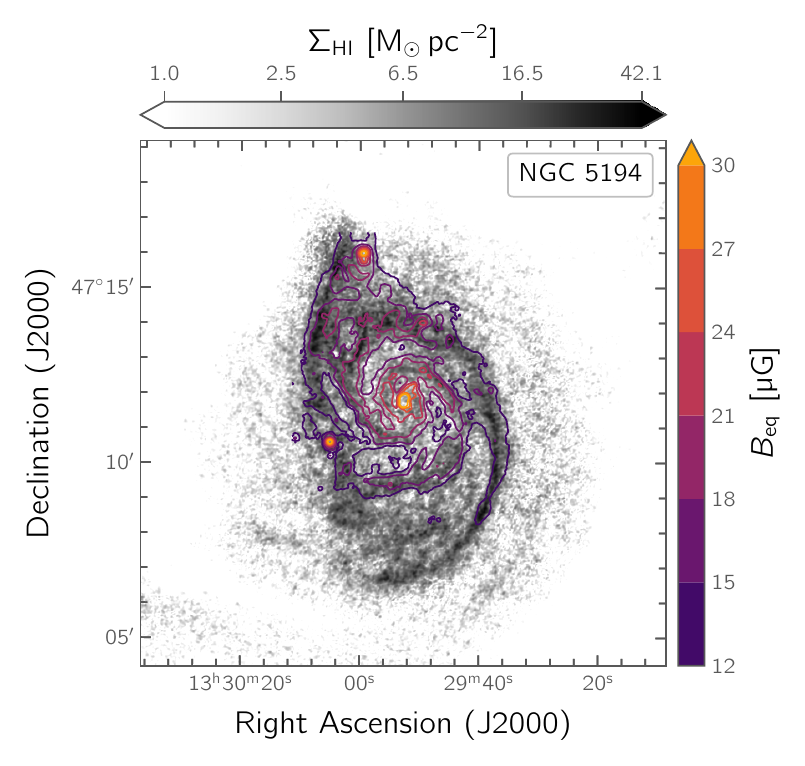}
    \end{subfigure}
    \begin{subfigure}[t]{0.02\textwidth}
        \textbf{(b)}    
    \end{subfigure}
    \begin{subfigure}[t]{0.47\linewidth}
        \vspace{0.7cm}
        \includegraphics[width=0.9\linewidth,valign=t]{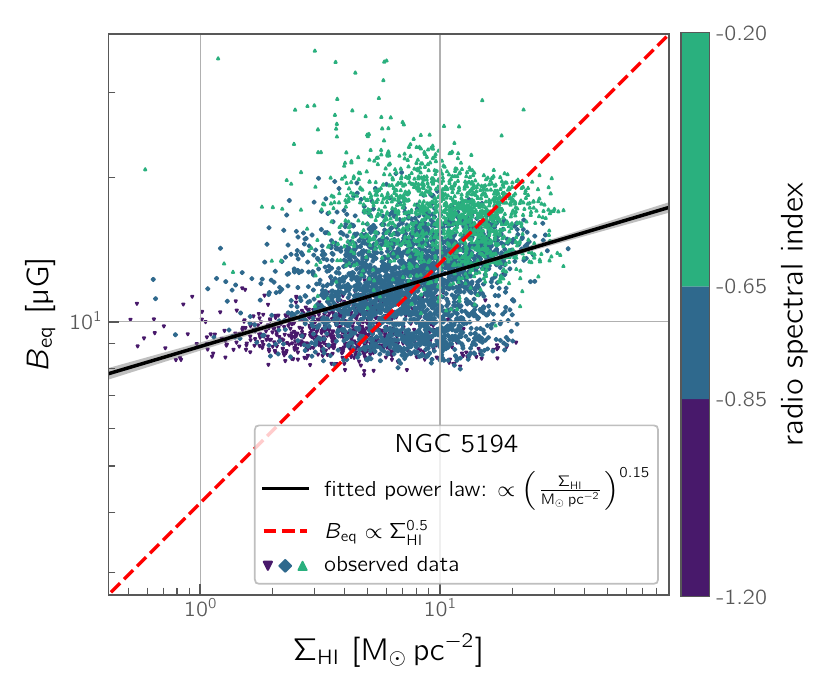}
    \end{subfigure}
    \\
    \begin{subfigure}[t]{0.02\textwidth}
        \textbf{(c)}    
    \end{subfigure}
    \begin{subfigure}[t]{0.47\linewidth}
        \includegraphics[width=0.9\linewidth,valign=t]{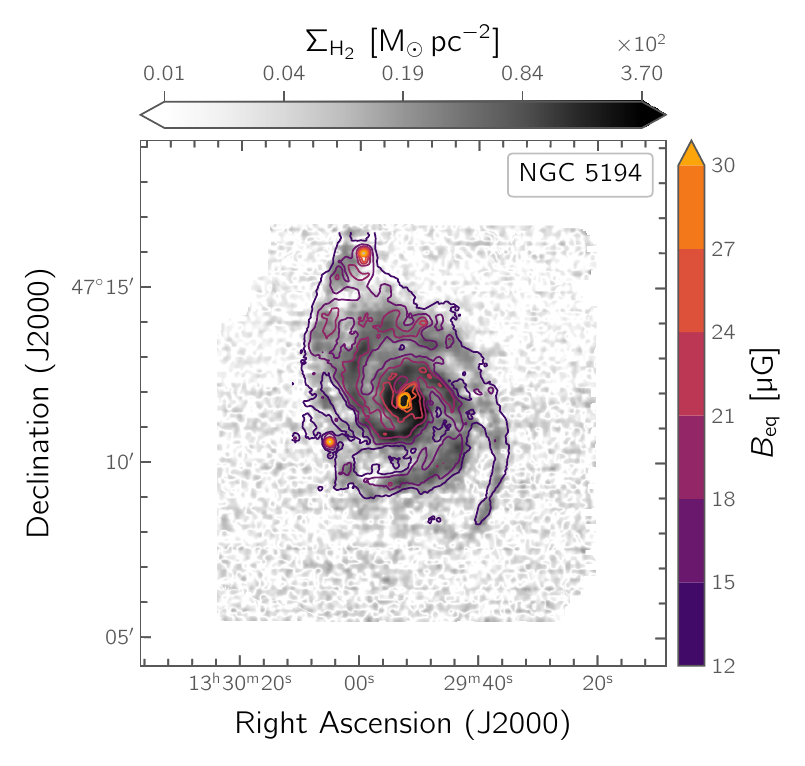}
    \end{subfigure}
    \begin{subfigure}[t]{0.02\textwidth}
        \textbf{(d)}    
    \end{subfigure}
    \begin{subfigure}[t]{0.47\linewidth}
        \vspace{0.7cm}
        \includegraphics[width=0.9\linewidth,valign=t]{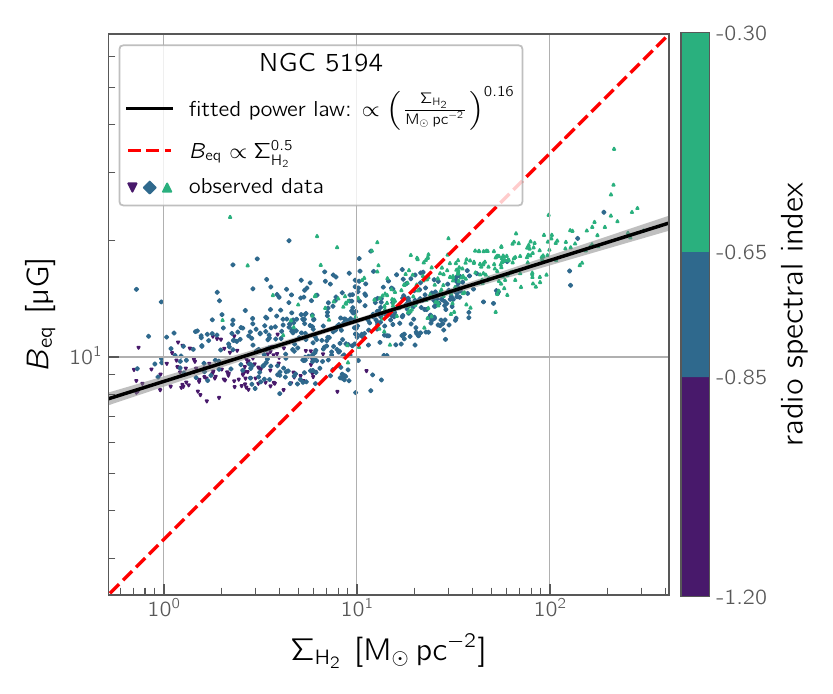}
    \end{subfigure}
    \\
    \begin{subfigure}[t]{0.02\textwidth}
        \textbf{(e)}    
    \end{subfigure}
    \begin{subfigure}[t]{0.47\linewidth}
        \includegraphics[width=0.9\linewidth,valign=t]{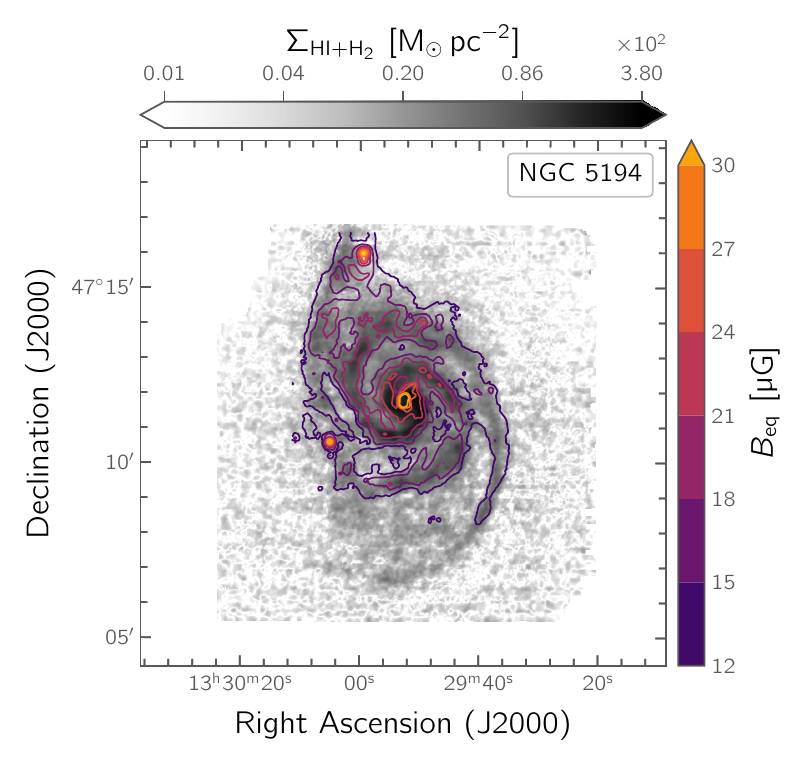}
    \end{subfigure}
    \begin{subfigure}[t]{0.02\textwidth}
        \textbf{(f)}    
    \end{subfigure}
    \begin{subfigure}[t]{0.47\linewidth}
        \vspace{0.7cm}
        \includegraphics[width=0.9\linewidth,valign=t]{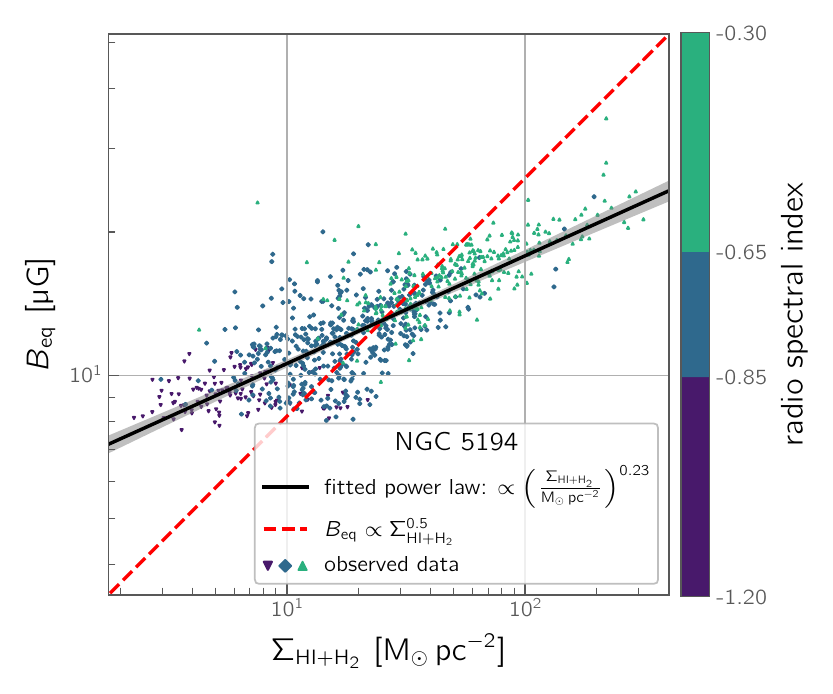}
    \end{subfigure}
    \caption{NGC 5194. \corr}
    \label{fig:n5194_corr}
\end{figure*}
\addcontentsline{toc}{subsection}{NGC 5194}

\begin{figure*}
	\centering
    \begin{subfigure}[t]{0.02\textwidth}
        \textbf{(a)}    
    \end{subfigure}
    \begin{subfigure}[t]{0.47\linewidth}
        \includegraphics[width=0.9\linewidth,valign=t]{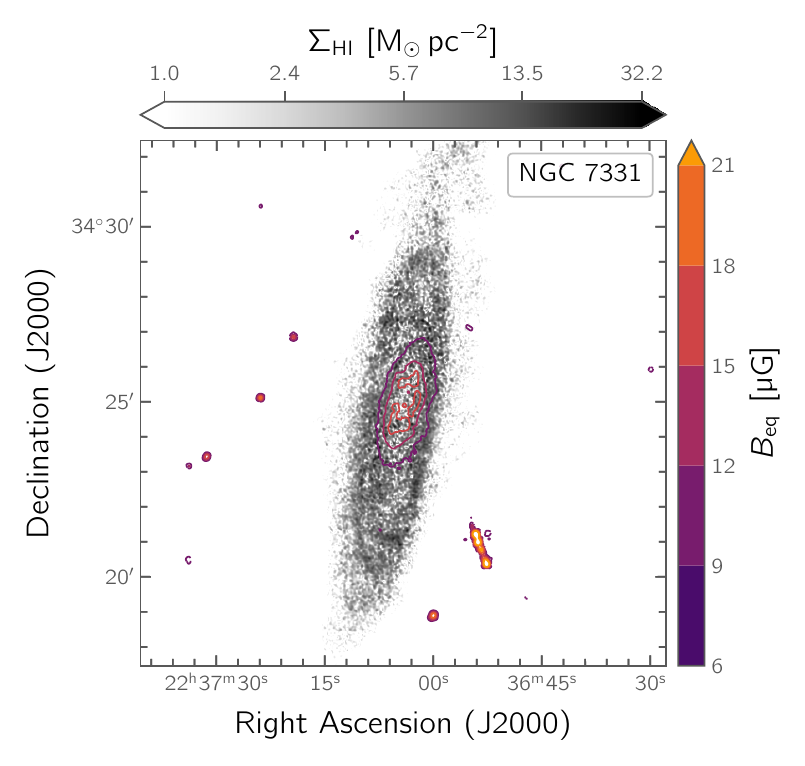}
    \end{subfigure}
    \begin{subfigure}[t]{0.02\textwidth}
        \textbf{(b)}    
    \end{subfigure}
    \begin{subfigure}[t]{0.47\linewidth}
        \vspace{0.7cm}
        \includegraphics[width=0.9\linewidth,valign=t]{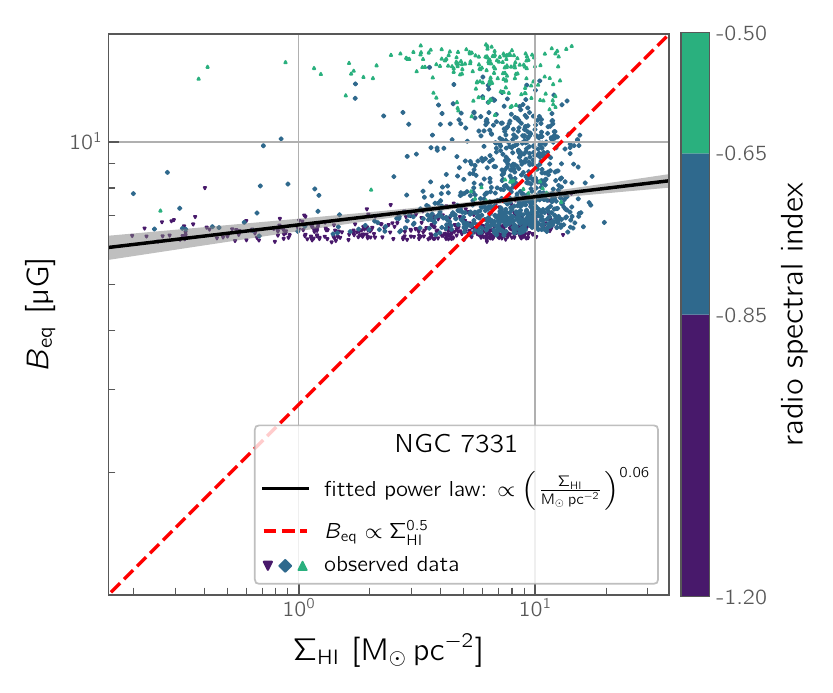}
    \end{subfigure}
    \\
    \begin{subfigure}[t]{0.02\textwidth}
        \textbf{(c)}    
    \end{subfigure}
    \begin{subfigure}[t]{0.47\linewidth}
        \includegraphics[width=0.9\linewidth,valign=t]{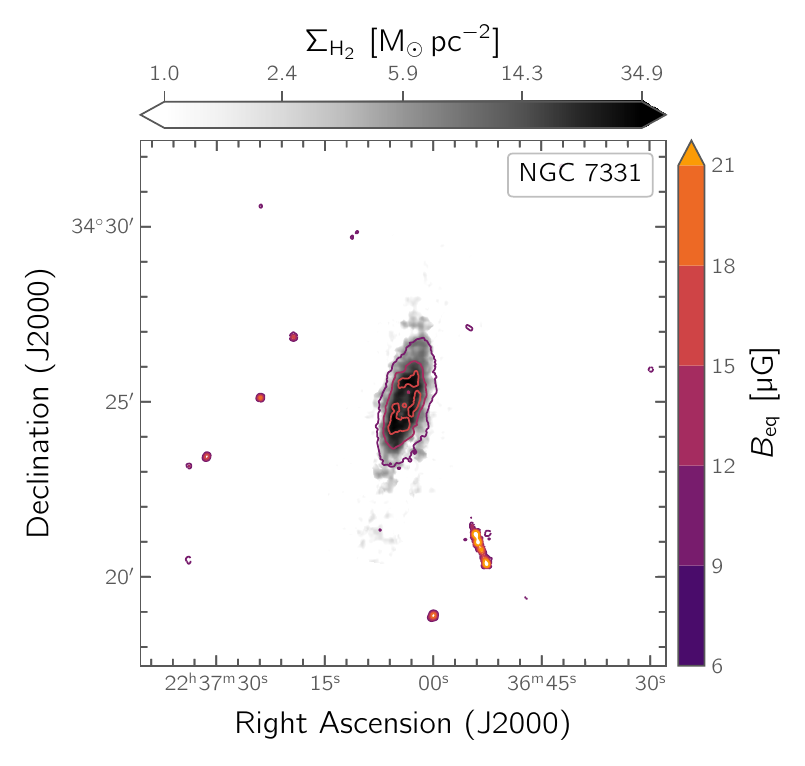}
    \end{subfigure}
    \begin{subfigure}[t]{0.02\textwidth}
        \textbf{(d)}    
    \end{subfigure}
    \begin{subfigure}[t]{0.47\linewidth}
        \vspace{0.7cm}
        \includegraphics[width=0.9\linewidth,valign=t]{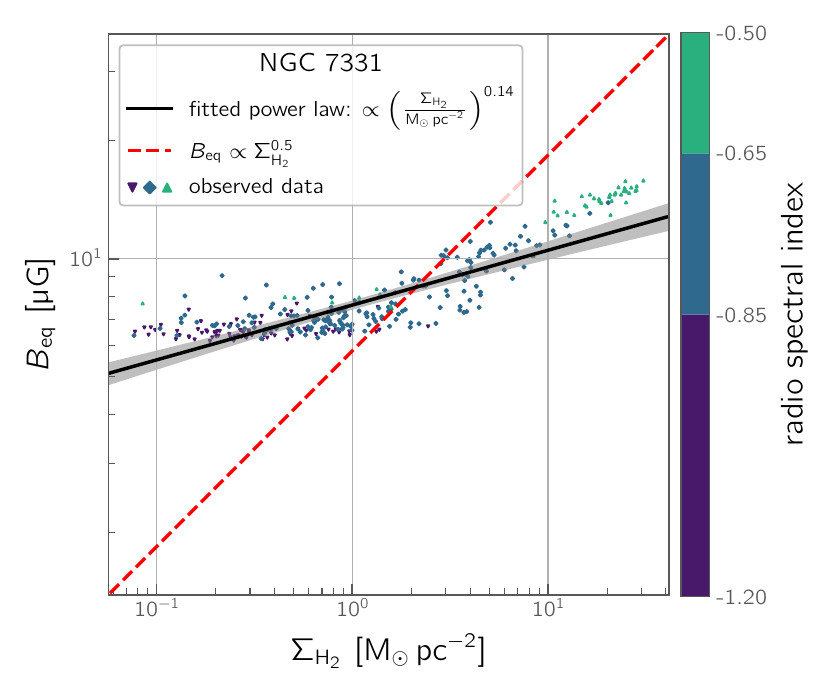}
    \end{subfigure}
    \\
    \begin{subfigure}[t]{0.02\textwidth}
        \textbf{(e)}    
    \end{subfigure}
    \begin{subfigure}[t]{0.47\linewidth}
        \includegraphics[width=0.9\linewidth,valign=t]{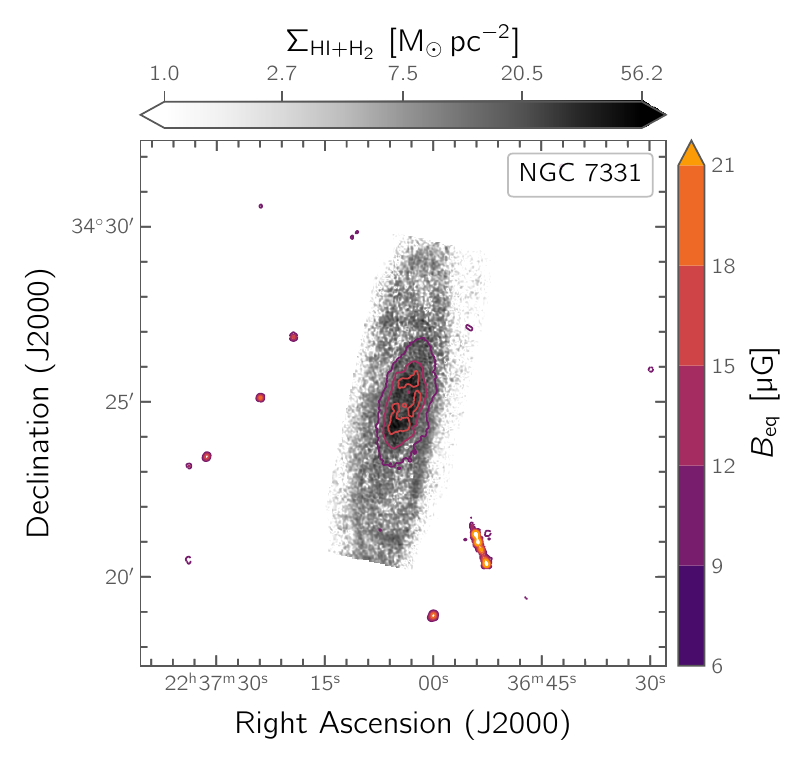}
    \end{subfigure}
    \begin{subfigure}[t]{0.02\textwidth}
        \textbf{(f)}    
    \end{subfigure}
    \begin{subfigure}[t]{0.47\linewidth}
        \vspace{0.7cm}
        \includegraphics[width=0.9\linewidth,valign=t]{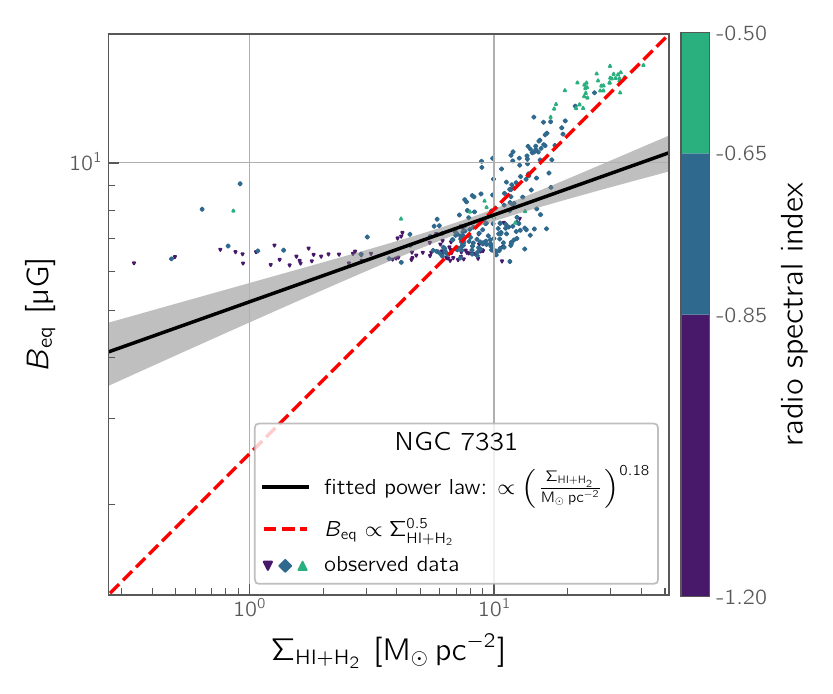}
    \end{subfigure}
    \caption{NGC 7331. \corr}
    \label{fig:n7331_corr}
\end{figure*}
\addcontentsline{toc}{subsection}{NGC 7331}

\section{Radial profiles of the magnetic field strengths}
\label{as:radial_profile}

To determine radial profiles of the magnetic field strength, we estimated the mean magnetic field within ellipsoidal annuli, where each annulus has a width of 6\,arcsec. Errors were estimated by adding in quadrature the mean magnetic field error and the error of the mean in each annulus. For the latter we divided the standard deviation of the magnetic field strength by the square-root of the number of beams. We disregarded pixels, where the magnetic field strength could not be determined. The position angle of the ellipses were taken from the literature, and the eccentricity of the ellipses were determined using the 3$\sigma$ contour lines from the 6-arcsec intensity maps \citep[see][for details]{Heesen2022}. These ellipses are hence bound by the ellipses presented already in the atlas of magnetic fields (Appendix~\ref{as:atlas_of_magnetic_fields_in_galaxies}).

\begin{figure*}
    \centering 
    \begin{subfigure}[t]{0.3\linewidth}
        \includegraphics[width=\linewidth,valign=t]{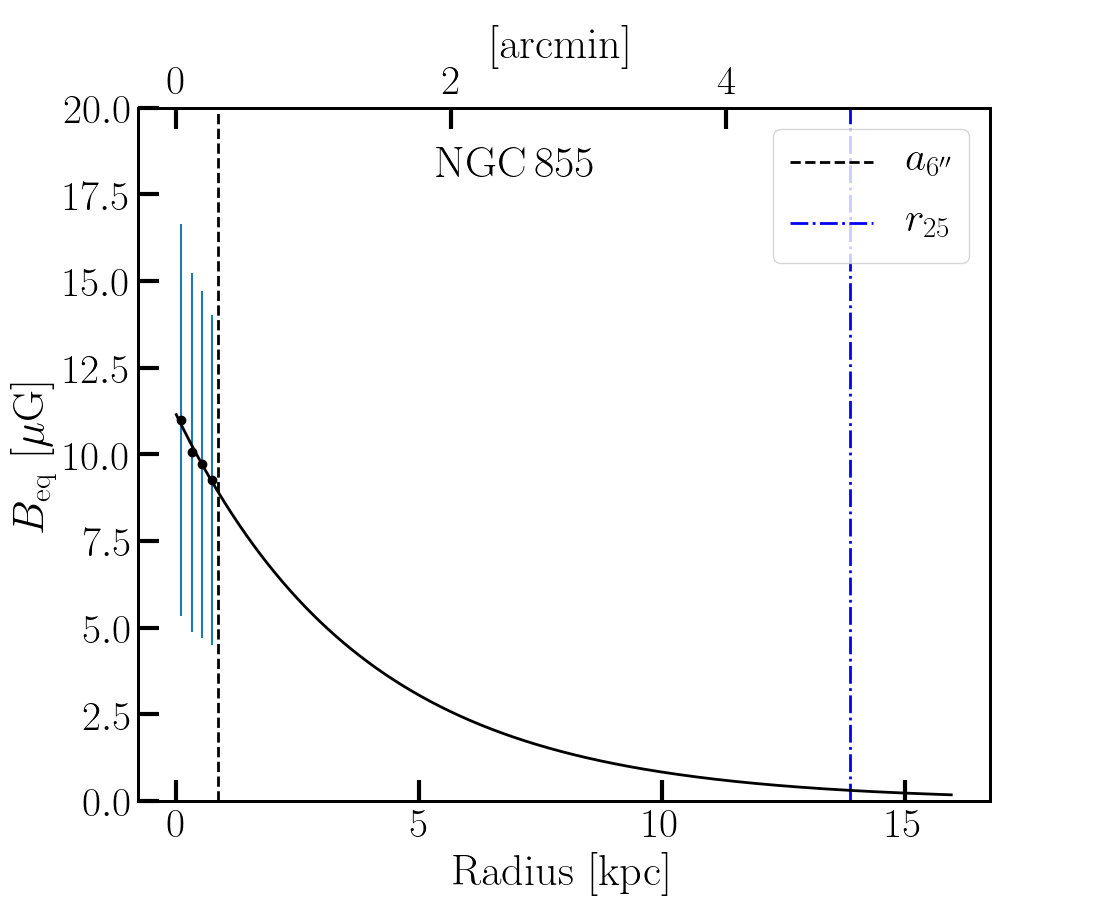}
    \end{subfigure}
    \begin{subfigure}[t]{0.3\linewidth}
        \includegraphics[width=\linewidth,valign=t]{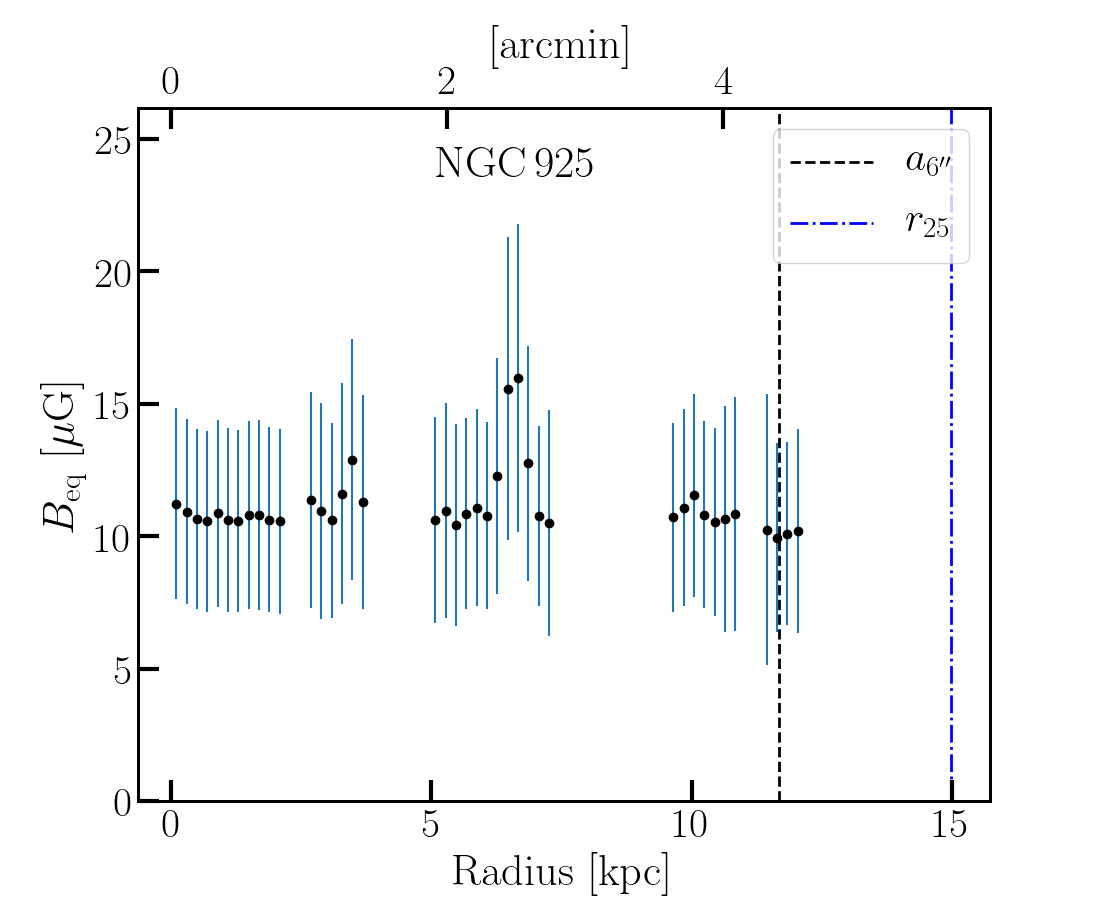}
    \end{subfigure}
    \begin{subfigure}[t]{0.3\linewidth}
        \includegraphics[width=\linewidth,valign=t]{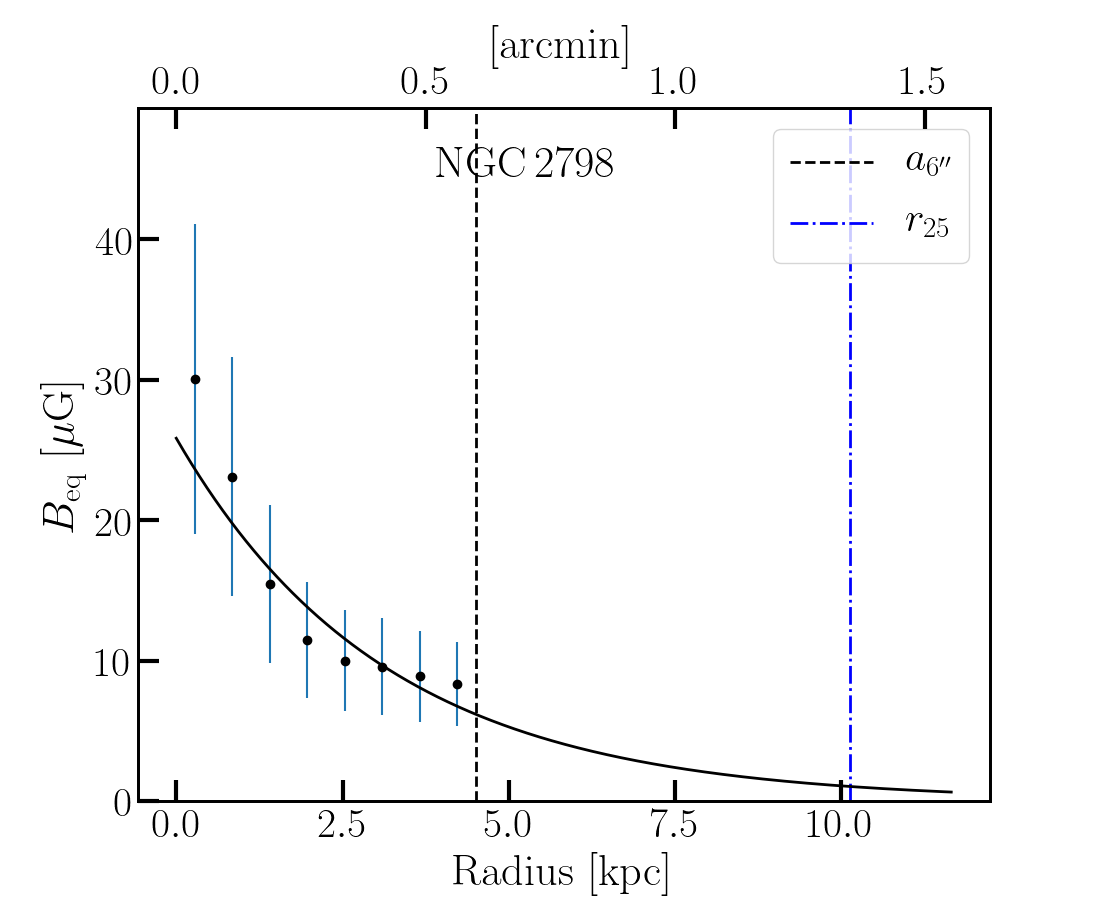}
    \end{subfigure}
    \\
    \begin{subfigure}[t]{0.3\linewidth}
        \includegraphics[width=\linewidth,valign=t]{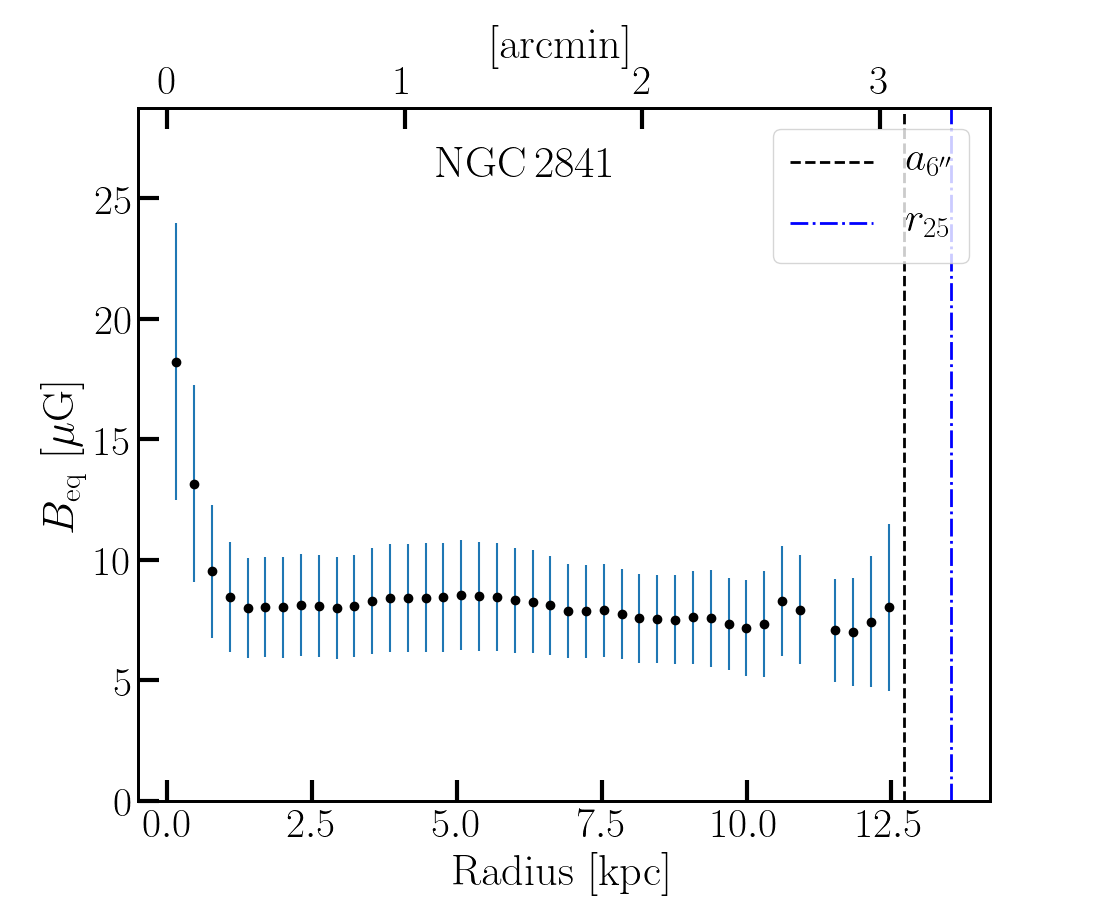}
    \end{subfigure}
    \begin{subfigure}[t]{0.3\linewidth}
        \includegraphics[width=\linewidth,valign=t]{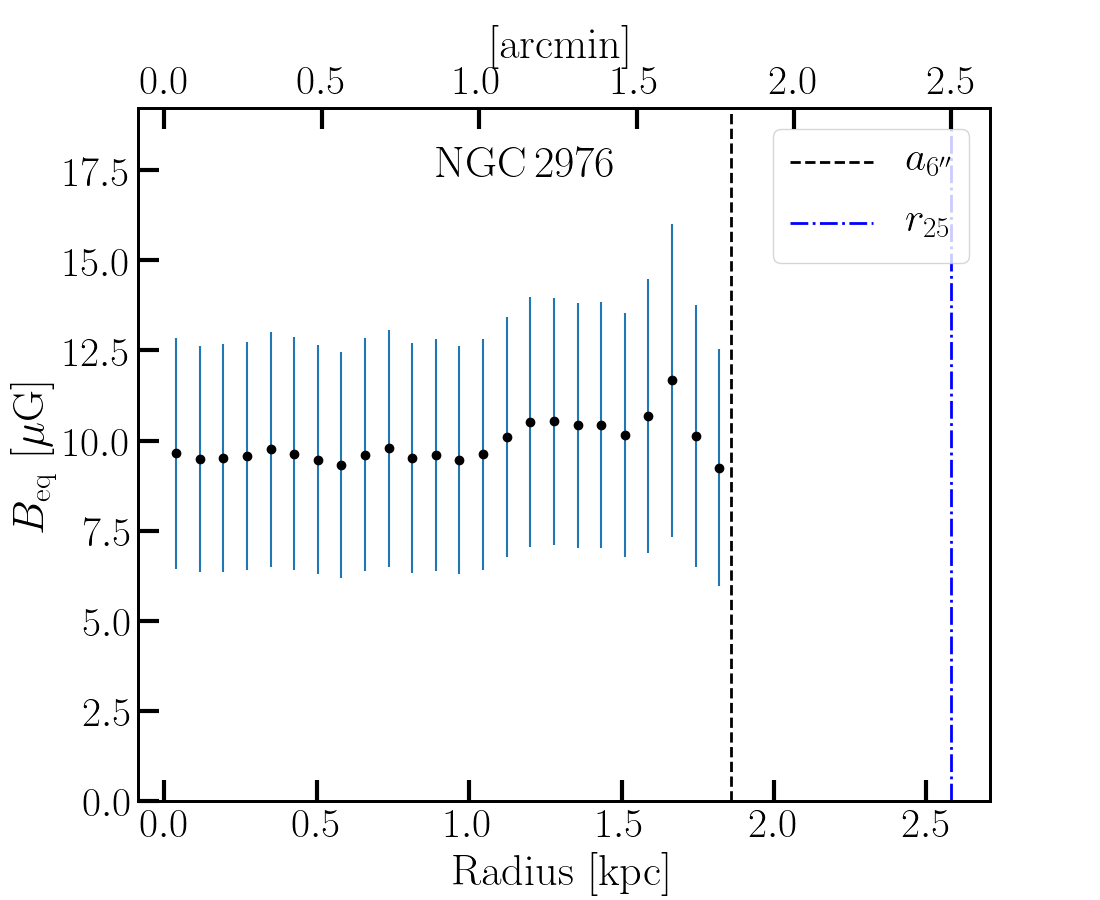}
    \end{subfigure}
    \begin{subfigure}[t]{0.3\linewidth}
        \includegraphics[width=\linewidth,valign=t]{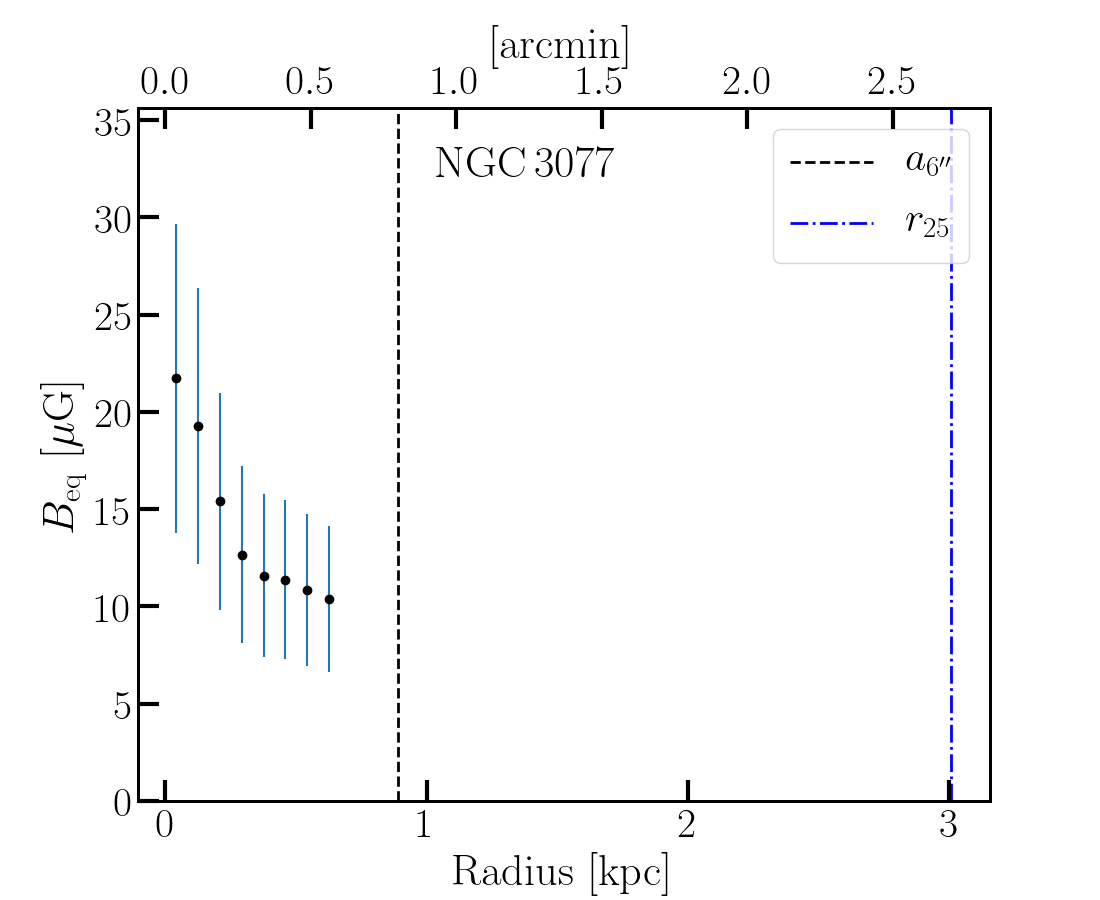}
    \end{subfigure}
    \\
    \begin{subfigure}[t]{0.3\linewidth}
        \includegraphics[width=\linewidth,valign=t]{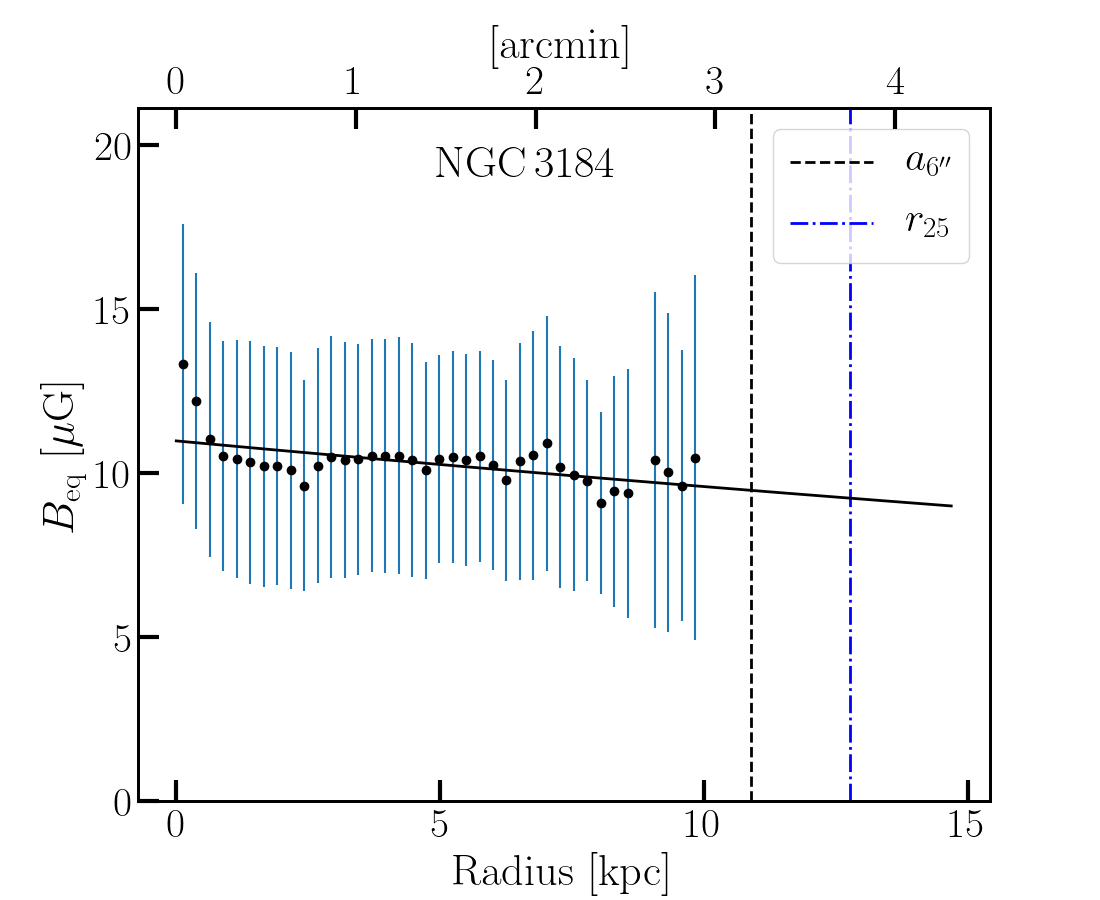}
    \end{subfigure}
    \begin{subfigure}[t]{0.3\linewidth}
        \includegraphics[width=\linewidth,valign=t]{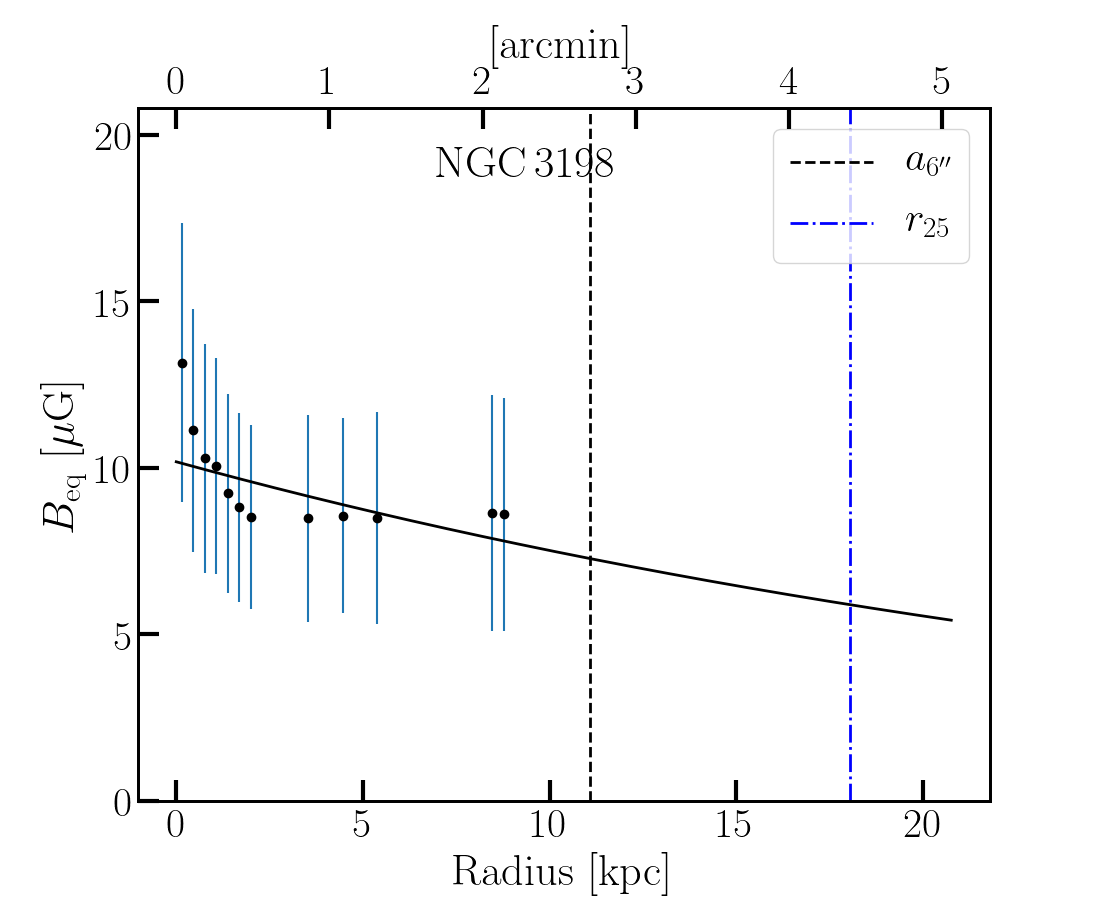}
    \end{subfigure}
    \begin{subfigure}[t]{0.3\linewidth}
        \includegraphics[width=\linewidth,valign=t]{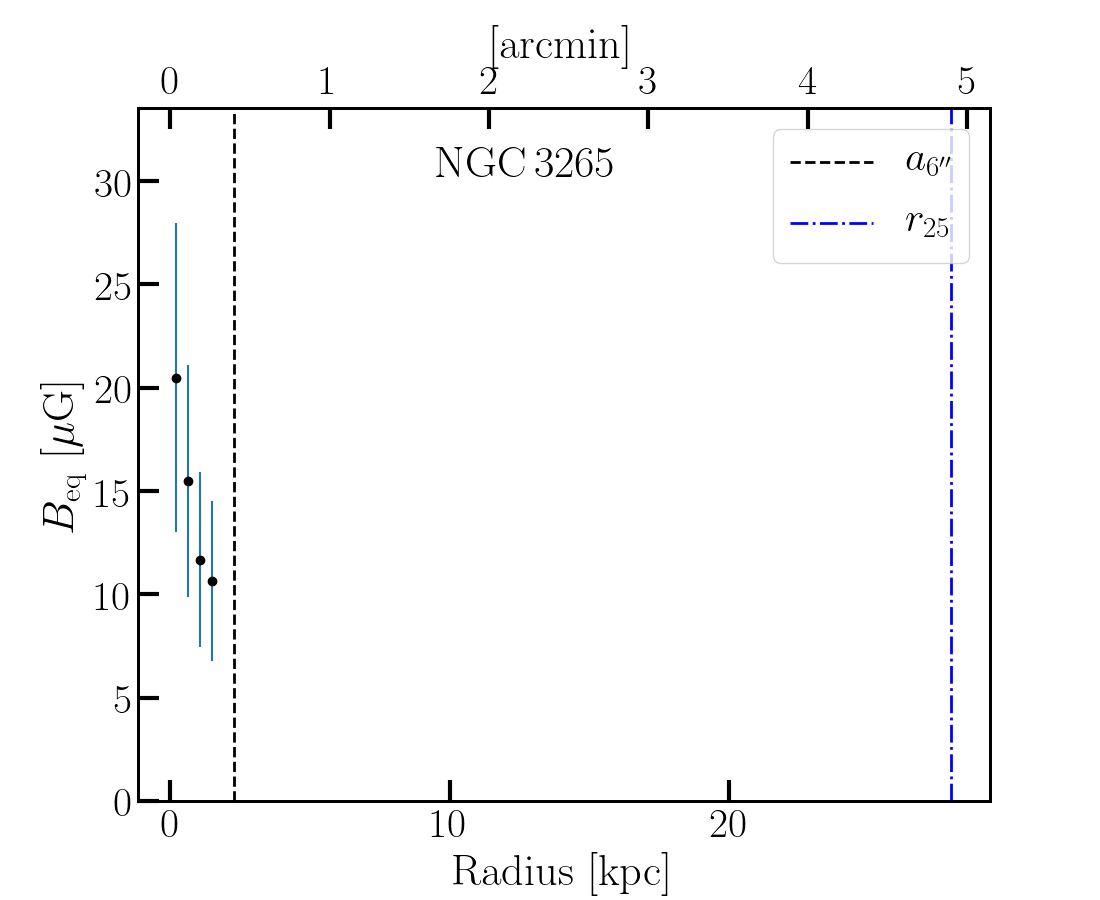}
    \end{subfigure}
    \\
    \begin{subfigure}[t]{0.3\linewidth}
        \includegraphics[width=\linewidth,valign=t]{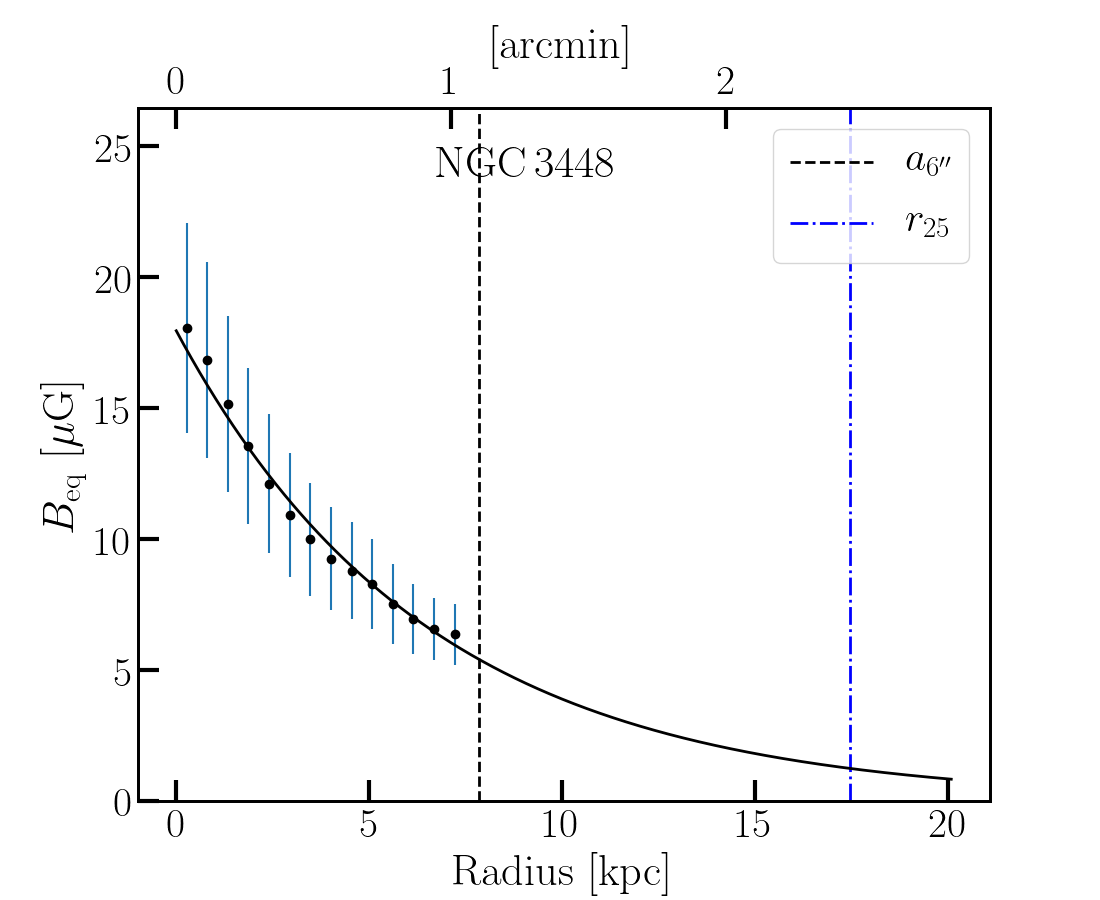}
    \end{subfigure}
    \begin{subfigure}[t]{0.3\linewidth}
        \includegraphics[width=\linewidth,valign=t]{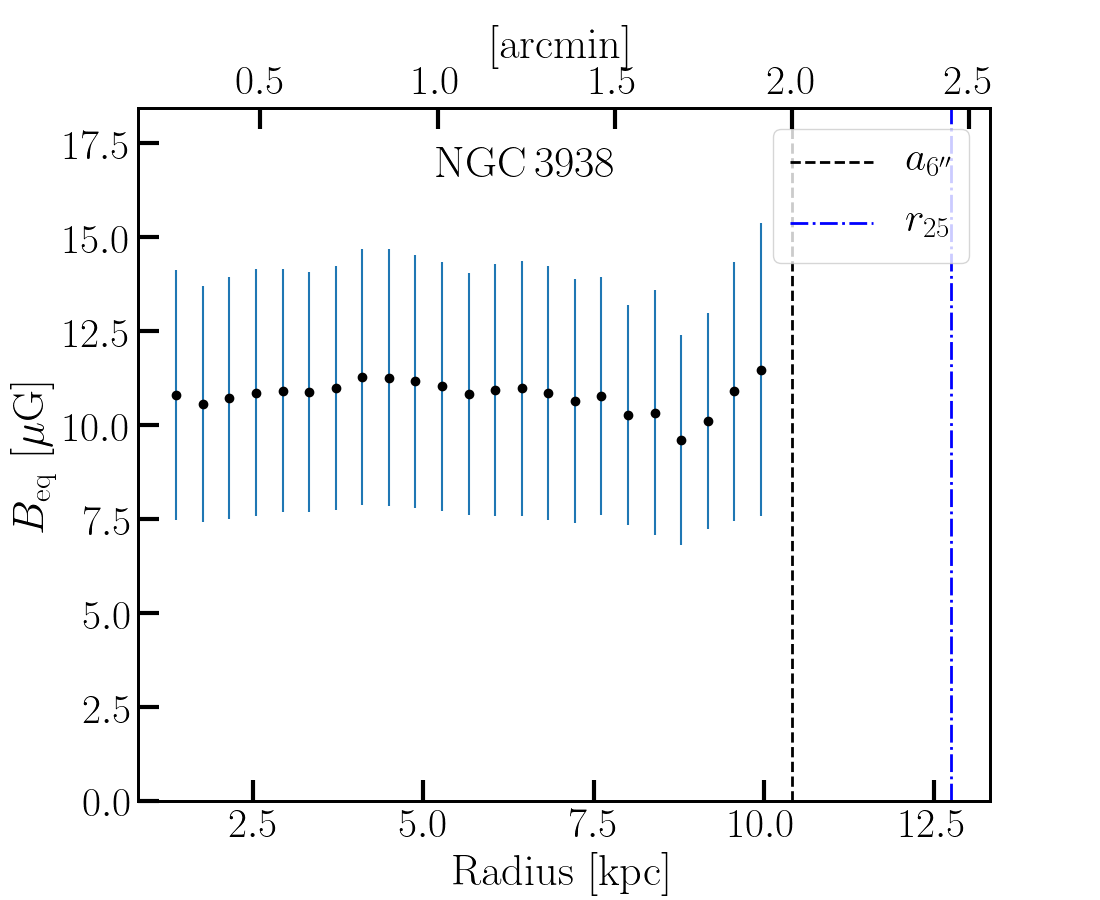}
    \end{subfigure}
    \begin{subfigure}[t]{0.3\linewidth}
        \includegraphics[width=\linewidth,valign=t]{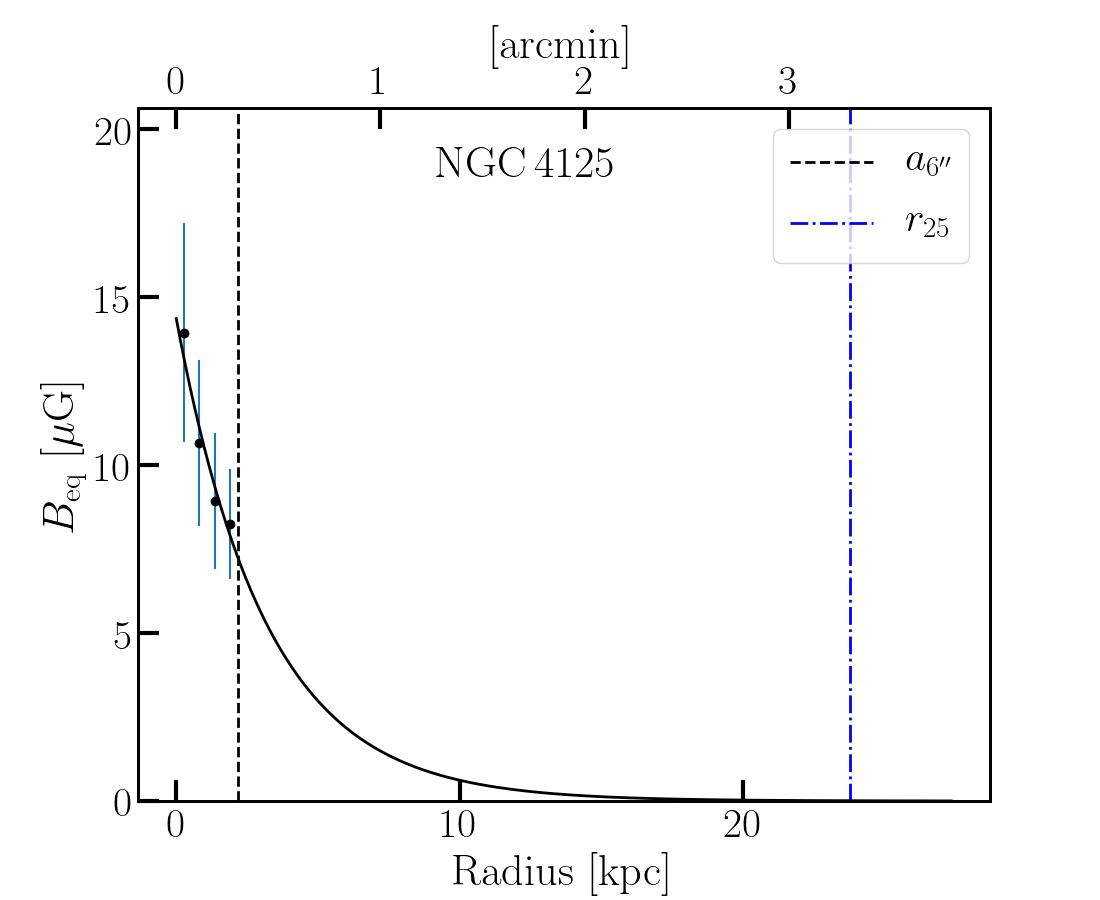}
    \end{subfigure}
    \\
     \begin{subfigure}[t]{0.3\linewidth}
        \includegraphics[width=\linewidth,valign=t]{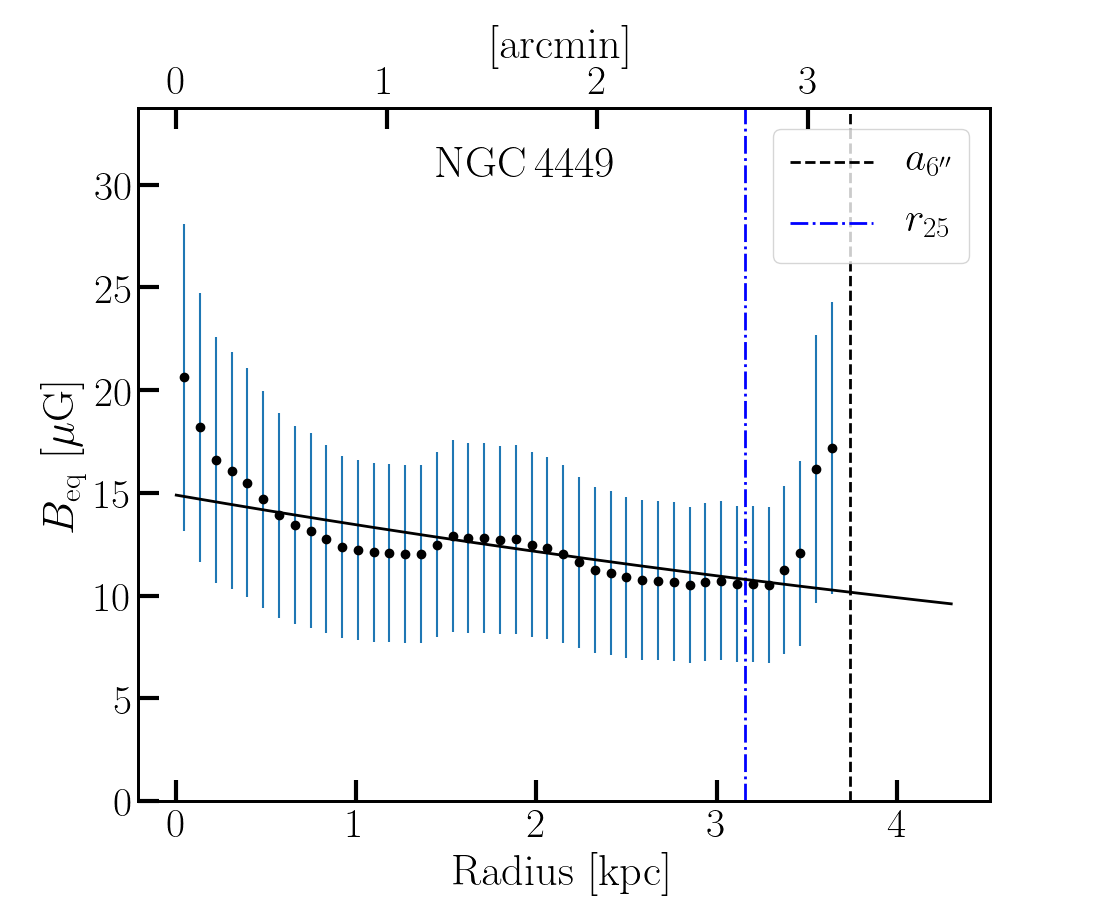}
    \end{subfigure}
    \begin{subfigure}[t]{0.3\linewidth}
        \includegraphics[width=\linewidth,valign=t]{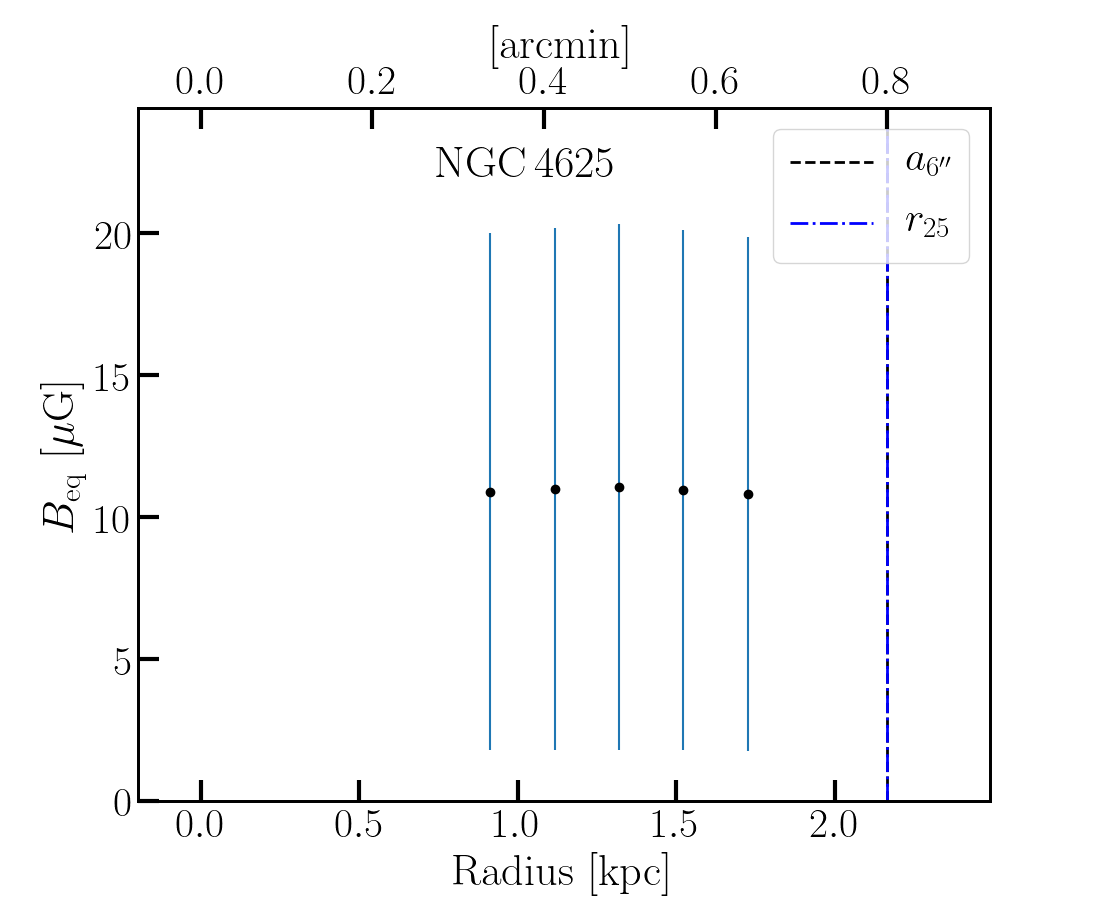}
    \end{subfigure}
    \begin{subfigure}[t]{0.3\linewidth}
        \includegraphics[width=\linewidth,valign=t]{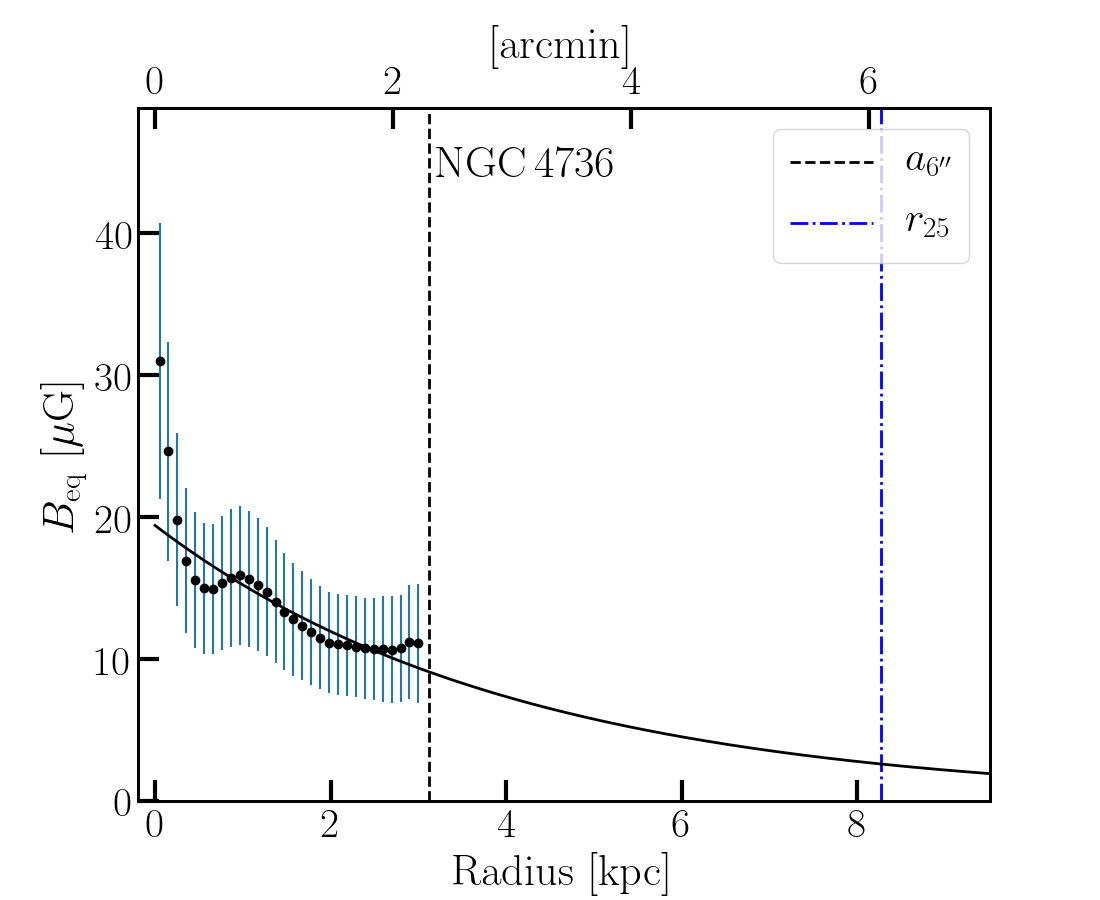}
    \end{subfigure}
    \caption{Radial profiles of the equipartition magnetic field strengths in our galaxy sample. Solid lines show best fitting exponential relations. Vertical lines show the extent of the radio disc ($r=a_{6\arcsec}$) and the extent of the optical stellar disc ($r=r_{25}$).}
\label{fig:radial_profile}
\end{figure*}

\newpage
\newpage

\begin{figure*}
    \centering 
    \begin{subfigure}[t]{0.3\linewidth}
        \includegraphics[width=\linewidth,valign=t]{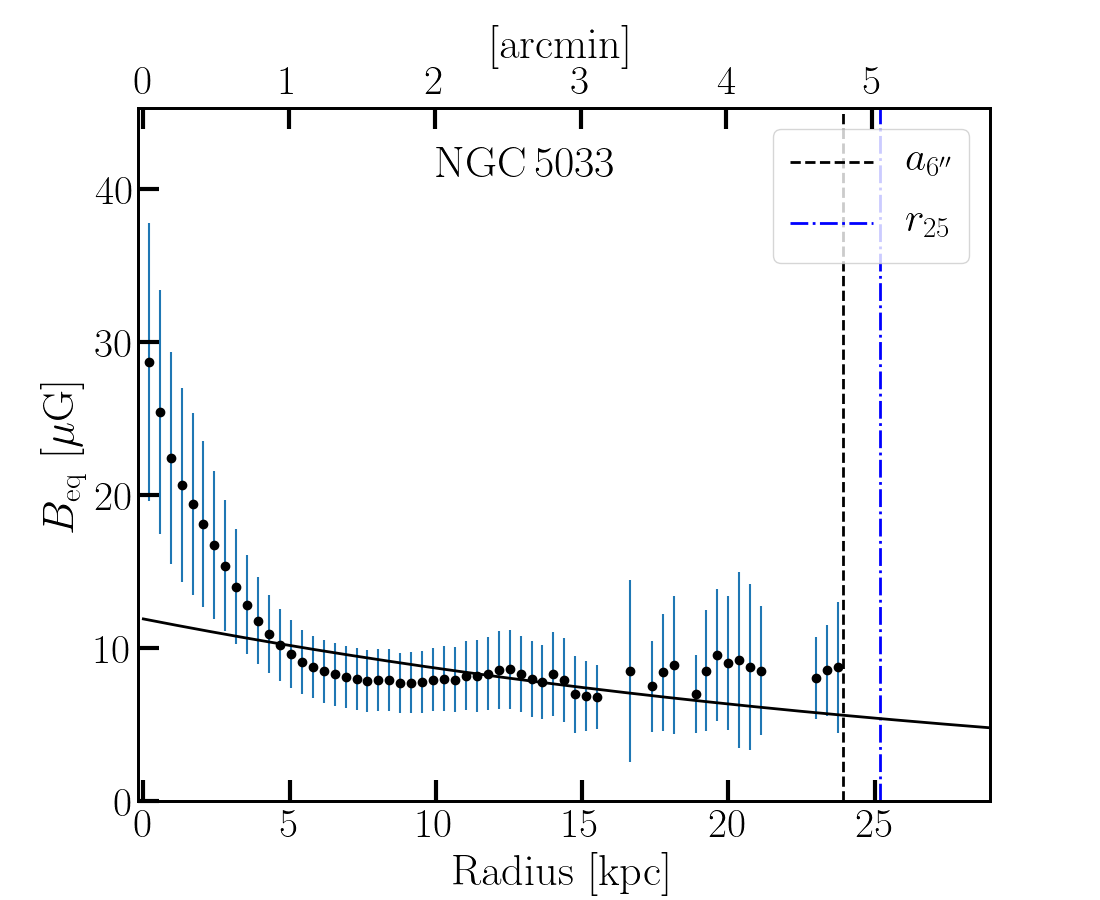}
    \end{subfigure}
    \begin{subfigure}[t]{0.3\linewidth}
        \includegraphics[width=\linewidth,valign=t]{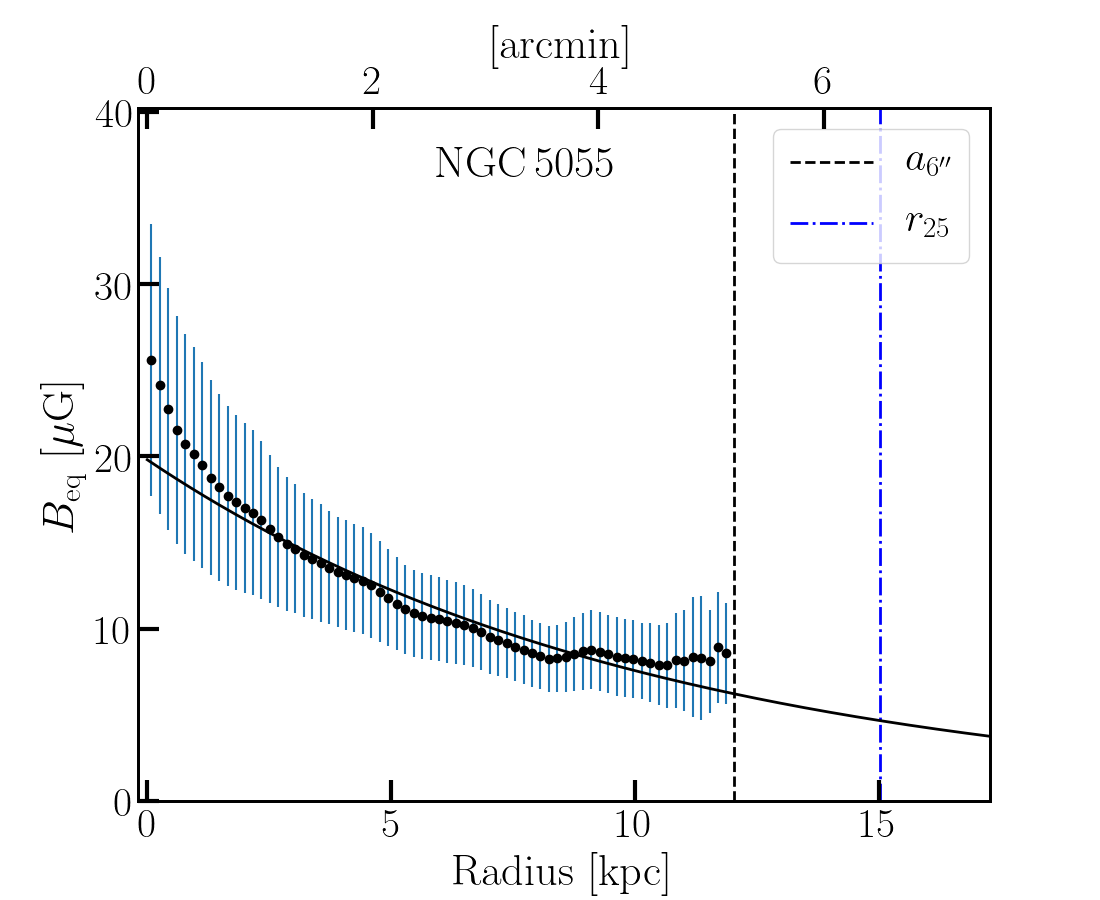}
    \end{subfigure}
    \begin{subfigure}[t]{0.3\linewidth}
        \includegraphics[width=\linewidth,valign=t]{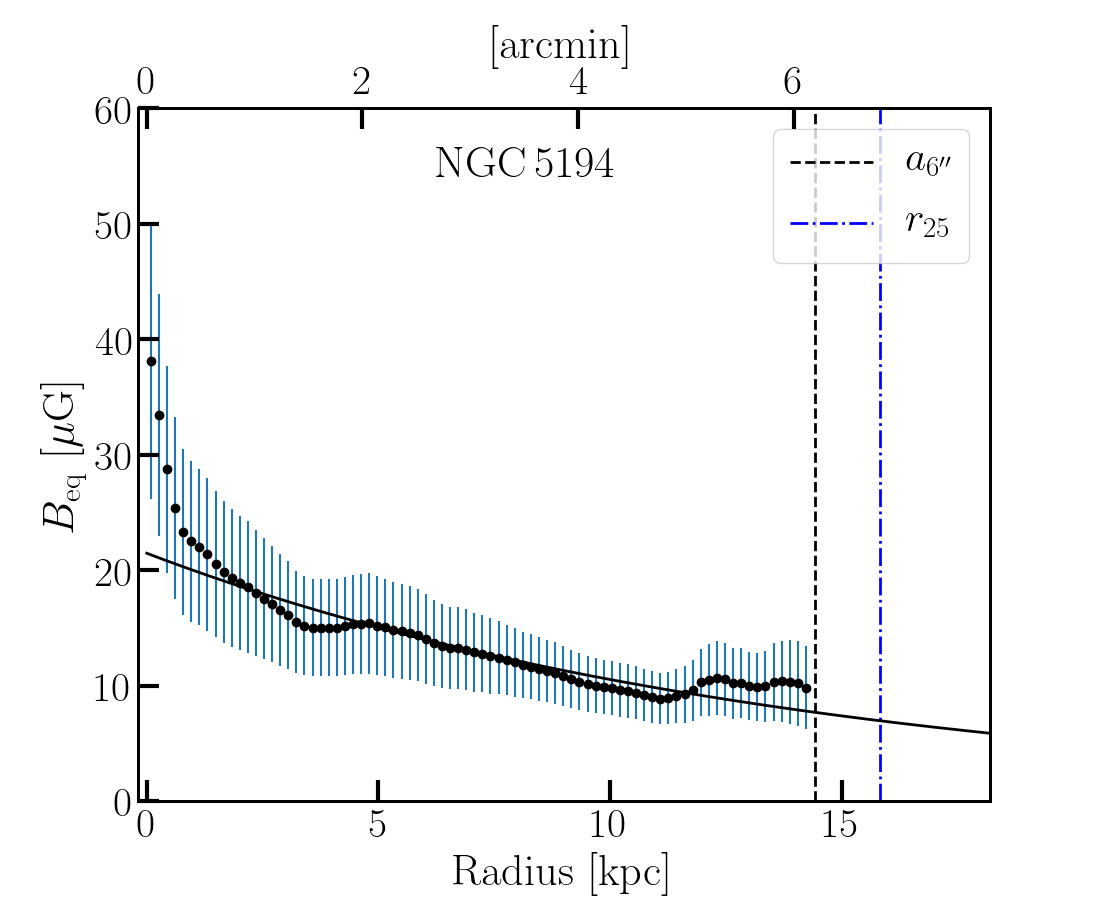}
    \end{subfigure}
    \\
    \begin{subfigure}[t]{0.3\linewidth}
        \includegraphics[width=\linewidth,valign=t]{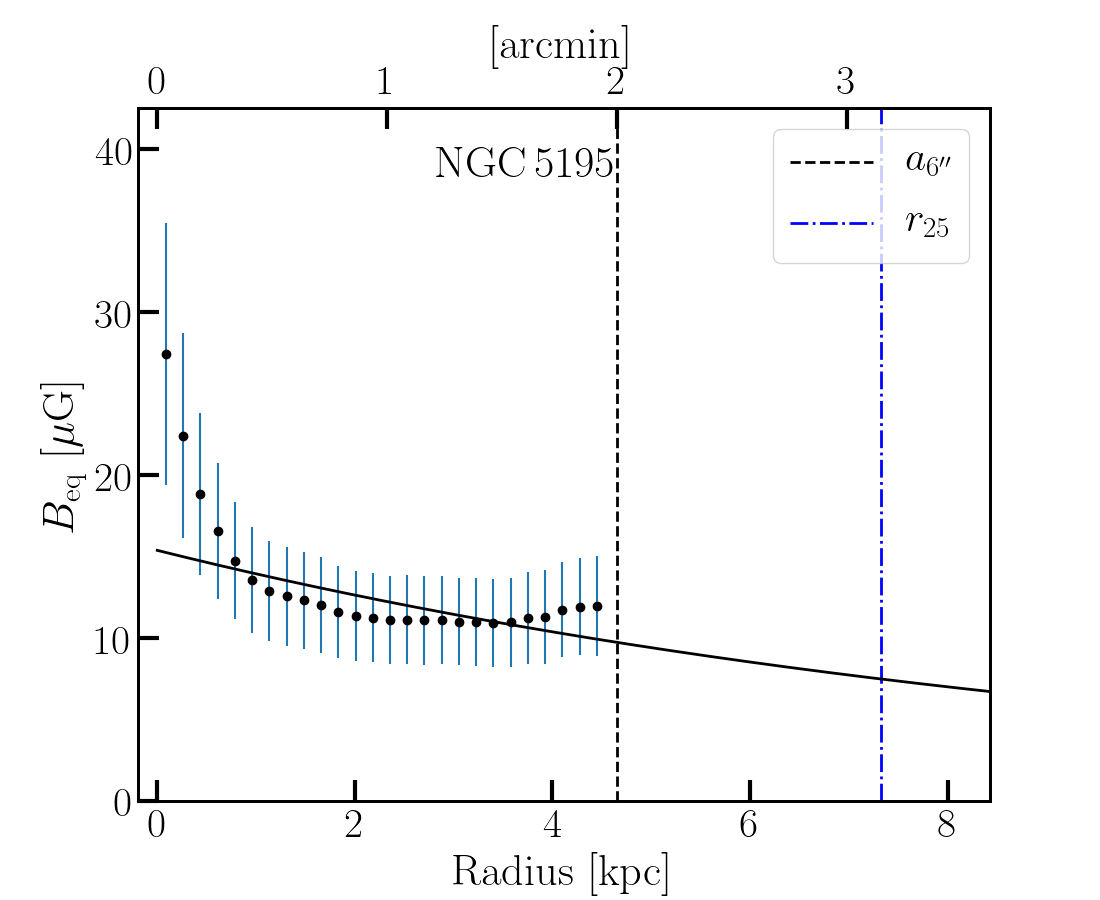}
    \end{subfigure}
    \begin{subfigure}[t]{0.3\linewidth}
        \includegraphics[width=\linewidth,valign=t]{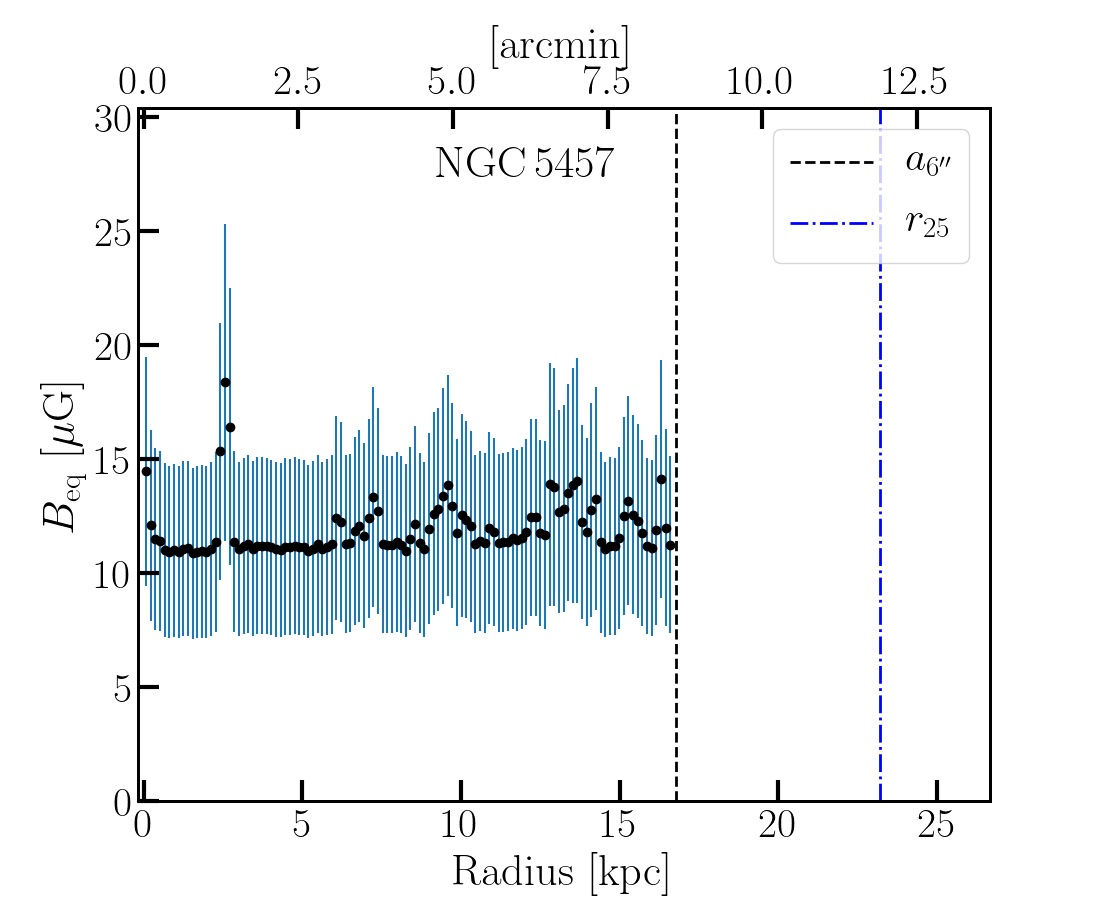}
    \end{subfigure}
    \begin{subfigure}[t]{0.3\linewidth}
        \includegraphics[width=\linewidth,valign=t]{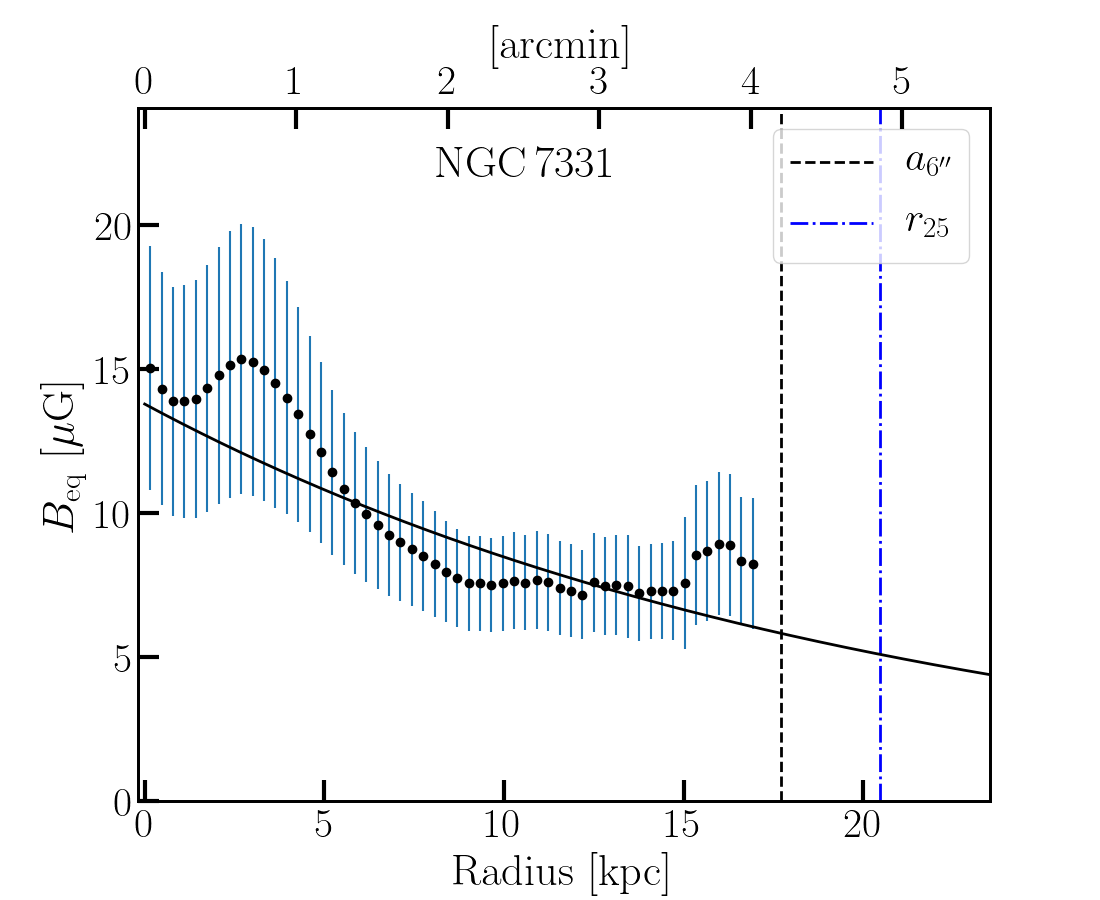}
    \end{subfigure}
    \caption{Radial profiles of the equipartition magnetic field strengths in our galaxy sample. Solid lines show best fitting exponential relations. Vertical lines show the extent of the radio disc ($r=a_{6\arcsec}$) and the extent of the optical stellar disc ($r=r_{25}$).}
\label{fig:radial_profile2}
\end{figure*}



}

\end{document}